\documentclass[10pt,a4paper]{amsart}
\usepackage[latin1]{inputenc}
\usepackage{graphicx}
\usepackage{mathrsfs}
\usepackage{dsfont}
\numberwithin{equation}{section}
\newtheorem{theorem}{Theorem}
\newtheorem{proposition}{Proposition}
\newtheorem{remark}{Remark}
\newtheorem{lemma}{Lemma}
\newtheorem{corollary}{Corollary}

\newcommand{\ii}{\infty}

\newcommand\R{{\ensuremath {\mathbb R} }}
\newcommand\C{{\ensuremath {\mathbb C} }}
\newcommand\N{{\ensuremath {\mathbb N} }}
\newcommand\Z{{\ensuremath {\mathbb Z} }}

\newcommand\1{{\ensuremath {\mathds 1} }}
\newcommand\cS{\mathcal{S}}
\renewcommand\phi{\varphi}

\newcommand{\gH}{\mathfrak{H}}
\newcommand{\gS}{\mathfrak{S}}

\newcommand{\wto}{\rightharpoonup}
\newcommand{\cP}{\mathcal{P}}
\newcommand{\cQ}{\mathcal{Q}}
\newcommand{\cC}{\mathcal{C}}

\newcommand{\cB}{\mathcal{B}}
\newcommand{\cK}{\mathcal{K}}
\newcommand{\cE}{\mathcal{E}}
\newcommand{\cF}{\mathcal{F}}
\newcommand{\cD}{\mathcal{D}}

\newcommand{\eps}{\epsilon}

\newcommand{\curv}{\mathscr{C}}
\newcommand\pscal[1]{{\ensuremath{\left\langle #1 \right\rangle}}}
\newcommand{\norm}[1]{ \left| \! \left| #1 \right| \! \right| }
\def\tr{{\rm Tr}}

\def\PPi{{\gamma^0_{\rm per}}}
\def\NN{\N}
\def\ZZ{\Z}
\def\RR{\R}

\newcommand \dps{\displaystyle }

\title[Local defects in periodic crystals]{A new approach to the modeling of local defects in crystals:\\ the reduced Hartree-Fock case}

\author[E. Cancès]{\'Eric CANC\`ES}
 \address{CERMICS, Ecole Nationale des Ponts et Chauss\'ees (Paris Tech) \& INRIA (Micmac Project), 6-8 Av. Pascal, 77455 Champs-sur-Marne, France.}
  \email{cances@cermics.enpc.fr}

\author[A. Deleurence]{Amélie DELEURENCE}
 \address{CERMICS, Ecole Nationale des Ponts et Chauss\'ees (Paris Tech) \& INRIA (Micmac Project), 6-8 Av. Pascal, 77455 Champs-sur-Marne, France.}
  \email{deleurence@cermics.enpc.fr}

\author[M. Lewin]{Mathieu LEWIN}
 \address{CNRS \& Laboratoire de Mathématiques UMR 8088, Université de Cergy-Pontoise, 2 Avenue Adolphe Chauvin, 95302 Cergy-Pontoise Cedex, France.}
  \email{Mathieu.Lewin@math.cnrs.fr}

\date{January 9, 2008. Final version to appear in \textit{Commun. Math. Phys.}}

\begin{document}
\maketitle
\begin{abstract}
This article is concerned with the derivation and the mathematical study of a new mean-field model for the description of interacting electrons in crystals with local defects. We work with a reduced Hartree-Fock model, obtained from the usual Hartree-Fock model by neglecting the exchange term.

First, we recall the definition of the self-consistent Fermi sea of the perfect crystal, which is obtained as a minimizer of some periodic problem, as was shown by Catto, Le Bris and Lions. We also prove some of its properties which were not mentioned before.

Then, we define and study in detail a nonlinear model for the electrons of the crystal in the presence of a defect. We use formal analogies between the Fermi sea of a perturbed crystal and the Dirac sea in Quantum Electrodynamics in the presence of an external electrostatic field. The latter was recently studied by Hainzl, Lewin, Séré and Solovej, based on ideas from Chaix and Iracane. This enables us to define the ground state of the self-consistent Fermi sea in the presence of a defect.

We end the paper by proving that our model is in fact the thermodynamic limit of the so-called supercell model, widely used in numerical simulations.
\end{abstract}


Describing the electronic state of crystals with local defects is a major issue in
solid-state physics, materials science and nano-electronics~\cite{Pisani,Kittel,Stoneham}.

In this article, we develop a theory based on formal analogies between the Fermi sea of a perturbed crystal and the polarized Dirac sea in Quantum Electrodynamics in the presence of an external electrostatic field. Recently, the latter model was extensively studied by Hainzl, Lewin, Séré and Solovej in the Hartree-Fock approximation \cite{HLS1,HLS2,HLSo,HLS3}, based on ideas from Chaix and Iracane \cite{CI} (see also \cite{CIL,BBHS}). This was summarized in the review \cite{HLSS}. Using and adapting these methods, we are able to propose a new mathematical approach for the self-consistent description of a crystal in the presence of local defects. 

We focus in this article on the {\it reduced Hartree-Fock} (rHF) model in which the so-called \emph{exchange term} is neglected.
To further simplify the mathematical formulas, we do not explicitly take the spin variable into account and
we assume that the host crystal is cubic with a single atom of charge $Z$ per unit
cell. The arguments below can be easily extended to the general case. 

In the whole paper, the main object of interest will be the so-called
\emph{density matrix} of the electrons. This is a self-adjoint operator
$0\leq\gamma\leq1$ acting on the one-body space $L^2(\R^3)$. When $\gamma$ has a
finite rank, it models a finite number of electrons. In the periodic
case, the ground state density matrix $\PPi$ has an infinite rank (it
describes infinitely many electrons) and 
commutes with the translations of the lattice. We will see in the sequel
that the ground state density matrix of a crystal with a local defect
can be written as $\gamma = \PPi + Q$, where $Q$ is a compact
perturbation of the periodic density matrix $\PPi$ of the reference
perfect crystal.

In each of the above three cases (finite number of electrons, perfect
crystal, defective crystal), the ground state density matrix can be
obtained by minimizing some nonlinear energy functional depending on a
set of admissible density matrices. In the case of a crystal with a
local defect, the perturbation $Q$ is a minimizer of some nonlinear
minimization problem set in the whole space $\R^3$, with a possible lack
of compactness at infinity. The main unusual feature compared to
standard variational problems is that $Q$ is a self-adjoint operator of
infinite rank. This was already the case in \cite{HLS1,HLS2,HLS3,HLSo}. 

The paper is organized as follows. In Section 1, we recall the
definition of the reduced Hartree-Fock model for a \emph{finite} number of
electrons, which serves as a basis for the theories of infinitely many
electrons in a (possibly perturbed) periodic nuclear distribution.  
Section 2 is devoted to the definition of the model for the
infinite periodic crystal, following mainly \cite{CLL-CRAS,CLL} (but we provide
some additional material compared to what was done in \cite{CLL-CRAS,CLL}). In
Section~3, we define a model for the crystal with local defects which takes
the perfect crystal as reference. In Section~4, we prove that this model
is the thermodynamic limit of the supercell model. 

For the convenience of the reader, we have gathered all the proofs in
Section~5. Often, the proofs follow the same lines as those in 
\cite{HLS1,HLS2,HLS3,HLSo} and we shall not detail identical
arguments. But there are many difficulties associated with the
particular model under study which do not appear in previous
works and which are addressed in detail here.

\section{The reduced Hartree-Fock model for $N$ electrons}\label{sec_rHF}
We start by recalling the definition of the reduced Hartree-Fock model
\cite{rHF} for a \emph{finite} number of electrons. Note that the \emph{reduced} Hartree-Fock model should not be confused with the \emph{restricted} Hartree-Fock model commonly used in numerical simulations (see e.g.~\cite{Dovesi}). We consider a system containing
$N$ nonrelativistic quantum electrons and a set of nuclei having a
density of charge $\rho_{\rm nuc}$. If for instance there are $K$
nuclei of charges $z_1,...,z_K \in \NN \setminus \left\{0\right\}$
located at $R_1,...,R_K\in\R^3$, then  
$$
\rho_{\rm nuc}(x):=\sum_{k=1}^K z_k \, m_k(x-R_k),
$$
where $m_1,...,m_K$ are positive measures on $\R^3$ of total mass
one. Point-like nuclei would correspond to $m_k=\delta$ (the Dirac measure) but for
convenience we shall deal with smeared nuclei in the sequel, i.e. we
assume that for all $k=1...K$, $m_k$ is a smooth nonnegative function
such that $\int_{\RR^3} m_k = 1$. The technical difficulties arising with point-like nuclei will be dealt with elsewhere.

The energy of the whole system in the reduced Hartree-Fock model reads \cite{rHF,CLL}
\begin{equation}
\mathcal{ E}^{\rm rHF}_{\rho_{\rm nuc}}(\gamma) = \tr \left( -
  \frac 1 2 \Delta \gamma \right) 
+ \frac 1 2 D \left( \rho_\gamma-\rho_{\rm nuc} , \rho_\gamma-\rho_{\rm nuc} \right).
\label{def_rHF}
\end{equation}
We have chosen a system of units such that $\hbar=m=e=\frac
1{4\pi\epsilon_0}=1$ where $m$ and $e$ are respectively the mass and the
charge of an electron, $\hbar$ is the reduced Planck constant and
 $\epsilon_0$ is the dielectric permittivity of the vacuum.
The first term in the right-hand side of \eqref{def_rHF} is the kinetic energy of the electrons
and $D(\cdot,\cdot)$ is the classical Coulomb interaction, which
reads for $f$ and $g$ in $L^{6/5}(\RR^3)$ as
\begin{equation}
D(f,g) = \int_{\RR^3} \int_{\RR^3} \frac{f(x) \, g(y)}{|x-y|}dx \, dy=4\pi\int_{\R^3}\frac{\overline{\widehat{f}(k)}\widehat{g}(k)}{|k|^2}dk.
\label{def_D_f_g} 
\end{equation}
where $\widehat{f}$ denotes the Fourier transform of $f$. In this mean-field model, the state of the $N$ electrons is described by the one-body density matrix $\gamma$, which is an element of the following class
\begin{equation*}
\mathcal{ P}^{N} = \bigg\{ \gamma \in \cS(L^2(\RR^3)) \; | \; \ 0 \le \gamma
  \le 1, \ \tr(\gamma) = N, \ \tr\left(\sqrt{-\Delta} \gamma\sqrt{-\Delta}\right) < \infty \bigg\}.
\end{equation*}
Here and below, $\cS(\gH)$ denotes the space of bounded self-adjoint operators acting on the Hilbert space $\gH$. Also we define $\tr((-\Delta)\gamma):=\tr(\sqrt{-\Delta}\gamma\sqrt{-\Delta})$ which makes sense when $\gamma\in\cP^N$.
The set $\cP^N$ is the closed convex hull of the set of orthogonal projectors of rank $N$ acting on $L^2(\R^3)$ and having a finite kinetic energy. Each such projector $\gamma=\sum_{i=1}^N|\phi_i\rangle\langle\phi_i|$ is the density matrix of a Hartree-Fock state
\begin{equation}
\Psi=\phi_1\wedge\cdots\wedge\phi_N
\label{Slater}
\end{equation}
in the usual $N$-body space of fermionic wavefunctions with finite kinetic energy $\bigwedge_{i=1}^NH^1(\R^3)$. 

The function $\rho_\gamma$ appearing in \eqref{def_rHF} is the density associated with the operator~$\gamma$, defined by $\rho_\gamma(x) = \gamma(x,x)$ where $\gamma(x,y)$ is the kernel of the trace class operator 
$\gamma$. Notice that for all $\gamma \in \mathcal{ P}^N$, one has $\rho_\gamma\geq0$ and $\sqrt{\rho_\gamma}\in H^1(\RR^3)$, hence the last term of \eqref{def_rHF} is well-defined, since $\rho_\gamma\in L^1(\RR^3) \cap L^3(\RR^3)\subset L^{6/5}(\R^3)$.  

It can be proved (see the appendix of \cite{rHF}) that if $N \le \sum_{k=1}^M z_k $ (neutral
or positively charged 
systems), the variational problem 
\begin{equation}
I_{\rm rHF}(\rho_{\rm nuc},N) = \inf \left\{ \mathcal{ E}^{\rm rHF}_{\rho_{\rm nuc}}(\gamma),
  \quad \gamma \in \mathcal{ P}^N \right\} 
\label{def_min_rHF}
\end{equation}
has a minimizer $\gamma$ and that the corresponding minimizing density $\rho_\gamma$ is
unique. 

The Hartree-Fock model \cite{LS} is the variational
approximation of the time-independent Schrödinger equation obtained by
restricting the set of fermionic wavefunctions under consideration to the subset of
functions of the form \eqref{Slater}. The HF functional reads
\begin{equation}
\mathcal{ E}^{\rm HF}_{\rho_{\rm nuc}}(\gamma) = \mathcal{ E}^{\rm rHF}_{\rho_{\rm nuc}}(\gamma)-\frac12\iint_{\R^6}\frac{|\gamma(x,y)|^2}{|x-y|}dx\,dy,
\label{def_HF}
\end{equation}
the last term being called the \emph{exchange energy}. As the
Hartree-Fock energy functional
is nonconvex, there is little hope to obtain rigorous thermodynamic
limits in this setting, at least with current state-of-the-art techniques.
For this reason, the exchange term is often neglected in mathematical studies.

\section{The reduced Hartree-Fock model for a perfect crystal}\label{sec_per}
In this article, we clamp
the nuclei on a periodic lattice, optimizing only over the state of the
electrons. More precisely we are interested in the change of the
electronic state of the crystal when a local defect is
introduced. To this end, we shall rely heavily on the rHF model for the
infinite perfect crystal (with no defect) which was studied by Catto, Le Bris and Lions in \cite{CLL-CRAS,CLL}. The latter can be obtained as the thermodynamical limit of the rHF model for finite systems which was introduced in the previous section. This will be explained in Section~\ref{thermo-lim} below.

\medskip

Let $\Gamma =
[-1/2,1/2)^3$ be the unit cell. We denote by 
$\Gamma^\ast = [-\pi,\pi)^3$ the first Brillouin zone of the lattice,
and by $\tau_k$ the translation operator on $L^2_{\rm loc}(\RR^3)$
defined by $\tau_k u (x) = u(x-k)$. We then introduce
\begin{multline*}
\mathcal{ P}_{\rm per}= 
\bigg\{ \gamma \in {\cS}(L^2(\RR^3)) \; | \; 0 \le \gamma \le 1, \ 
\forall k \in \ZZ^3,\ \tau_k \gamma = \gamma \tau_k, \\ 
\int_{\Gamma^\ast}
\tr_{L^2_\xi(\Gamma)}((1-\Delta_\xi)^{1/2} \gamma_\xi(1-\Delta_\xi)^{1/2}) \, d\xi < \infty  \; \bigg\}
\end{multline*}
where $(\gamma_\xi)_{\xi \in \Gamma^\ast}$ is the Bloch waves decomposition of
$\gamma$, see \cite{RS,CLL}:
$$
\gamma = \frac{1}{(2\pi)^3}  \int_{\Gamma^\ast} \gamma_\xi \, d\xi,\qquad \gamma_\xi \in \mathcal{ S}(L^2_\xi(\Gamma)),
$$
$$L^2_\xi(\Gamma) = \left\{ u \in L^2_{\rm 
  loc}(\RR^3)\ |\ \tau_k u = e^{-i k \cdot \xi} u, \; \forall k \in \ZZ^3 \right\}$$
which corresponds to the decomposition in fibers $L^2(\R^3) =
\int^{\oplus}_{\Gamma^*} L^2_\xi(\Gamma)d\xi$. 
For any $\gamma\in\cP_{\rm per}$, we denote by $\gamma_\xi(x,y)$ the
integral kernel of $\gamma_\xi$. The density of $\gamma$ is
then the nonnegative $\Z^3$-periodic function of $L^1_{\rm loc}(\R^3)
\cap L^3_{\rm loc}(\R^3)$ defined as 
$$\rho_\gamma(x):=\frac{1}{(2\pi)^3}\int_{\Gamma^*} \gamma_\xi(x,x) \, d\xi.$$
Notice that for any $\gamma\in\cP_{\rm per}$
$$\int_\Gamma\rho_\gamma(x)dx=\frac{1}{(2\pi)^3}  \int_{\Gamma^\ast}
\tr_{L^2_\xi(\Gamma)}(\gamma_\xi) \, d\xi,$$
i.e. this gives the number of electrons per unit cell. Later we shall add the constraint that the system is neutral and restrict to states $\gamma\in\cP_{\rm per}$ satisfying
$$\int_\Gamma\rho_\gamma(x)dx=Z$$
where $Z$ is the total charge of the nuclei in each unit cell.

We also introduce the $\ZZ^3$-periodic Green kernel of the Poisson
interaction~\cite{LS2}, denoted by $G_1$ and uniquely defined by 
$$
\left\{ 
\begin{array}{l}
\dps -\Delta G_1 = 4\pi \left( \sum_{k \in \ZZ^3} \delta_k - 1 \right)
\\
\displaystyle G_1 \  \ZZ^3\mbox{-periodic},\ \mathop{\mbox{min}}_{\RR^3} G_1 = 0.
\end{array} \right.
$$
The Fourier expansion of $G_1$ is 
$$
G_1(x) = c + \sum_{k \in 2\pi \Z^3 \setminus \left\{0\right\}}
\frac{4\pi}{|k|^2} e^{i k \cdot x} 
$$
with $c = \int_\Gamma G_1 > 0$. The electrostatic potential
associated with a $\ZZ^3$-periodic density $\rho\in L^1_{\rm loc}(\R^3)\cap L^3_{\rm loc}(\R^3)$
 is the $\ZZ^3$-periodic function defined as
$$
(\rho \star_\Gamma G_1)(x) := \int_\Gamma G_1(x-y) \, \rho(y) \, dy.
$$
We also set for any $\Z^3$-periodic functions $f$ and $g$
\begin{equation*}
D_{G_1}(f,g)  :=  \int_\Gamma \int_\Gamma G_1(x-y) \, f(x) \, g(y) dx \,
dy.
\end{equation*}
Throughout this article, we will denote by $\chi_I$ the
characteristic function of the set $I\subset\R$ and by $\chi_{I}(A)$ the
spectral projector on $I$ of the self-adjoint operator $A$.

The periodic density of the nuclei is given by
\begin{equation}
\mu_{\rm per}(x)=\sum_{R\in\Z^3}Z\,m(x-R).
\label{def_rho_nuc}
\end{equation}
We assume for simplicity that $m$ is a nonnegative function of $\cC_c^\ii(\R^3)$ with support in $\Gamma$, and that $\int_{\R^3}m(x)dx=1$. Hence $\int_\Gamma\mu_{\rm per}(x)dx=Z$, the total charge of the nuclei in each unit cell.
The periodic rHF energy is then defined for $\gamma\in\cP_{\rm per}$ as
\begin{equation}
\fbox{$\displaystyle\mathcal{ E}^0_{\rm per}(\gamma) = \frac{1}{(2\pi)^3}
\int_{\Gamma^\ast} \tr_{L^2_\xi(\Gamma)}
\left( -\frac 1 2 \Delta \gamma_\xi \right) d\xi + 
\frac 1 2 D_{G_1}\left(\rho_\gamma - \mu_{\rm per}, \rho_\gamma -
\mu_{\rm per} \right).$}
\end{equation}
Introducing
\begin{equation}
\cP^Z_{\rm per}:=\left\{\gamma\in\cP_{\rm per}\ |\
  \int_\Gamma\rho_\gamma=Z\right\}, 
\label{def_PZper}
\end{equation}
the periodic rHF ground state energy (per unit cell) is given by
\begin{equation}
\fbox{$\displaystyle I_{\rm per}^0=\inf\left\{\cE^0_{\rm per}(\gamma),\  \gamma \in
    \cP^Z_{\rm per}\right\}.$}
\label{def_I0per}
\end{equation}
It was proved by Catto, Le Bris and Lions in \cite{CLL} that there
exists a minimizer $\gamma^0_{\rm per}\in \cP^Z_{\rm per}$ to the
minimization problem \eqref{def_I0per}, and that all the minimizers of
\eqref{def_I0per} share the same density $\rho_{\gamma^0_{\rm per}}$. 
We give in Appendix \ref{proof_thm_periodic} the proof of the following
\begin{theorem}[Definition of the periodic rHF minimizer]\label{thm_per}
Let $Z \in \N\setminus\{0\}$. The minimization problem \eqref{def_I0per}
admits a \emph{unique minimizer} $\gamma^0_{\rm per}$.
Denoting by 
\begin{equation}
H^0_{\rm per}:=-\frac{\Delta}{2}+(\rho_{\gamma^0_{\rm per}}-\mu_{\rm
  per}) \star_\Gamma  G_1,
\label{def_H0_per}
\end{equation}
the corresponding periodic mean-field Hamiltonian, $\gamma^0_{\rm per}$
is solution to the self-consistent equation
\begin{equation}
\gamma^0_{\rm per}=\chi_{(-\ii,\epsilon_{F}]}(H^0_{\rm per}),
\label{SCF_per}
\end{equation}
where $\epsilon_{F}$ is a Lagrange multiplier called \emph{Fermi level}, which can be interpreted
as a chemical potential.

Additionally, for any $\epsilon_{F}\in\R$ such that \eqref{SCF_per} holds,
$\gamma^0_{\rm per}$ is the unique minimizer on $\cP_{\rm
  per}$ of the energy functional
$$\gamma \mapsto \cE^0_{\rm
  per}(\gamma)-\epsilon_{F}\int_\Gamma\rho_\gamma.$$
\end{theorem}


Theorem \ref{thm_per} contains three main results that were not
present in \cite{CLL}: first $\gamma^0_{\rm per}$ is unique, second it
is a projector, and third it satisfies Equation \eqref{SCF_per}. These
three properties are crucial for a proper construction of the model for the
crystal with a defect. 

It can easily be seen that $(\rho_{\gamma^0_{\rm per}}-\mu_{\rm
  per}) \star_\Gamma  G_1$ belongs to $L^2_{\rm loc}(\R^3)$. By a result of Thomas \cite{Thomas} this implies that the spectrum of $H^0_{\rm per}$ is purely absolutely continuous. This is an essential property for the proof of the uniqueness of $\gamma^0_{\rm per}$. 
Let $(\lambda_n(\xi))_{n\geq1}$ denote the nondecreasing sequence of
the eigenvalues of $(H^0_{\rm per})_\xi$. Then
$$\sigma(H^0_{\rm per})=\bigcup_{n\geq1}\lambda_n(\Gamma^*),\qquad H^0_{\rm per}=\frac{1}{(2\pi)^3}\int_{\Gamma^*}(H^0_{\rm per})_\xi \; d\xi.$$

The projector $\gamma^0_{\rm per}$ represents the state of the \emph{Fermi sea}, i.e.~of the infinite system of all the electrons in the periodic crystal. Of course, it is an infinite rank projector, meaning that
$$\gamma^0_{\rm per}=\sum_k|\phi_k\rangle\langle\phi_k|$$
should be interpreted as the one-body matrix of a formal infinite Slater determinant
$$\Psi=\phi_1\wedge\phi_2\wedge\cdots\wedge\phi_k\wedge\cdots.$$
The fact that $\gamma^0_{\rm per}$ is additionally a spectral projector associated with the continuous spectrum of an operator leads to the obvious analogy with the \emph{Dirac sea} which is the projector on the negative spectral subspace of the Dirac operator \cite{HLS1,HLS2,HLS3,HLSo}.

\medskip

Most of our results will hold true for insulators (or semi-conductors) only. When necessary, we shall take $Z\in\N\setminus\{0\}$ and make the following assumption:

\medskip

\noindent
\textnormal{\textbf{(A1)}} {\it There is a gap between the $Z$-th and the $(Z+1)$-st bands, i.e. $\Sigma^+_Z<\Sigma^-_{Z+1}$, where $\Sigma^+_Z$ and $\Sigma^-_{Z+1}$ are respectively the maximum and the minimum of the $Z$-th and the $(Z+1)$-st bands of $H^0_{\rm per}$.}

\medskip

We emphasize that Assumption \textnormal{\textbf{(A1)}} is a condition on the solution $\gamma^0_{\rm per}$ of the \emph{nonlinear} problem \eqref{def_I0per}. Note that under \textnormal{\textbf{(A1)}}, one has $\gamma^0_{\rm per}=\chi_{(-\ii,\epsilon_{F}]}(H^0_{\rm per})$ for any $\epsilon_{F}\in(\Sigma^+_Z,\Sigma^-_{Z+1})$.

\section[The rHF model for a crystal with a defect]{The reduced Hartree-Fock model for a crystal with a defect}\label{sec:defect_model}
In this section, we define the reduced Hartree-Fock model describing the
behavior of the Fermi sea and possibly of a finite number of bound
electrons (or holes) close to a local defect. Our model is an obvious
transposition of the Bogoliubov-Dirac-Fock model which was proposed by
Chaix and Iracane \cite{CI} to describe the polarized Dirac sea
(and a finite number of relativistic electrons) in the presence of an
external potential. Our mathematical definition of the reduced energy
functional follows mainly ideas from \cite{HLS1,HLS2}. 
We shall prove in Section~4 that this model can be obtained as the thermodynamic limit of the so-called supercell model. An analogous result was proved in  \cite{HLSo} for the Bogoliubov-Dirac-Fock (BDF) model.

Assume that the periodic nuclear density $\mu_{\rm per}$ defined in
\eqref{def_rho_nuc} is replaced by a locally perturbed nuclear density
$\mu_{\rm per}+ \nu$. The defect $\nu$ can model a vacancy, an
interstitial atom, or an impurity, with possible local rearrangement of
the neighboring atoms.
The main idea underlying the model is to define a \emph{finite} energy by subtracting the \emph{infinite} energy of the periodic Fermi sea $\gamma^0_{\rm per}$ defined in the previous section, from the \emph{infinite} energy of the perturbed system under consideration. For the BDF model, this was proposed first in \cite{HLSo}. Formally, one obtains for a test state $\gamma$
\begin{multline}
\cE^{\rm rHF}_{\mu_{\rm per}+\nu}(\gamma)-\cE^{\rm rHF}_{\mu_{\rm per}+\nu}(\gamma^0_{\rm per}) \ ``="\  \tr\left(H^0_{\rm per}(\gamma-\gamma^0_{\rm per})\right)\\
-\int_{\RR^3}\int_{\RR^3}  \frac{\nu(x)\rho_{[{\gamma-\gamma^0_{\rm
        per}}]}(y)}{|x-y|}dx\,dy+\frac12 \int_{\RR^3} \int_{\RR^3}
\frac{\rho_{[\gamma-\gamma^0_{\rm per}]}(x)\rho_{[\gamma-\gamma^0_{\rm
      per}]}(y)}{|x-y|}dx\,dy. 
\label{formal}
\end{multline}
Of course the two terms in the left-hand side of \eqref{formal} are not
well-defined because $\mu_{\rm per}$ is periodic and because $\gamma$
and $\gamma^0_{\rm per}$ have infinite ranks, but we shall be able to
give a mathematical meaning to the right-hand side, exploiting the fact
that $Q:=\gamma-\gamma^0_{\rm per}$ induces a small perturbation of the
reference state $\gamma^0_{\rm per}$. The formal computation
\eqref{formal} will be justified by means of thermodynamic limit
arguments in Section \ref{thermo-lim}. 

\subsection{Definition of the reduced Hartree-Fock energy of a defect}
\label{sec:def_rHF_def}
We now define properly the reduced Hartree-Fock energy of the Fermi sea in the presence of the defect
$\nu$. We denote by $\gS_p$
the Schatten class of operators $Q$ acting on $L^2(\R^3)$ having a
finite $p$ trace, i.e. such that $\tr(|Q|^p)<\ii$. Note that $\gS_1$ is
the space of trace-class operators, and that $\gS_2$ is the space of
Hilbert-Schmidt operators. Let $\Pi$ be an orthogonal projector on
$L^2(\R^3)$ such that both $\Pi$ and $1-\Pi$ have infinite ranks. A self-adjoint compact
operator $Q$ is said to be $\Pi$-trace class
($Q\in\gS_1^\Pi$) when $Q\in\gS_2$ and $\Pi
Q\Pi,(1-\Pi)Q(1-\Pi)\in\gS_1$. Its $\Pi$-trace is then defined as
$\tr_\Pi(Q)=\tr(\Pi Q\Pi+(1-\Pi)Q(1-\Pi))$. Notice that if $Q\in\gS_1$, then
$Q\in\gS_1^\Pi$ for any $\Pi$ and $\tr_\Pi(Q)=\tr(Q)$. 
See \cite[Section 2.1]{HLS1} for general properties related to this definition.
In the following, we use the shorthand notation
$$Q^{--}:=\gamma^0_{\rm per}Q\gamma^0_{\rm per},\quad 
Q^{++}:=(1-\gamma^0_{\rm per})Q(1-\gamma^0_{\rm per}),
$$
$$
\gS_1^0:=\gS_1^{\gamma^0_{\rm per}}=\left\{ Q \in \gS_2 \; \big| \; 
Q^{++} \in \gS_1, \; Q^{--} \in \gS_1 \right\}\quad \text{and}\quad \tr_0(Q):=\tr_{\gamma^0_{\rm per}}(Q).
$$ 

We also introduce the Banach space
\begin{equation*}
\mathcal{Q}  =  \left\{Q\in\gS_1^0\ \big|\ Q^*=Q,\ |\nabla|Q\in\gS_2,\ 
 |\nabla|Q^{++}|\nabla|\in\gS_1,\ |\nabla|Q^{--}|\nabla|\in\gS_1\right\},
\end{equation*}
endowed with its natural norm
\begin{multline}
\norm{Q}_{\cQ}:=\norm{Q}_{\gS_2}+\norm{Q^{++}}_{\gS_1}+\norm{Q^{--}}_{\gS_1}\\
+\norm{|\nabla|Q}_{\gS_2}+ \norm{|\nabla|Q^{++}|\nabla|}_{\gS_1}+\norm{|\nabla|Q^{--}|\nabla|}_{\gS_1}.
\end{multline}
The convex set on which the energy will be defined is
\begin{equation}
\cK:=\big\{Q\in\cQ\ |\  -\gamma^0_{\rm per}\leq Q\leq 1-\gamma^0_{\rm
  per}\big\}. 
\label{def_cK}
\end{equation}
Notice that $\cK$ is the closed convex hull of states $Q\in\cQ$ of the special form $Q=\gamma-\gamma^0_{\rm per}$, $\gamma$ being an orthogonal projector on $L^2(\R^3)$.
Besides, the number $\tr_0(Q)$ can be interpreted as the charge of the system measured with respect to that of the unperturbed Fermi sea. It can be proved \cite[Lemma 2]{HLS1} that $\tr_0(Q)$ is always an integer if $Q$ is a Hilbert-Schmidt operator of the special form $Q=\gamma-\gamma^0_{\rm per}$, with $\gamma$ an orthogonal projector. Additionally, in this case, $\tr_0(Q)=0$ when $\|Q\|<1$.

Note that the constraint $-\gamma^0_{\rm per}\leq Q\leq 1-\gamma^0_{\rm per}$ in \eqref{def_cK} is equivalent \cite{BBHS,HLS1} to the inequality
\begin{equation}
Q^2\leq Q^{++}-Q^{--}
\label{equivalent_condition} 
\end{equation}
and implies in particular that $Q^{++}\geq 0$ and $Q^{--}\leq0$ for any $Q\in\cK$.

In order to define properly the energy of $Q$, we need to associate a
density $\rho_Q$ with any state $Q\in\cK$. We shall see that $\rho_Q$
can in fact be defined for any $Q\in\cQ$. This is not obvious \emph{a
  priori} since $\cQ$ does not only contain trace-class
operators. Additionally we need to check that the last two terms of
\eqref{formal} are well-defined. For this purpose, we introduce the
so-called Coulomb space 
$$\cC:=\{\rho \in \mathcal{ S}'(\RR^3) \ | \  D(\rho,\rho)<\ii\}$$
where $D(f,g)=4\pi\int_{\R^3}|k|^{-2}\overline{\widehat{f}(k)}\widehat{g}(k)dk$
was already defined before in \eqref{def_D_f_g}.
The dual space of $\cC$ is the Beppo-Levi space
$\cC':= \left\{ V \in  L^6(\RR^3) \ | \ \nabla V \in L^2(\RR^3) \right\}.$
We now use a duality argument to define $\rho_Q$:
\begin{proposition}[Definition of the density $\rho_Q$ for
  $Q\in\cQ$]\label{cont_rho} Assume that $Q\in\cQ$. Then $QV\in \gS_1^0$
  for any $V=V_1+V_2\in\cC'+\left(L^2(\R^3)\cap L^\ii(\R^3)\right)$ and moreover
  there exists a constant $C$ (independent of $Q$ and $V$) such that
$$|\tr_0(QV)|\leq C\norm{Q}_{\cQ}(\norm{V_1}_{\cC'}+\norm{V_2}_{L^2(\R^3)}).$$
Thus the linear form $V\in\cC'+\left(L^2(\R^3)\cap L^\ii(\R^3)\right)\mapsto \tr_0(QV)$ can be continuously extended to $\cC'+L^2(\R^3)$ and there exists a uniquely defined function $\rho_Q\in\cC\cap L^2(\R^3)$ such that
$$\forall V=V_1+V_2\in\cC'+\left(L^2(\R^3)\cap L^\ii(\R^3)\right),\quad \pscal{\rho_Q,V_1}_{\cC,\cC'}+\int_{\R^3}\rho_QV_2=\tr_0(QV).$$
The linear map $Q\in\cQ\mapsto\rho_Q\in\cC\cap L^2(\R^3)$ is continuous:
$$\norm{\rho_Q}_{\cC}+\norm{\rho_Q}_{L^2(\R^3)}\leq C\norm{Q}_\cQ.$$
Eventually when $Q\in\gS_1\subset\gS_1^0$, then $\rho_Q(x)=Q(x,x)$ where $Q(x,y)$ is the integral kernel of $Q$.
\end{proposition}
The proof of Proposition \ref{cont_rho} is given in Section \ref{proof_cont_rho}.

Assuming that \textnormal{\textbf{(A1)}} holds true, we are now in a position to give a rigorous sense to the right-hand side
of \eqref{formal} for $\gamma-\gamma^0_{\rm per} = Q\in\cK$. In the sequel, we
use the following \emph{notation} for any $Q\in\cQ$: 
\begin{equation}
\tr_0(H^0_{\rm per}Q):=\tr(|H^0_{\rm per}-\kappa|^{1/2}(Q^{++}-Q^{--})|H^0_{\rm per}-\kappa|^{1/2})+\kappa\tr_0(Q)
\label{def_kinetic}
\end{equation}
where $\kappa$ is an arbitrary real number in the gap
$(\Sigma_Z^+,\Sigma_{Z+1}^-)$ (this expression will be proved to be
independent of $\kappa$, see Corollary~\ref{cor_def_kinetic} below). Then
we define the energy of any state $Q\in\cK$ as 
\begin{equation}
\fbox{$\displaystyle \cE^\nu(Q):=\tr_0(H^0_{\rm per}Q)-D(\rho_Q,\nu)+\frac12 D(\rho_Q,\rho_Q).$}
\label{def_rHFd}
\end{equation}

The function $\nu$ is an external density of charge representing the nuclear charge of the defect. For the sake of simplicity, we shall assume that $\nu\in L^1(\R^3)\cap L^2(\R^3)\subset\cC$ throughout the paper, although some of our results are true with a weaker assumption. We shall need the following
\begin{lemma}\label{estim_H0per}Assume that \textnormal{\textbf{(A1)}} holds true.
For any fixed $\kappa$ in the gap $(\Sigma^+_Z,\Sigma^-_{Z+1})$, there exist two constants $c_1,c_2>0$ such that
\begin{equation}
c_1(1-\Delta)\leq |H^0_{\rm per}-\kappa|\leq c_2(1-\Delta)
\label{estim_H0per_eq}
\end{equation}
as operators on $L^2(\R^3)$. In particular 
$$\norm{|H^0_{\rm per}-\kappa|^{1/2}(1-\Delta)^{-1/2}}\leq \sqrt{c_2},\quad \norm{|H^0_{\rm per}-\kappa|^{-1/2}(1-\Delta)^{1/2}}\leq 1/\sqrt{c_1}.$$
Similarly, $(H^0_{\rm per}-\kappa)(1-\Delta)^{-1}$ and its inverse are bounded operators.
\end{lemma}
The proof of the above lemma is elementary; it will be given in Section \ref{proof_estim_H0per}. By the definition of $\cQ$ and Lemma \ref{estim_H0per}, it is clear that the right-hand side of \eqref{def_kinetic} is a well-defined quantity for any $Q\in\cQ$ and any $\kappa\in(\Sigma^+_Z,\Sigma^-_{Z+1})$. By Proposition \ref{cont_rho} which states that $\rho_Q\in\cC$ for any $Q\in\cQ$, we deduce that \eqref{def_rHFd} is a well-defined functional.

We shall need the following space of more regular operators
\begin{equation}
\cQ_{\rm r}:=\{Q\in\cQ\ |\  (-\Delta) Q^2(-\Delta)\in\gS_1,\  (-\Delta) (Q^{++}-Q^{--})(-\Delta)\in\gS_1\} 
\end{equation}
and the associated convex set
$$\cK_{\rm r}:=\cK\cap\cQ_{\rm r}.$$
The following result will be useful (its proof will be given below in Section \ref{proof_lemma_dense}):
\begin{lemma}\label{lemma_dense}
The space $\cQ_{\rm r}$ (resp. the convex set $\cK_{\rm r}$) is dense in $\cQ$
(resp. in $\cK$) for the topology of $\cQ$. 
\end{lemma}
\begin{corollary}\label{cor_def_kinetic}Assume that \textnormal{\textbf{(A1)}} holds true.
When $Q\in\cQ_{\rm r}$, then $H^0_{\rm per}Q\in\gS_1^0$.
For any $Q\in\cQ$, the expression \eqref{def_kinetic} for $\tr_0(H^0_{\rm per}Q)$ does not depend on $\kappa\in(\Sigma^+_Z,\Sigma^-_{Z+1})$. If $Q\in\cK$, then
\begin{eqnarray}
0 & \leq&  c_1\tr((1-\Delta)^{1/2}Q^2(1-\Delta)^{1/2})\label{coercive}\\
 & \leq&  c_1\tr((1-\Delta)^{1/2}(Q^{++}-Q^{--})(1-\Delta)^{1/2})\nonumber\\
&\leq& \tr_0(H^0_{\rm per}Q)-\kappa\tr_0(Q)\nonumber\\
&\leq& c_2\tr((1-\Delta)^{1/2}(Q^{++}-Q^{--})(1-\Delta)^{1/2})\nonumber
\end{eqnarray}
where $c_1$ and $c_2$ are given by Lemma \ref{estim_H0per}.
\end{corollary}
\begin{proof}
Let $Q\in\cQ_{\rm r}$ and $\kappa\in(\Sigma^+_Z,\Sigma^-_{Z+1})$. Then
$((H_{\rm per}^0-\kappa)Q)^{++}=|H^0_{\rm per}-\kappa|Q^{++}=|H^0_{\rm
  per}-\kappa|(1-\Delta)^{-1}(1-\Delta)Q^{++}\in\gS_1$ by Lemma
\ref{estim_H0per} and the definition of $\cQ_{\rm r}$. A similar argument for
$((H_{\rm per}^0-\kappa)Q)^{--}$ proves that $H^0_{\rm per}Q\in\gS_1^0$. Then
for any $Q\in\cQ_{\rm r}$, \eqref{coercive} is a straightforward consequence of
\eqref{estim_H0per_eq} and \eqref{equivalent_condition}. We conclude
using the density of $\cQ_{\rm r}$ in $\cQ$ and the density of $\cK_{\rm r}$ in $\cK$.
\end{proof}
The following is an adaptation of \cite[Thm 1]{HLS1}:
\begin{corollary}\label{bd_below}
 Let $\nu\in L^1(\R^3)\cap L^2(\R^3)$, $Z\in \NN \setminus \left\{0\right\}$ and assume
 that \textnormal{\textbf{(A1)}} holds. For any  $\kappa\in(\Sigma^+_Z,\Sigma^-_{Z+1})$, one has for some $d_1,d_2>0$
\begin{multline}
\forall Q\in\cK,\quad \cE^\nu(Q)-\kappa\tr_0(Q) \geq 
d_1 \bigg( \norm{Q^{++}}_{\gS_1} +  \norm{Q^{--}}_{\gS_1} \\   
+\norm{|\nabla|Q^{++}|\nabla| }_{\gS_1}
 + \norm{|\nabla|Q^{--}|\nabla|
  }_{\gS_1} \bigg) + 
d_2 \left( \norm{Q}_{\gS_2}^2 +\norm{|\nabla|Q}_{\gS_2}^2
\right)-\frac12 D(\nu,\nu)  
\label{estim_below}
\end{multline}
Hence
$\cE^\nu-\kappa\tr_0$ is bounded from below and coercive on $\cK$.
Additionally, when $\nu\equiv0$, $Q\mapsto \cE^0(Q)-\kappa\tr_0(Q)$ is nonnegative, 0 being its unique minimizer.
\end{corollary}

\begin{proof}
Inequality \eqref{estim_below} is a straightforward consequence of
\eqref{coercive} and the fact that $D(\cdot,\cdot)$ defines a scalar
product on $\cC$. The rest of the proof is obvious.
\end{proof}

\begin{remark}\rm
The energy $\cE^\nu(Q)$ measures the energy of a state $\gamma=\gamma^0_{\rm per}+Q$ with respect to that of $\gamma^0_{\rm per}$. Thus the last statement of Corollary \ref{bd_below} is another way of expressing the fact that $\gamma^0_{\rm per}$ is the state of lowest energy of the periodic system when there is no defect.
\end{remark}

\subsection{Existence of minimizers with a chemical potential}
\label{sec:min_chem_pot}
In view of Corollary~\ref{bd_below}, it is natural to introduce the following minimization problem
\begin{equation} \label{eq:Emunu}
\mbox{\fbox{$\displaystyle E^\nu_{\epsilon_F}:=\inf\{\cE^\nu(Q)-\epsilon_F\tr_0(Q),\
    Q\in\cK\}>-\ii$}}
\end{equation}
for any Fermi level $\epsilon_F\in(\Sigma_Z^{+},\Sigma_{Z+1}^-)$. 
The following result is proved in Section \ref{proof_exists_min}, following ideas from \cite{HLS2}:
\begin{theorem}[Existence of minimizers with a chemical
  potential]\label{exists_min} Let $\nu\in L^1(\R^3)\cap L^2(\R^3)$, $Z\in \NN \setminus
  \left\{0\right\}$ and assume that \textnormal{\textbf{(A1)}} holds. Then for any
  $\epsilon_F\in(\Sigma^+_Z,\Sigma^-_{Z+1})$, there exists a minimizer
  $\bar Q\in\cK$ for~\eqref{eq:Emunu}. Problem~\eqref{eq:Emunu} may have
  several minimizers, but they all share the same density $\bar
  \rho = \rho_{\bar Q}$. Any minimizer $\bar Q$ of~\eqref{eq:Emunu}
  satisfies the self-consistent equation  
\begin{equation}
\left\{\begin{array}{l}
\bar Q=\chi_{(-\ii,\epsilon_F)}(H_{\bar Q})-\gamma^0_{\rm per}+\delta,\smallskip\\
H_{\bar Q}=H^0_{\rm per}+(\rho_{\bar Q}-\nu)\ast|\cdot|^{-1}
\end{array}\right.
\label{scf_defauts}
\end{equation}
where $\delta$ is a finite rank self-adjoint operator satisfying $0\leq\delta\leq1$ and ${\rm Ran}(\delta)\subseteq \ker(H_{\bar Q}-\epsilon_F)$.
\end{theorem}

\begin{remark}\rm
It is easily seen that $(\rho_{\bar Q}-\nu)\ast|\cdot|^{-1}$ is a compact perturbation of $H^0_{\rm per}$, implying that $H_{\bar Q}$ is self-adjoint on $\cD(H^0_{\rm per})=\cD(-\Delta)=H^2(\R^3)$ and that 
$\sigma_{\rm ess}(H_{\bar Q})=\sigma(H^0_{\rm per}).$
Thus the discrete spectrum of $H_{\bar Q}$ is composed of isolated eigenvalues of finite multiplicity, possibly accumulating at the ends of the bands.
\end{remark}

Recall that the charge of the minimizing state $\bar Q$ obtained in
Theorem \ref{exists_min} is defined as $\tr_0(\bar Q)$. Similarly to
\cite{HLS1,HLS2}, it can be proved by perturbation theory that for any
fixed $\epsilon_F$, there exists a constant $C(\epsilon_F)$ such that when
$D(\nu,\nu)\leq C(\epsilon_F)$, one has $\ker(H_{\bar Q}-\epsilon_F)=\{0\}$ and  
$\tr_0(\bar Q)=0$,
i.e. the minimizer of the energy with chemical potential $\epsilon_F$ is a neutral perturbation of the periodic Fermi sea. 

\begin{figure}
\centering
\includegraphics{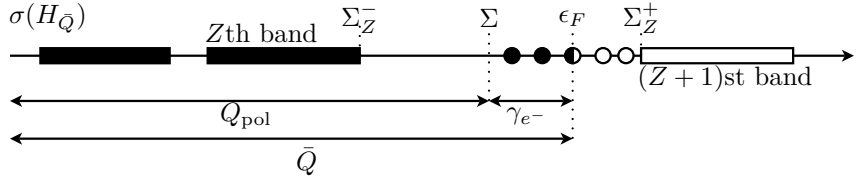}
\caption{Decomposition $\bar Q=Q_{\rm pol}+\gamma_{e^-}$ for not too
  strong a positively charged nuclear defect ($\nu \ge 0$).}
\end{figure}

For a fixed external density $\nu$ and an adequately chosen chemical
potential $\epsilon_F$, one can have $\tr_0(\bar Q)\neq 0$ meaning either that
electron-hole pairs have been created from the Fermi sea, and/or that
the system of lowest energy contains a finite number of bound electrons
or holes close to the defect. In the applications, one will usually have
for a positively charged nuclear defect ($\nu \ge 0$) that the spectrum
of $H_{\bar Q}$ 
contains a sequence of eigenvalues converging to the bottom
$\Sigma_{Z+1}^-$ of the lowest unfilled band (conduction band), and that $\epsilon_F$ is chosen such that exactly $q$ eigenvalues are filled, corresponding to $q$ bound electrons:
\begin{equation}
\bar Q=\left(\chi_{(-\ii,\Sigma)}(H_{\bar Q})-\gamma^0_{\rm per}\right)+\left(\chi_{[\Sigma,\epsilon_F)}(H_{\bar Q})+\delta\right):=Q_{\rm pol}+\gamma_{\rm e^-},
\label{decomp_solution}
\end{equation}
where we have chosen as a reference the center of the gap
$$\Sigma:=\frac{\Sigma^+_Z+\Sigma^-_{Z+1}}{2}.$$
For not too strong a defect density $\nu$, one has $\ker(H_{\bar Q}-\Sigma)=\{0\}$ and $\tr_0(Q_{\rm pol})=0$. Hence $\tr(\gamma_{\rm e^-})=q$. Let us assume for simplicity that $\delta=0$ and that $q\in\N\setminus\{0\}$. Then 
\begin{equation}
\gamma_{\rm e^-}=\chi_{[\Sigma,\epsilon_F)}(H_{\bar Q})=\sum_{n=1}^q|\phi_n\rangle\langle\phi_n|
\label{form_gamma_e}
\end{equation}
where $(\phi_n)$ are eigenfunctions of $H_{\bar Q}$ corresponding to its $q$ first eigenvalues in $[\Sigma,\epsilon_F)$:
\begin{equation}
H_{\bar Q}\phi_n=\lambda_n\phi_n.
\label{eq_scf_electrons}
\end{equation}
Notice that
\begin{equation}
H_{\bar Q}=-\Delta/2+(\rho_{\gamma_{\rm e^-}}-\nu)\ast|\cdot|^{-1}+V_{\rm pol}
\label{MF_scf_electrons}
\end{equation}
where
$$V_{\rm pol}=(\rho_{\gamma^0_{\rm per}}-\mu_{\rm per}) \star_\Gamma G_1+\rho_{Q_{\rm pol}}\ast|\cdot|^{-1}$$
is the polarization potential created by the self-consistent Fermi sea and seen by the $q$ electrons.
Thus the $q$ electrons solve a usual reduced Hartree-Fock equation \eqref{eq_scf_electrons} in which the mean-field operator \eqref{MF_scf_electrons} additionally contains the self-consistent polarization of the medium.

The interpretation given in the previous paragraph is different if the positive density of charge $\nu$ of the defect is strong enough to create an electron-hole pair from the Fermi sea.

\medskip

We end this section by specifying the regularity of solutions of \eqref{scf_defauts}. The proof is given in Section \ref{proof_regul_solution}.
\begin{proposition}\label{regul_solution} Let $\nu\in L^1(\R^3)\cap L^2(\R^3)$, $Z\in \NN
  \setminus \left\{ 0 \right\}$ and assume that \textnormal{\textbf{(A1)}} holds. Any $Q\in\cK$ solution of the self-consistent equation \eqref{scf_defauts} belongs to $\cK_{\rm r}$.
\end{proposition}

\begin{remark}\label{rem_S_1}\rm
Notice that it is natural to wonder whether $Q\in\gS_1$, which would in particular imply that $\rho_Q\in L^1(\R^3)$. This is known to be false for the Bogoliubov-Dirac-Fock model studied in \cite{HLS1,HLS2,HLS3,HLSo}. We do not answer this question for our model in the present paper. 
\end{remark}

\subsection{Existence of minimizers under a charge constraint}
In the previous section, we stated the existence of  minimizers
for any chemical potential in the gap of the periodic operator
$H^0_{\rm per}$, but of course the total charge $\tr_0(\bar Q)$ of the
obtained solution was unknown \emph{a priori}. Here we tackle the more
subtle problem of minimizing the energy while \emph{imposing a charge
  constraint}. Mathematically this is more difficult because although
the energy $\cE^\nu(Q)$ is convex on $\cK$ and weakly lower
semi-continuous (wlsc) for the weak-$\ast$ topology of $\cQ$ (as will be shown
in the proof of Theorem \ref{exists_min}), the $\gamma^0_{\rm
  per}$-trace functional $Q\in\cK\mapsto\tr_0(Q)$ is continuous but
\emph{not} wlsc for the weak-$\ast$ topology of $\cQ$: in principle it is possible that a (positive or negative) part of the charge
of a minimizing sequence for the charge-constrained minimization problem
escapes to infinity, leaving at the limit a state of a different (lower or higher)
charge. In fact, we can prove that a minimizer exists under a charge
constraint, if and only if some binding conditions hold, the role of which being to
prevent the lack of compactness. 

 As explained above, imposing $\tr_0(Q)=q$ should intuitively lead (for a sufficiently weak defect density $\nu$) to a system of $q$ electrons
 coupled to a polarized Fermi sea. Notice that we do \emph{not} impose that
 $q=\int_{\R^3}\nu$, i.e. our model allows \emph{a priori} to treat
 defects with non-zero total charge.

As usual in reduced Hartree-Fock theories, we consider the case of a real charge constraint $q\in\R$:
\begin{equation} \label{eq:minEnuq}
\mbox{\fbox{$\displaystyle E^\nu(q):=\inf\{\cE^\nu(Q),\ Q\in\cK,\
  \tr_0(Q)=q\}.$}}
\end{equation}
When no defect is present, $E^0(q)$ can be computed explicitly:
\begin{proposition}[Defect-free charge-constrained
  energy]\label{energy_free} Let $Z\in \NN \setminus
  \left\{0\right\}$ and assume that \textnormal{\textbf{(A1)}} holds. Then one has 
$$E^0(q)=\left\{\begin{array}{ll}
\Sigma^-_{Z+1}q & {\rm when\ } q\geq0\\
\Sigma^+_Zq & {\rm when\ } q\leq0.\\
\end{array}\right.$$
The minimization problem \eqref{eq:minEnuq} with $\nu\equiv0$ has no solution except when $q=0$.
\end{proposition}
The proof of Proposition \ref{energy_free} is given in Section \ref{proof_energy_free}.
We now state the main result of this section, which is directly inspired
from~\cite{HLS3}: 
\begin{theorem}[Existence of minimizers under a charge
  constraint]\label{HVZ} Let $\nu\in L^1(\R^3)\cap L^2(\R^3)$, $Z\in \NN \setminus
  \left\{0\right\}$ and assume that \textnormal{\textbf{(A1)}} holds. The following
  assertions are equivalent: 

\smallskip

\textnormal{\textbf{(a)}} Problem \eqref{eq:minEnuq} admits a minimizer $\bar Q$;

\smallskip

\textnormal{\textbf{(b)}}  Every minimizing sequence for \eqref{eq:minEnuq} is precompact in $\cQ$ and converges towards a minimizer $\bar Q$ of \eqref{eq:minEnuq};

\smallskip

\textnormal{\textbf{(c)}} $\forall q'\in\R\setminus\{0\},\quad E^\nu(q)<E^\nu(q-q')+E^0(q')$.

\smallskip

Assume that the equivalent conditions \textnormal{\textbf{(a)}}, \textnormal{\textbf{(b)}} and \textnormal{\textbf{(c)}} above are
fulfilled. In this case, the minimizer $\bar Q$ is not necessarily
unique, but all the minimizers share the same density $\bar \rho =
\rho_{\bar Q}$. Besides, there exists
$\epsilon_F\in[\Sigma^+_Z,\Sigma^-_{Z+1}]$ such that the obtained
minimizer $\bar Q$ is a global minimizer for $E^\nu_{\epsilon_F}$ defined in \eqref{eq:Emunu}. It solves Equation \eqref{scf_defauts} for some $0\leq\delta\leq1$ with ${\rm Ran}(\delta)\subseteq\ker(H_{\bar Q}-\epsilon_F)$. The operator $\delta$ is finite rank if $\epsilon_F\in(\Sigma^+_Z,\Sigma^-_{Z+1})$ and trace-class if $\epsilon_F\in\{\Sigma^+_Z,\Sigma^-_{Z+1}\}$.

Additionally the set of $q$'s in $\R$ satisfying the above equivalent conditions is a non-empty closed interval $I\subseteq\R$. This is the largest interval on which $q\mapsto E^\nu(q)$ is strictly convex.
\end{theorem}

\begin{remark}\rm
One has $I=\left\{\tr_0(\bar Q),\ \bar Q\text{ min. of } E^\nu_{\epsilon_F},\ \epsilon_F\in[\Sigma_Z^+,\Sigma_{Z+1}^-]\right\}$. Hence $I\neq\emptyset$ by Theorem \ref{exists_min}.
\end{remark}

Theorem \ref{HVZ} is proved in Section \ref{proof_HVZ}. Many of the
above statements are very common in reduced Hartree-Fock theories and
not all the details will be given (see, e.g. \cite{rHF}). The difficult
part is the proof that \textbf{(b)} is 
equivalent to \textbf{(c)}, for which we use ideas from~\cite{HLS3}. 

Conditions like \textbf{(c)} appear classically when analyzing the
compactness properties of minimizing sequences, for instance by using
the concentration-compactness principle of P.-L. Lions \cite{Lions-84}. They are also
very classical for linear models in which the bottom of the essential
spectrum has the form of the minimum with respect to $q'$ of the right hand side of \textbf{(c)}, as expressed by the HVZ Theorem \cite{H,V,Zhislin-60}. 
Assume for simplicity that $q>0$ and that $\bar Q$ can be written as in \eqref{decomp_solution} and \eqref{form_gamma_e}. When $0<q'\leq q$, \textbf{(c)} means that it is not favorable to let $q'$ electrons escape to infinity, while keeping $q-q'$ electrons near the defect. When $q'<0$, it means that it is not favorable to let $\vert q'\vert$ holes escape to infinity, while keeping $q+\vert q'\vert$ electrons near the defect. When $q'>q$, it means that it is not favorable to let $q'$ electrons escape to infinity, while keeping $q'-q$ holes near the defect. 

In this article we do not show when \textbf{(c)} holds true. Proving \textbf{(c)} usually requires some decay property of the density of charge $\rho_{\bar Q}$ for a solution $\bar Q$ of the nonlinear equation \eqref{scf_defauts}. In particular, knowing that $\rho_{\bar Q}\in L^1(\R^3)$ would be very useful (see Remark \ref{rem_S_1}). We plan to investigate more closely the decay properties of $\rho_{\bar Q}$ and the validity of \textbf{(c)} in the near future. At present, the validity of a condition similar to \textbf{(c)} for the Bogoliubov-Dirac-Fock model was only proved in the nonrelativistic limit or in the weak coupling limit, see \cite{HLS3}.

\section{Thermodynamic limit of the supercell model}\label{thermo-lim}
As mentioned before, we shall now justify the model of the previous
section by proving that it is the thermodynamic limit of the supercell model. 

Let us emphasize that there are several ways of performing thermodynamic
limits. In \cite{CLL}, the authors consider a box of size $L$,
$\Lambda_L:=[-L/2,L/2)^3$, and assume that the nuclei are located on
$\Z^3\cap\Lambda_L$. Then they consider the rHF model of Section \ref{sec_rHF} for $N$ electrons
living \emph{in the whole space}, with $N = ZL^3$ chosen to impose
neutrality. Denoting by $\rho_L$ the ground state electronic density of
the latter problem, it is proved in \cite[Thm 2.2]{CLL} that the energy per unit cell converges to $I^0_{\rm per}$, and that the following holds:
\begin{equation}
 \sqrt{L^{-3}\sum_{k\in\Z^3\cap \Lambda_{L}}\rho_{{L}}(x-k)}
 \rightarrow \sqrt{\rho_{\gamma^0_{\rm per}}}
\label{limit_CLL}
\end{equation}
weakly in $H^1_{\rm loc}(\R^3)$, strongly in $L^p_{\rm loc}(\R^3)$ for all $2\leq p<6$ and almost everywhere on $\R^3$ when $L\to\ii$. Let us recall that $I^0_{\rm per}$ and $\gamma^0_{\rm per}$ are defined in Section \ref{sec_per}.

Another way for performing thermodynamic limits is to confine the nuclei \emph{and} 
the electrons in a domain $\Omega_L$ with $|\Omega_L|\to\ii$, by means of Dirichlet boundary
conditions for the electrons. The latter approach was chosen for the Schrödinger model with quantum nuclei  in the
canonical and grand canonical ensembles~\cite{Ruelle} by Lieb and Lebowitz in the seminal paper \cite{LiLe} (see also \cite{Lieb}), where the existence of a limit for the energy per unit volume is proved.
The crystal case in the Schrödinger model was tackled by Fefferman
\cite{F} in the same spirit. We do not know whether Fefferman's proof can be
adapted to treat the Hartree-Fock case.

Another possibility, perhaps less satisfactory from a physical viewpoint but
more directly related to practical calculations (see
e.g.~\cite{Dovesi}), is to take $\Omega_L=\Lambda_L$ and to impose periodic boundary conditions on the
box $\Lambda_L$. Usually the Coulomb interaction is also replaced by a
$(L\Z^3)$-periodic Coulomb potential, leading to the so-called
\emph{supercell model} which will be described in detail below. This
approach has the advantage of respecting the symmetry of the system in the
crystal case. It was used by Hainzl, Lewin and Solovej in \cite{HLSo} to
justify the Hartree-Fock approximation of no-photon Quantum Electrodynamics. The supercell limit of a linear model for photonic crystals is studied in \cite{Soussi}.

Of course the conjecture is that the final results (the energy per unit
cell and the ground state density of the crystal) should not depend on
the chosen thermodynamic limit procedure. This is actually the case for the reduced Hartree-Fock model of the crystal. See \cite{HSo} for a result in this direction for a model with quantum nuclei.

\medskip

Let us now describe the supercell model. For $L \in \NN \setminus
\left\{0\right\}$, we introduce
the supercell $\Lambda_L=[-L/2,L/2)^3$ and the Hilbert space
$$
L^2_{\rm per}(\Lambda_L) = \left\{ \phi \in L^2_{\rm loc}(\RR^3) \ | \
\phi \mbox{ $(L\Z^3)$-periodic} \right\}.
$$
We also introduce the $L\Z^3$-periodic Coulomb potential $G_L$ defined as the
unique solution to
$$
\left\{ 
\begin{array}{l}
\dps -\Delta G_L = 4\pi \left( \sum_{k \in L \, \ZZ^3} \delta_k -
  \frac{1}{L^3} \right) 
\\
G_L \  L \, \ZZ^3\mbox{-periodic},\ \ \mathop{\mbox{min}}_{\RR^3} G_L = 0.
\end{array} \right.
$$
It is easy to check that $G_L(x) = L^{-1}G_1(x/L)$ and that
$$
G_L(x) = \frac c L + 
\sum_{k \in \frac{2\pi}L \Z^3 \setminus \left\{0\right\}}
\frac{4\pi}{|k|^2} \, \frac{1}{L^3} \, e^{ik \cdot x}.
$$
For any $L\Z^3$-periodic function $g$, we define
$$
\left( g \star_{\Lambda_L} G_L \right)(x) := 
\int_{\Lambda_L} G_L(x-y) \, g(y) \, dy,
$$
$$
D_{G_L}(f,g)  :=  \int_{\Lambda_L} \int_{\Lambda_L} G_L(x-y) \, f(x) \,
g(y) \, dx \, dy.
$$
An admissible electronic state
is then described by a one-body density matrix $\gamma$ in
\begin{equation*}
\mathcal{ P}_{{\rm sc},L}  =  \bigg\{ \gamma \in \gS_1(L^2_{\rm
  per}(\Lambda_L)) \  | \ \gamma^\ast = \gamma, \ 
 0  \le \gamma \le 1,\ 
\tr_{L^2_{\rm per}(\Lambda_L)}(-\Delta \gamma)<+\infty \bigg\}.
\end{equation*}
Any $\gamma\in\cP_{{\rm sc},L}$ has a well-defined density of charge
$\rho_{\gamma}(x)=\gamma(x,x)$ where $\gamma(x,y)$ is the kernel of the
operator $\gamma$. Notice that $\gamma(x+Lz,y+Lz')=\gamma(x,y)$ for any
$z,z'\in\Z^3$, which implies that $\rho_\gamma$ is $L\Z^3$-periodic. 

Throughout this section, we use the subscript `sc' to indicate that we consider the thermodynamic limit of the supercell model.

\subsection{Thermodynamic limit without defect} \label{sec:TL_wd}
Because our model with defect uses the defect-free density matrix of the
Fermi sea as a reference, we need to start with the study of the
thermodynamic limit without defect. We are going to prove for the
supercell model a result analogous to \cite[Thm 2.2]{CLL}.

The reduced Hartree-Fock energy functional of the supercell model is
defined for $\gamma\in\cP_{{\rm sc},L}$ as 
$$
\fbox{$\displaystyle \mathcal{ E}^0_{{\rm sc},L}(\gamma) = \tr_{L^2_{\rm per}(\Lambda_L)} 
\left(- \frac 1 2 \Delta \gamma \right) + 
\frac 1 2 D_{G_L}\left(\rho_\gamma - \mu_{\rm per}, \rho_\gamma -
\mu_{\rm per} \right)$}
$$
where we recall that $\mu_{\rm per}(x)=\sum_{R\in\Z^3}Z m(x-R)$ is a $\Z^3$- (thus $L\Z^3$-) periodic function. The reduced Hartree-Fock ground state energy for a neutral system in the box of size $L$ is then given by
\begin{equation} \label{eq:supercell2}
\fbox{$\displaystyle I^0_{{\rm sc},L} = \inf \left\{ \mathcal{ E}^0_{{\rm sc},L}(\gamma), \;  \gamma
    \in \mathcal{ P}_{{\rm sc},L},\
    \int_{\Lambda_L}\rho_{\gamma}=\int_{\Lambda_L}\mu_{\rm per} = ZL^3
    \right\}.$}
\end{equation}
Let us recall that $I^0_{\rm per}$, $\gamma^0_{\rm per}$ and $H^0_{\rm per}$ are defined in Section \ref{sec_per}. In Section~\ref{sec:proof_TLSCP}, we prove the following

\begin{theorem}[Thermodynamic limit of the defect-free supercell
  model]\label{thm:supercell_0} Let $Z>0$.

\smallskip

\noindent i) For all $L \in \NN\setminus\{0\}$, the minimizing problem $I^0_{{\rm
 sc},L}$ has at least one minimizer, and all the minimizers share the
 same density. This density is
 $\ZZ^3$-periodic. Besides, there is one minimizer $\gamma^0_{{\rm
 sc},L}$ of \eqref{eq:supercell2} which commutes with the translations
 $\tau_k$, $k\in\Z^3$. 

\medskip

\noindent ii) The following thermodynamic limit properties hold true:

\smallskip
\noindent $\bullet$ {\rm (Convergence of the energy per unit cell)}.
$$\lim_{L\to\ii}\frac{I^0_{{\rm sc},L}}{L^3} = I^0_{\rm per};$$
 $\bullet$ {\rm (Convergence of the density)}.
\begin{equation}
\sqrt{\rho_{\gamma^0_{{\rm sc},L}}}\wto \sqrt{\rho_{\gamma^0_{\rm per}}} \quad \text{weakly in}\ H^1_{\rm loc}(\RR^3),
 \label{limit_density_nous}
\end{equation}
$${\rho_{\gamma^0_{{\rm sc},L}}}\to {\rho_{\gamma^0_{\rm per}}} \quad \text{strongly in\ }L^p_{\rm loc}(\RR^3)\ \text{ for}\ 1 \le p <3 \text{\ and a.e.};$$
$\bullet$ {\rm (Convergence of the mean-field Hamiltonian and its spectrum)}. Let 
$$
H^0_{{\rm sc},L} = - \frac{\Delta}2 + (\rho_{\gamma^0_{{\rm sc},L}} -
\mu_{\rm per}) \star_\Gamma G_1
$$
seen as an operator acting on $L^2(\R^3)$. 
Then, for all
$L\in\N\setminus\{0\}$, $H^0_{{\rm sc},L} - H^0_{\rm per}$ is a bounded operator and  
$$ \lim_{L\to\ii}\norm{H^0_{{\rm sc},L} - H^0_{\rm per}}=0.$$
Denoting by $(\lambda_n^L(\xi))_{n \in \NN\setminus\{0\}}$ the nondecreasing sequence of eigenvalues of $(H^0_{{\rm sc},L})_\xi$ for $\xi \in \Gamma^\ast$, one has 
\begin{equation} \label{eq:gap_unif}
\lim_{L\to\ii}\sup_{n \in \NN \setminus \left\{0\right\}} 
\sup_{\xi \in \Gamma^\ast} \left|
  \lambda_{n}^L(\xi) - \lambda_{n}(\xi) \right|=0
\end{equation}
where $(\lambda_n(\xi))_{n\geq1}$ are the eigenvalues of $(H^0_{\rm per})_\xi$ introduced in Theorem~\ref{thm_per}.

\medskip

\noindent iii) Assume in addition that \textnormal{\textbf{(A1)}} holds. Fix some $\epsilon_F\in(\Sigma^+_Z,\Sigma^-_{Z+1})$. Then for $L$ large enough, the minimizer $\gamma^0_{{\rm sc},L}$ of $I^0_{{\rm sc},L}$ is unique. It is also the unique  minimizer of the following problem
\begin{equation}
I^0_{{\rm sc},L,\epsilon_F}:=\inf \left\{\cE^0_{{\rm
      sc},L}(\gamma)-\epsilon_F\tr_{L^2_{\rm per}(\Lambda_L)}(\gamma),\quad \gamma\in\cP_{{\rm sc},L}\right\}.
\label{def_min_free_mu}
\end{equation}
\end{theorem}

\begin{remark}\rm
Notice that some of the above assertions are more precise for the
supercell model than for the thermodynamic limit procedure considered in
\cite[Thm 2.2]{CLL} (compare for instance \eqref{limit_density_nous}
with \eqref{limit_CLL}). This is because the supercell model respects
the symmetry of the system, allowing in particular to have a minimizer
$\gamma^0_{{\rm sc},L}$ in the box of size $L^3$ which is periodic for
the lattice $\Z^3$. For an insulator, the uniqueness of $\gamma^0_{{\rm
    sc},L}$ for large $L$ and the convergence properties of
\textit{iii)} are also very interesting for computational purposes. 
\end{remark}

\subsection{Thermodynamic limit with defect}
We end this section by considering the thermodynamic limit of the
supercell model with a defect. Recall that $\nu\in L^1(\R^3)\cap L^2(\R^3)\subset\cC$ is the density of charge of the
defect. First we need to periodize this function with respect to the large box $\Lambda_L$, for instance by defining
$$\nu_L(x):=\sum_{z\in\Z^3}(\1_{\Lambda_L}\nu)(\cdot-Lz).$$ 
The reduced Hartree-Fock energy functional of the supercell model with
defect is then defined for $\gamma\in\cP_{{\rm sc},L}$ as
$$
\fbox{$\displaystyle \mathcal{ E}^\nu_{{\rm sc},L}(\gamma) = \tr_{L^2_{\rm per}(\Lambda_L)} 
\left(- \frac 1 2 \Delta \gamma \right) + 
\frac 1 2 D_{G_L}\left(\rho_\gamma - \mu_{\rm per}-\nu_L, \rho_\gamma -
\mu_{\rm per}-\nu_L \right).$}
$$
For $\epsilon_F\in(\Sigma^+_Z,\Sigma^-_{Z+1})$, we consider the following minimization problem
\begin{equation} \label{eq:supercell_defect}
\fbox{$\displaystyle I^\nu_{{\rm sc},L,\epsilon_F} = \inf \left\{ \mathcal{ E}^\nu_{{\rm
    sc},L}(\gamma)-\epsilon_F\tr_{L^2_{\rm per}(\Lambda_L)}(\gamma), \;  \gamma
    \in \mathcal{ P}_{{\rm sc},L}\right\}.$}
\end{equation}
We recall that $\gamma^0_{\rm per}$ is defined in Section~\ref{sec_per},
that $E_\kappa^\nu$ and $\bar Q$ are defined in Section~\ref{sec:min_chem_pot}, and that $I^0_{{\rm sc},L,\epsilon_F}$ is defined in
Section~\ref{sec:TL_wd}.
In Section~\ref{sec:proof_TLSCD}, we prove the
\begin{theorem}[Thermodynamic limit of the supercell model with
  defect]\label{thermo_defect} Let $Z \in \NN \setminus
  \left\{0\right\}$. Assume that \textnormal{\textbf{(A1)}} holds and fix some
  $\epsilon_F\in(\Sigma^+_Z,\Sigma^-_{Z+1})$. Then one has 
\begin{equation}
\lim_{L\to\ii}\left(I^\nu_{{\rm sc},L,\epsilon_F}-I^0_{{\rm
      sc},L,\epsilon_F}\right)=E_{\epsilon_F}^\nu - \int_{\R^3}\nu\left((\rho_{\gamma^0_{\rm
      per}}-\mu_{\rm per})\star_\Gamma G_1\right) + \frac 12 D(\nu,\nu).
\label{eq_thermo_limit_defect}
\end{equation}

Additionally, if $\gamma_{{\rm sc},L}^\nu$ denotes a minimizer for
\eqref{eq:supercell_defect}, then one has, up to extraction of a
subsequence, 
$$(\gamma^\nu_{{\rm sc},L}-\gamma^0_{{\rm sc},L})(x,y)\to \bar Q(x,y)$$
weakly in $H^1_{\rm loc}(\R^3\times\R^3)$ and strongly in $L^2_{\rm loc}(\R^3\times\R^3)$, where $\bar Q$ is a minimizer of \eqref{eq:Emunu}, as obtained in Theorem~\ref{exists_min}. Besides, 
$$\rho_{\gamma^\nu_{{\rm sc},L}}-\rho_{\gamma^0_{{\rm sc},L}}\to \bar \rho$$
weakly in $L^2_{\rm loc}(\R^3)$,
where $\bar \rho$ is the common density of all the minimizers of~\eqref{eq:Emunu}. 
\end{theorem}

\begin{remark}\rm
In numerical simulations, the right-hand side
of~\eqref{eq_thermo_limit_defect} is approximated by $I^\nu_{{\rm
    sc},L,\epsilon_F}-I^0_{{\rm sc},L,\epsilon_F}$ for a given value of $L$. This
approach has several drawbacks. First, the values of $L$ that lead to
tractable numerical simulations are in many cases much too small to
obtain a correct estimation of the limit $L \to \infty$. Second, it is
not easy to extend this method for computing $E^\nu_{\eps_F}$, to the
direct evaluation of $E^\nu(q)$ for a given $q$ (i.e. the energy
of a defect with a prescribed total charge). The formalism introduced in
the present article (problems~\eqref{eq:Emunu} and~\eqref{eq:minEnuq})
suggests an alternative way for computing energies of defects in
crystalline materials. A work in this direction was already started \cite{CDL2}.
\end{remark}

\section{Proof of the main results}

Unless otherwise stated, the operators used in the following proofs are
considered as operators on $L^2(\RR^3)$.

\subsection{Useful estimates}
We gather in this section some results which we shall need throughout the proofs. We start with the

\subsubsection{Proof of Lemma \ref{estim_H0per}}\label{proof_estim_H0per}
Recall that $H^0_{\rm per}=-\Delta/2+V_{\rm per}$ with $V_{\rm
  per}:=(\rho_{\gamma^0_{\rm per}}-\mu_{\rm per}) \star_\Gamma G_1\in L^\ii(\R^3)$. Thus $|H^0_{\rm per}-\kappa|\geq H^0_{\rm per}-\kappa\geq -\Delta/2- C$
for some large enough constant $C$. On the other hand, as $\kappa \in (\Sigma_{Z}^+,\Sigma_{Z+1}^-)$, there exists
  $\alpha > 0$ such that $\left| H^0_{\rm per} - \kappa \right| \ge \alpha$. This implies that
$$
|H^0_{\rm per}-\kappa|\geq \max(-\Delta/2-C,\alpha)\geq c_1(1-\Delta)
$$
for some constant $c_1>0$. The proof of the upper bound in \eqref{estim_H0per_eq} is straightforward.

Then 
$(-\Delta/2+c)^{-1}(H^0_{\rm per}-\kappa+c) = 1+(-\Delta/2+c)^{-1}(V_{\rm per}-\kappa)$
is a bounded invertible operator for $c$ large enough, since 
$$\norm{(-\Delta/2+c)^{-1}(V_{\rm per}-\kappa)}\leq \frac{\norm{V_{\rm per}}_{L^\ii}+|\kappa|}{c}.$$
Thus $(H^0_{\rm per}-\kappa+c)^{-1}(-\Delta/2+c)$ is bounded for a well-chosen $c\gg1$, which clearly implies that
$$(H^0_{\rm per}-\kappa)^{-1}(-\Delta+1)=\frac{H^0_{\rm per}-\kappa+c}{H^0_{\rm per}-\kappa}(H^0_{\rm per}-\kappa+c)^{-1}(-\Delta/2+c)\frac{-\Delta+1}{-\Delta/2+c}$$
is also bounded, together with its inverse.\qed

\subsubsection{Some commutator estimates}
Throughout this paper, we shall use Cauchy's formula to express the projector $\PPi$:
\begin{equation}
\PPi = \frac{1}{2 i \pi} \int_\curv (z-H^0_{\rm per})^{-1} \, dz,
\label{Cauchy}
\end{equation}
where $\curv$ is a fixed regular bounded closed contour enclosing the
lowest $Z$ bands of the spectrum of $H^0_{\rm per}$.

The following result will be a useful tool to replace the resolvent $(z-H^0_{\rm per})^{-1}$ with the operator $(-\Delta+1)^{-1}$ which will be easier to manipulate. Its proof is the same as the one of Lemma \ref{estim_H0per}.
\begin{lemma}\label{resolv_bd}The operator $B(z):=(z-H^0_{\rm
    per})^{-1}(-\Delta+1)$ and its inverse are bounded uniformly in
  $z\in\curv$.  
\end{lemma}

The next result provides some useful properties of commutators:
\begin{lemma}\label{lem:commut}The operators $[\gamma^0_{\rm per},\Delta]$ and $(1-\Delta)\,[\gamma^0_{\rm per},|\nabla|]\,(1-\Delta)$ are bounded.
\end{lemma}
\begin{proof} The boundedness of $[\gamma^0_{\rm per},\Delta]$ follows
  from \eqref{Cauchy} and from the fact that $[(z-H^0_{\rm per})^{-1},\Delta]$ is bounded uniformly in $z\in\curv$ by Lemma \ref{resolv_bd}.

Using again \eqref{Cauchy}, it suffices to prove that
$(1-\Delta)[(z-H^0_{\rm per})^{-1},|\nabla|](1-\Delta)$ is bounded
uniformly in $z\in\curv$ to infer that 
$(1-\Delta)\,[\gamma^0_{\rm per},|\nabla|]\,(1-\Delta)$ is bounded. In
order to prove the uniform boundedness of $(1-\Delta)[(z-H^0_{\rm
  per})^{-1},|\nabla|](1-\Delta)$, we use the formal equality
\begin{equation} 
 [(z-A)^{-1},B] =  (z-A)^{-1} [A,B] (z-A)^{-1}. 
\label{identite_commutateurs} 
\end{equation} 
We thus obtain
\begin{equation}
(1-\Delta)[(z-H^0_{\rm per})^{-1},|\nabla|](1-\Delta)  =  B(z)^* \left[|\nabla| ,V_{\rm per}\right] B(z). \end{equation}
Using \eqref{Cauchy} and lemma~\ref{resolv_bd}, we obtain
$$\|(1-\Delta)[\PPi,|\nabla|](1-\Delta)\|\leq C\norm{\left[|\nabla| ,V_{\rm per}\right]}\leq C\norm{\nabla V_{\rm per}}_{L^\ii(\R^3)}.$$
\end{proof}

\begin{lemma}\label{lem:commut1}
Let $V=V_1+V_2$ with $V_1\in \cC'$ and $V_2\in L^2(\R^3)$. Then $[\PPi,V]\in \gS_2$ and there exists a positive real constant $C$ such that
$$
\| [\PPi,V] \|_{\gS_2} \le C(\| V_1 \|_{\cC'}+\norm{V_2}_{L^2(\R^3)}).
$$
\end{lemma}
\begin{proof}
Formulas \eqref{Cauchy} and \eqref{identite_commutateurs} lead to
\begin{eqnarray*}
[\PPi,V_2] & = & - \frac{1}{4 i \pi} \int_\curv B(z)(-\Delta+1)^{-1}
[\Delta,V_2](-\Delta+1)^{-1}B(z)^*\, dz \\
& = & -\frac{1}{4 i \pi} \int_\curv B(z)((-\Delta+1)^{-1} \Delta)
V_2(-\Delta+1)^{-1}B(z)^*\, dz \\
& & 
\qquad\qquad\qquad+ \frac{1}{4 i \pi} \int_\curv B(z)(-\Delta+1)^{-1} V_2 (\Delta(-\Delta+1)^{-1}) B(z)^*\, dz.
\end{eqnarray*}
As $(-\Delta +1)^{-1}$ and $(-\Delta+1)^{-1}\Delta$ are bounded
operators, we obtain, using Lemma~\ref{resolv_bd},
$$\norm{[\PPi,V_2]}_{\gS_2}\leq C \norm{(-\Delta+1)^{-1} V_2}_{\gS_2},$$
for some constant $C$ independent of $V_2$.
Likewise,
\begin{eqnarray*}
[\PPi,V_1] & = & -\frac{1}{4 i \pi} \int_\curv B(z)(-\Delta+1)^{-1}
[\Delta,V_1](-\Delta+1)^{-1}B(z)^*\, dz \\
& = & -\sum_{i=1}^3 \frac{1}{4 i \pi} \int_\curv B(z)
\left((-\Delta+1)^{-1} \partial_{x_i} \right)
\frac{\partial V_1}{\partial{x_i}} (-\Delta+1)^{-1}B(z)^*\, dz \\
& & \qquad+ \sum_{i=1}^3 \frac{1}{4 i \pi} \int_\curv
B(z)(-\Delta+1)^{-1} \frac{\partial V_1}{\partial{x_i}} \left( 
\partial_{x_i} (-\Delta+1)^{-1} \right)) B(z)^*\, dz,
\end{eqnarray*}
which implies
$$\norm{[\PPi,V_1]}_{\gS_2}\leq C \norm{(-\Delta+1)^{-1} \nabla
  V_1}_{\gS_2}$$
for some constant $C$ independent of $V_1$.
We then use the Kato-Seiler-Simon inequality (see \cite{SeSi} and
\cite[Thm 4.1]{Simon}) 
\begin{equation}
\forall p\geq2,\qquad \norm{f(-i\nabla)g(x)}_{\gS_p}\leq (2\pi)^{-3/p}
\norm{g}_{L^p(\R^3)}\norm{f}_{L^p(\R^3)}
\label{KSS}
\end{equation}
to infer
\begin{equation}
\norm{[\PPi,V_2]}_{\gS_2}\leq C' \norm{V_2}_{L^2(\R^3)},
\label{estim_L2}
\end{equation}
\begin{equation}
\norm{[\PPi,V_1]}_{\gS_2}\leq C'\norm{\nabla
  V_1}_{L^2(\R^3)}=C'\norm{V_1}_{\cC'}. 
\label{estim_L1}
\end{equation}
\end{proof}

\begin{lemma}\label{lem:commut2}
The operator $[\, |H^0_{\rm per}-\kappa|,|\nabla|\,]$ is bounded for any $\kappa$ in the gap $(\Sigma^+_Z,\Sigma^-_{Z+1})$.
\end{lemma}
\begin{proof}
We have 
$|H^0_{\rm per}-\kappa|=-(H^0_{\rm per}-\kappa)\PPi+(H^0_{\rm per}-\kappa)(1-\PPi)$
and thus
\begin{eqnarray*}
[ \, |H^0_{\rm per}-\kappa|,|\nabla|\,] &= & -2(H^0_{\rm per}-\kappa)[\PPi,|\nabla|]+[|\nabla|,V_{\rm per}](2\PPi-1)\\
 & = & 2(B(\kappa)^*)^{-1}(1-\Delta)[\PPi,|\nabla|]+[|\nabla|,V_{\rm per}](2\PPi-1)
\end{eqnarray*}
which gives the result since $\norm{[|\nabla|,V_{\rm per}]}\leq \norm{\nabla V_{\rm per}}_{L^\ii(\R^3)}$ and $(1-\Delta)[\PPi,|\nabla|]$ is bounded by Lemma \ref{lem:commut}.
\end{proof}

\subsection{Proof of Proposition \ref{cont_rho}}\label{proof_cont_rho}
Let $V=V_1+V_2$ where $V_1\in\cC'$ and $V_2\in L^2(\R^3)\cap
L^\ii(\R^3)$, and $Q\in\cQ$. Notice that
$$(QV)^{++}=Q^{++} V(\gamma_{\rm per}^0)^\perp+Q^{+-}[\gamma_{\rm per}^0, V](\gamma_{\rm per}^0)^\perp,$$
\begin{equation}
(QV)^{--}=Q^{--} V\gamma_{\rm per}^0-Q^{-+}[\gamma_{\rm per}^0, V]\gamma_{\rm per}^0,
\label{Q--_term}
\end{equation}
where $(\gamma_{\rm per}^0)^\perp=1-\gamma_{\rm per}^0$.
We only treat the $(QV)^{--}$ term, the argument being the same for $(QV)^{++}$.

First we write
$
Q^{--} V\gamma_{\rm per}^0  =  Q^{--}(1+|\nabla|)(1+|\nabla|)^{-1}V\gamma^0_{\rm per}
$
and notice that $(1+|\nabla|)^{-1}V$ is bounded since $V_2\in L^\ii(\R^3)$ by assumption and 
$$\norm{(1+|\nabla|)^{-1}V_1}_{\gS_6}\leq C\norm{V_1}_{L^6}\leq C'\norm{\nabla V_1}_{L^2}=C'\norm{V_1}_{\cC'}$$
by the Kato-Seiler-Simon inequality \eqref{KSS} and the critical Sobolev embedding of $H^1(\R^3)$ in $L^6(\R^3)$. This proves that $Q^{--} V\gamma_{\rm per}^0$ is a trace-class operator. Thus the following is true:
\begin{eqnarray*}
|\tr(Q^{--} V\gamma_{\rm per}^0)| & = & |\tr(Q^{--} V)|\\
 & = & |\tr((1+|\nabla|)Q^{--}(1+|\nabla|)(1+|\nabla|)^{-1} V(1+|\nabla|)^{-1})|\\
 & \leq & \norm{Q}_{\cQ}\norm{(1+|\nabla|)^{-1} V(1+|\nabla|)^{-1}}_{\gS_\ii}.
\end{eqnarray*}
Then
\begin{eqnarray*}
\norm{(1+|\nabla|)^{-1} V_1 (1+|\nabla|)^{-1}}_{\gS_\ii} & \leq &
\norm{(1+|\nabla|)^{-1} V_1}_{\gS_6} \norm{(1+|\nabla|)^{-1}}
\\
& \leq & C \norm{V_1}_{L^6} \le C'\norm{V_1}_{\cC'},
\end{eqnarray*}
and
\begin{eqnarray*}
\norm{(1+|\nabla|)^{-1} V_2(1+|\nabla|)^{-1}}_{\gS_\ii} & \leq &
\norm{(1+|\nabla|)^{-1} |V_2| (1+|\nabla|)^{-1}}_{\gS_\ii} \\
& \leq & \norm{(1+|\nabla|)^{-1} |V_2|^{1/2}}_{\gS_4}^2 \leq
C\norm{V_2}_{L^2}.
\end{eqnarray*}
Hence, 
$$
|\tr(Q^{--} V\gamma_{\rm per}^0)| \le C \, \norm{Q}_{\cQ}
(\norm{V_1}_{\cC'}+\norm{V_2}_{L^2}).
$$
For the second term of \eqref{Q--_term}, we just use Lemma \ref{lem:commut1} which tells us that $Q^{-+}[\gamma_{\rm per}^0, V]\gamma_{\rm per}^0\in\gS_1$ since $Q^{-+}\in\gS_2$ and $[\gamma_{\rm per}^0, V]\in\gS_2$. Additionally
$$|\tr(Q^{-+}[\gamma_{\rm per}^0, V]\gamma_{\rm per}^0)|\leq\norm{Q^{-+}[\gamma_{\rm per}^0, V]\gamma_{\rm per}^0}_{\gS_1}\leq C\norm{Q^{-+}}_{\gS_2}(\norm{V_1}_{\cC'}+\norm{V_2}_{L^2}).$$
The end of the proof of Proposition \ref{cont_rho} is then obvious.
\qed

\subsection{Proof of Lemma \ref{lemma_dense}}\label{proof_lemma_dense}
Let $Q \in \cQ$. For $\epsilon > 0$, we introduce the following
regularization operator 
\begin{equation}
R_\epsilon:=(1+\epsilon|H^0_{\rm per}-\Sigma|)^{-1}
\label{def_R_epsilon} 
\end{equation}
and set 
$$Q_\epsilon:= R_\epsilon Q R_\epsilon.$$
Notice first that $Q_\epsilon\in\cQ_{\rm r}$. Indeed, using the same notation
as in Lemma~\ref{resolv_bd}, we obtain
$$(1-\Delta)R_\epsilon=(1-\Delta)(H^0_{\rm per}-\Sigma)^{-1}\frac{H^0_{\rm per}-\Sigma}{1+\epsilon|H^0_{\rm per}-\Sigma|}=-B(\Sigma)^*\frac{H^0_{\rm per}-\Sigma}{1+\epsilon|H^0_{\rm per}-\Sigma|}$$
which shows that
$\|(1-\Delta)R_\epsilon\|\leq \epsilon^{-1}\|B(\Sigma)^*\|.$
Similarly, $\|R_\epsilon(1-\Delta)\|\leq \epsilon^{-1}\|B(\Sigma)\|$.
As $R_\epsilon$ commutes with $\gamma^0_{\rm per}$, we infer
\begin{equation*}
(1-\Delta) Q_\epsilon^{--} (1-\Delta) = 
 (1-\Delta) R_\epsilon Q^{--} R_\epsilon (1-\Delta)\in \gS_1.
\end{equation*}
Likewise, $(1-\Delta) Q_\epsilon^{++} (1-\Delta) \in \gS_1$.
Then we show that $Q_\epsilon\in\cK_{\rm r}\subset\cK$ when $Q\in\cK$. To
prove this, we use the fact that $-\gamma^0_{\rm per}\leq Q\leq
1-\gamma^0_{\rm per}$ is equivalent to $Q^2\leq
Q^{++}-Q^{--}$ (see Section~\ref{sec:def_rHF_def}). As $\| R_\epsilon \|
\le 1$, we obtain 
\begin{equation}
(Q_\epsilon)^2  =  R_\epsilon Q(R_\epsilon)^2Q R_\epsilon \leq   R_\epsilon Q^2 R_\epsilon\\
 \leq  R_\epsilon(Q^{++}-Q^{--})R_\epsilon=(Q_\epsilon)^{++}-(Q_\epsilon)^{--}
\end{equation}
where we have used that $(R_\epsilon)^2\leq 1$ and that $\gamma^0_{\rm per}$ commutes with $R_\epsilon$.
Hence, it only remains to prove that $Q_\epsilon\to Q$ for the $\cQ$-topology as $\epsilon\to0$, for any fixed $Q\in\cQ$. We shall need the
\begin{lemma}\label{lemma_approx}
For any $1\leq p<\ii$ and any fixed $Q\in\gS_p$, one has
\begin{equation}
 \lim_{\epsilon\to0}\norm{R_\epsilon Q-Q}_{\gS_p}=0.
\label{limit_regul}
\end{equation}
\end{lemma}
\begin{proof}
Notice that 
$$R_\epsilon-1=-\frac{\epsilon|H^0_{\rm per}-\Sigma|}{1+\epsilon|H^0_{\rm per}-\Sigma|}$$
satisfies $\|R_\epsilon-1\|\leq 1$. Hence 
$\norm{(R_\epsilon-1)Q}_{\gS_p}\leq \norm{Q}_{\gS_p}.$
By linearity and density of ``smooth'' finite rank operators in $\gS_p$ for any $1\leq p<\ii$, it suffices to prove \eqref{limit_regul} for $Q=|f\rangle\langle f|$ with $f\in H^2(\R^3)$. Then
$$\norm{(R_\epsilon-1)|f\rangle\langle f|}_{\gS_1} \leq \norm{(R_\epsilon-1)f}_{L^2}\norm{f}_{L^2}\leq \epsilon\norm{|H^0_{\rm per}-\Sigma|f}_{L^2}\norm{f}_{L^2}$$
which converges to 0 as $\epsilon\to0$ and controls $\norm{(R_\epsilon-1)|f\rangle\langle f|}_{\gS_p}$ for $1\leq p< \ii$.
\end{proof}

We are now able to complete the proof of Lemma \ref{lemma_dense}. First, by \eqref{identite_commutateurs}
$$[R_\epsilon,|\nabla|]=-\epsilon R_\epsilon\; \left[|H^0_{\rm per}-\Sigma|,|\nabla|\right]\; R_\epsilon$$
and therefore by Lemma \ref{lem:commut2} there exists a constant $C>0$ such that
\begin{equation}
\|\;[R_\epsilon,|\nabla|]\;\|\leq C\epsilon.
\label{estim_R_epsilon_commut} 
\end{equation}
Hence, $\lim_{\epsilon\to0}\|\;[R_\epsilon,|\nabla|]\;\|=0$. Compute now for instance
\begin{eqnarray*}
|\nabla|(R_\epsilon QR_\epsilon -Q)^{--}|\nabla| & = &  |\nabla|((R_\epsilon -1)Q^{--}R_\epsilon +Q^{--}(R_\epsilon-1))|\nabla|\\
 & = &  [|\nabla|,R_\epsilon]Q^{--}[R_\epsilon,|\nabla|]+[|\nabla|,R_\epsilon]Q^{--}|\nabla|R_\epsilon\\
 & & \quad+(R_\epsilon-1)|\nabla|Q^{--}[R_\epsilon,|\nabla|] + |\nabla|Q^{--}[R_\epsilon,|\nabla|]\\
 & & \quad+(R_\epsilon-1)|\nabla|Q^{--}|\nabla|R_\epsilon+|\nabla|Q^{--}|\nabla|(R_\epsilon-1).
\end{eqnarray*}
Applying either \eqref{estim_R_epsilon_commut} or Lemma \ref{lemma_approx} to each term of the previous expression allows to conclude that 
$$\lim_{\epsilon\to0}\norm{|\nabla|(Q_\epsilon^{--}-Q^{--})|\nabla|}_{\gS_1}=0.$$ 
The proof is the same for the other terms in the definition of $\norm{\cdot}_\cQ$. 
\qed

\subsection{Proof of Proposition \ref{regul_solution}: regularity of solutions}\label{proof_regul_solution}
Let $Q\in\cQ$ be of the form
$$Q=\chi_{(-\ii,\epsilon_F)}(H^0_{\rm per}+V)-\PPi+\delta$$
where $0\leq\delta\leq 1$ is a finite rank operator with $\text{Ran}(\delta)\subseteq\ker(H^0_{\rm per}+V-\epsilon_F)$ and $V=\rho\ast|\cdot|^{-1}$ for some $\rho\in L^1(\R^3)\cap L^2(\R^3)$ (in our case $\rho=\rho_Q-\nu$). Note that 
$V\in \cC'\cap L^\ii(\R^3)$ since
\begin{eqnarray*}
\norm{V}_{L^\ii}&\leq&(2\pi)^{-3/2}\int_{\R^3}|\widehat{V}(k)|dk=C\int_{\R^3}\frac{|\widehat{\rho}(k)|}{|k|^2}dk\\
&\leq&C\left(\int_{\R^3}\frac{|\widehat{\rho}(k)|^2(1+|k|^2)}{|k|^2}dk\right)^{1/2}
\left(\int_{\R^3}\frac{dk}{|k|^2(1+|k|^2)}\right)^{1/2} < \infty.
\end{eqnarray*}

Since $\ker(H^0_{\rm per}+V-\epsilon_F)\subseteq \cD(H^0_{\rm per}+V)=\cD(H^0_{\rm per})=H^2(\R^3)$, it is clear that the finite rank operator $\delta$ satisfies $(1-\Delta)\delta(1-\Delta)\in\gS_1$. Thus, up to a change of $\epsilon_F$, we can assume that $\ker(H^0_{\rm per}+V-\epsilon_F)=\{0\}$ and that $\delta=0$:
$$Q=\chi_{(-\ii,\epsilon_F)}(H^0_{\rm per}+V)-\PPi.$$
Then we remark that $Q^2=Q^{++}-Q^{--}$, hence we only have to prove that $(1-\Delta)Q\in\gS_2$.

Let $\curv$ be a smooth curve enclosing the whole spectrum of $H^0_{\rm per}+V$ below $\epsilon_F$. Since $V\in L^\ii(\R^3)$ and $|H^0_{\rm per}+V-z|\geq c>0$ uniformly in $z\in\curv$,
we can mimic the proof of Lemma \ref{resolv_bd} and find that 
\begin{equation}
 \sup_{z\in\curv}\|(1-\Delta)(H^0_{\rm per}+V-z)^{-1}\|<\ii.
\label{estim_Cauchy_regul}
\end{equation}
We then use Cauchy's formula \eqref{Cauchy} and iterate the resolvent
formula 
$$
(z-H^0_{\rm per} - V)^{-1} = 
(z-H^0_{\rm per})^{-1} + (z-H^0_{\rm per} - V)^{-1} V (z-H^0_{\rm per})^{-1}
$$
to obtain
$$
Q  =  \frac{1}{2i\pi}\int_\curv\bigg((z-H^0_{\rm per}-V)^{-1}-(z-H^0_{\rm per})^{-1}\bigg)dz\\
 =  Q_1+Q_2+Q_3
$$
with
$$ Q_1=\frac{1}{2i\pi}\int_\curv(z-H^0_{\rm per})^{-1}V(z-H^0_{\rm per})^{-1}dz,$$
$$Q_2=\frac{1}{2i\pi}\int_\curv(z-H^0_{\rm per})^{-1}V(z-H^0_{\rm per})^{-1}V(z-H^0_{\rm per})^{-1}dz,$$
$$Q_3=\frac{1}{2i\pi}\int_\curv(z-H^0_{\rm per}-V)^{-1}V(z-H^0_{\rm per})^{-1}V(z-H^0_{\rm per})^{-1}V(z-H^0_{\rm per})^{-1}dz.$$
Notice that $(1-\Delta)Q_3\in\gS_2$ by \eqref{estim_Cauchy_regul} and
the estimate 
$$\norm{V(z-H^0_{\rm per})^{-1}}_{\gS_6}\leq \norm{V(1-\Delta)^{-1}}_{\gS_6}\norm{B(z)^*}\leq C\norm{V}_{L^6}$$
where we have used \eqref{KSS} and Lemma \ref{resolv_bd}.

Let us now prove that $(1-\Delta)Q_1\in\gS_2$. First we notice that
\begin{multline*}
\int_\curv(z-H^0_{\rm per})^{-1}\PPi V\PPi(z-H^0_{\rm per})^{-1}dz\\
=\int_\curv(z-H^0_{\rm per})^{-1}(\PPi)^\perp V(\PPi)^\perp(z-H^0_{\rm per})^{-1}dz=0 
\end{multline*}
by the residuum formula. Then we have
\begin{multline*}
(1-\Delta)\int_\curv(z-H^0_{\rm per})^{-1}\PPi V(\PPi)^\perp(z-H^0_{\rm per})^{-1}dz\\
=\int_\curv B(z)^*[\PPi,V](\PPi)^\perp(z-H^0_{\rm per})^{-1}dz 
\end{multline*}
which belongs to $\gS_2$ by Lemmas \ref{resolv_bd} and \ref{lem:commut1}.
The proof is the same for $Q_2$.\qed

\subsection{Proof of Theorem \ref{exists_min}: existence of a minimizer with chemical potential}\label{proof_exists_min}
Let $(Q_n)_{n \in \NN}$ be a minimizing sequence for
\eqref{eq:Emunu}. It follows from~\eqref{estim_below} that
$(Q_n)_{n \in \NN}$ is bounded in $\cQ$. By Proposition~\ref{cont_rho},
$(\rho_{Q_n})_{n \in \NN}$ is bounded in $\cC\cap L^2(\R^3)$. Up
to extraction, we can assume that there exists $\bar Q$ in the convex
set $\cK$ such that 

\begin{tabular}{ll}
i) & $Q_n\wto \bar Q$ and $|\nabla|Q_n\wto|\nabla|\bar Q$ weakly in $\gS_2$;\\
ii) & $|H^0_{\rm per}-{\epsilon_{F}}|^{1/2}Q_n^{++}|H^0_{\rm per}-{\epsilon_{F}}|^{1/2}\wto |H^0_{\rm per}-{\epsilon_{F}}|^{1/2}\bar Q^{++}|H^0_{\rm per}-{\epsilon_{F}}|^{1/2}$,\\
 & $|H^0_{\rm per}-{\epsilon_{F}}|^{1/2}Q_n^{--}|H^0_{\rm per}-{\epsilon_{F}}|^{1/2}\wto |H^0_{\rm per}-{\epsilon_{F}}|^{1/2}\bar Q^{--}|H^0_{\rm per}-{\epsilon_{F}}|^{1/2}$\\
 &  for the weak-$\ast$ topology of $\gS_1$;\\
iii) & $\rho_{Q_n}  \rightharpoonup \rho_{\bar Q}$ in $\cC\cap L^2(\R^3)$.
\end{tabular}

\noindent Recall that $\gS_1$ is the dual of the space of compact operators \cite[Thm VI.26]{RS1}. Thus here $A_n\wto A$ for the weak-$\ast$ topology of $\gS_1$ means $\tr(A_nK)\to\tr(AK)$ for any compact operator $K$. 

Then, as $D(\cdot,\cdot)$ defines a scalar product on $\cC$, 
$$
D(\rho_{\bar Q}-\nu,\rho_{\bar Q}-\nu) \le \liminf_{n \rightarrow +\infty}
D(\rho_{Q_n}-\nu,\rho_{Q_n}-\nu).
$$
Now since $Q_n^{++}\geq0$, $|H^0_{\rm per}-{\epsilon_{F}}|^{1/2}Q_n^{++}|H^0_{\rm per}-{\epsilon_{F}}|^{1/2}$ is also a nonnegative operator for any $n$. Thus Fatou's Lemma \cite{Simon} yields
\begin{equation*}
  \tr(|H^0_{\rm per}-{\epsilon_{F}}|^{1/2}\bar Q^{++}|H^0_{\rm per}-{\epsilon_{F}}|^{1/2})\\
\leq  \liminf_{n\to\ii}\tr(|H^0_{\rm per}-{\epsilon_{F}}|^{1/2}Q_n^{++}|H^0_{\rm per}-{\epsilon_{F}}|^{1/2}).
\end{equation*}
The same argument for the term involving $-\bar Q^{--}\geq0$ yields
$$\cE^\nu(\bar Q)-{\epsilon_{F}}\tr_0(\bar Q)\leq \liminf_{n\to\ii}(\cE^\nu(Q_n)-{\epsilon_{F}}\tr_0(Q_n))=E^\nu_{\epsilon_{F}},$$
i.e. $\bar Q\in\cK$ is a minimizer. 

The proof that $\bar Q$ satisfies the self-consistent equation \eqref{scf_defauts} is classical: writing that $\cE^\nu((1-t)\bar Q+tQ)\geq\cE^\nu(\bar Q)$ for any $Q\in\cK_{\rm r}$ and $t\in[0,1]$, one deduces that $\bar Q$ minimizes the following linear functional
\begin{equation}
Q\in\cK \mapsto F(Q):= \tr(|H^0_{\rm per}-{\epsilon_{F}}|^{1/2}(Q^{++}-Q^{--})|H^0_{\rm per}-{\epsilon_{F}}|^{1/2})+D(\rho_{\bar Q}-\nu,\rho_Q).\label{linearized_fn}
\end{equation}
Notice that when $Q\in\cK_{\rm r}\subseteq\cK$, one has
$$F(Q)=\tr_0((H_{\bar Q}-{\epsilon_{F}})Q)$$
where we have used the definition of $\rho_Q$ in
Proposition~\ref{cont_rho} to infer 
$$D(\rho_{\bar Q}-\nu,\rho_Q)=\tr_0 \left( \left( (\rho_{\bar Q}-\nu) \star
  |\cdot|^{-1} \right)  Q \right),$$
since $(\rho_{\bar Q}-\nu)\star|\cdot|^{-1}\in\cC'$ when $\rho_{\bar Q}-\nu\in\cC$. Minimizers of the functional \eqref{linearized_fn} are easily proved to be of the form \eqref{scf_defauts}. They belong to $\cK_{\rm r}$ by Proposition \ref{regul_solution}. \qed

\subsection{Proof of Proposition \ref{energy_free}: the value of $E^0(q)$}\label{proof_energy_free}
Clearly
\begin{equation}
E^0(q)\geq \inf\{\tr_0(H^0_{\rm per}Q),\ Q\in\cK,\ \tr_0(Q)=q\}:=\tilde E^0(q)
\label{estim_below_linear}
\end{equation}
since $D(\rho_Q,\rho_Q)\geq0$. It can be easily proved that 
$$\tilde E^0(q)=\left\{\begin{array}{ll}
\Sigma^-_{Z+1}q & {\rm when\ } q\geq0\\
\Sigma^+_Zq & {\rm when\ } q\leq0,\\
\end{array}\right.$$
see, e.g., the proof of Lemma 13 in \cite{HLS3}.

Thus it remains to prove that $E^0(q)\leq \tilde E^0(q)$ which we do by
a kind of scaling argument. Let us deal with the case $q\geq0$, the
other case being similar. We can assume that  
$\Sigma_{Z+1}^-=\min_{\xi\in\Gamma^*}\lambda_{Z+1}(\xi)=\lambda_{Z+1}(\xi_0)$
since each $\lambda_n(\xi)$ is known to be continuous on $\Gamma^*$. For simplicity, we also assume that $\xi_0$ is in the interior of $\Gamma^*$ (the proof can be easily adapted if this is not the case). Let us denote by $u_{Z+1}(\xi,\cdot)\in L^2_{\xi}$ an eigenvector associated with the eigenvalue $\lambda_{Z+1}(\xi)$ for any $\xi\in\Gamma^*$. It will be convenient to extend it on $\R^3\times\R^3$ by $u_{Z+1}(\xi,x)=0$ when $\xi\in\R^3\setminus\Gamma^*$.
Since $H^0_{\rm per}u_{Z+1}(\xi,x)=\lambda_{Z+1}(\xi)u_{Z+1}(\xi,x)$ for
any $\xi\in\Gamma^*$, it is clear that  
$$\sup_{\xi\in\R^3}\norm{\Delta u_{Z+1}(\xi,\cdot)}_{L^2_{\xi}(\Gamma)}<\ii,$$ 
i.e. $u_{Z+1}\in L^\ii(\R^3,H^2_{\rm loc}(\R^3))$ and $u_{Z+1}\in
    L^\ii(\R^3\times\R^3)$ by the $\Z^3$-periodicity (resp. the
    $2\pi\Z^3$-periodicity) of $e^{-i\xi\cdot
    x}u_{Z+1}(\xi,x)$ with respect to~$x$ (resp.~to~$\xi$). 

Consider now a fixed space $V$ of dimension $d=[q]+1$, consisting of
$\cC^\ii_0$ functions $f \; : \; \R^3 \rightarrow \C$ with support in
the unit ball $B(0,1)$ of $\R^3$. For any $\lambda\geq1$, we introduce the
following subspace of $L^2(\R^3)$: 
$$W_\lambda:=\left\{g_\lambda := \lambda^{3/2}\int_{\R^3} f(\lambda(\xi-\xi_0))u_{Z+1}(\xi,\cdot)d\xi,\quad f\in V\right\}$$
which has the same dimension as $V$ by the properties of the Bloch
decomposition, when $\lambda$ is large enough such that the ball
$B(\xi_0,\lambda^{-1})$ is contained in $\Gamma^*$. Noting that for any
$g_\lambda \in W_\lambda$ arising from some $f\in V$
\begin{eqnarray*}
|g_\lambda(x)| & \leq & \lambda^{-3/2}\int_{\R^3}
 |f(\xi')u_{Z+1}(\xi_0+\lambda^{-1} \xi' ,x)|d\xi'\\
 & \leq & \lambda^{-3/2} \norm{u_{Z+1}}_{L^\ii(\R^3\times\R^3)}\int_{B(0,1)}|f(\xi)|d\xi,\\
 & \leq & C\lambda^{-3/2} \norm{u_{Z+1}}_{L^\ii(\R^3\times\R^3)}\norm{f}_{L^2(\R^3)},
\end{eqnarray*}
we deduce by interpolation that
$$\forall\ 2<p\leq\ii,\qquad \lim_{\lambda\to\ii}\sup_{\substack{g\in W_\lambda\\ \norm{g}_{L^2(\R^3)}=1}}\norm{g}_{L^p(\R^3)}=0.$$
By construction one also has for any fixed $f\in V$ with associated
$g_\lambda \in W_\lambda$
$$\pscal{g_\lambda,(H^0_{\rm per}-\Sigma_{Z+1}^-)g_\lambda}=\lambda^{3}\int_{\R^3} |f(\lambda(\xi-\xi_0))|^2(\lambda_{Z+1}(\xi)-\Sigma_{Z+1}^-)d\xi\to_{\lambda\to\ii}0.$$

Take now an orthonormal basis $(\phi_1^\lambda,...,\phi_{[q]+1}^\lambda)$ of $W_\lambda$ and introduce the operator
$$Q^\lambda:=\sum_{n=1}^{[q]}|\phi_n^\lambda\rangle\langle\phi_n^\lambda|+(q-[q])|\phi_d^\lambda\rangle\langle\phi_d^\lambda|.$$
By construction $\gamma^0_{\rm per}\phi_n^\lambda=0$ for any $n=1,...,[q]+1$ and $\tr_0(Q^\lambda)=\tr(Q^\lambda)=q$, thus $Q^\lambda\in\cK$ satisfies the charge constraint. Then
$$\tr_0(H^0_{\rm
  per}Q^\lambda)=\sum_{n=1}^{[q]}\pscal{\phi_n^\lambda,H^0_{\rm
    per}\phi_n^\lambda}+(q-[q])\pscal{\phi_{[q]+1}^\lambda,H^0_{\rm
    per}\phi_{[q]+1}^\lambda}\to \Sigma^-_{Z+1}q$$ 
as $\lambda\to\ii$. Besides, $(\rho_{Q^\lambda}^{1/2})_{\lambda \ge 1}$
  is a bounded 
  family in $H^1(\R^3)$ which converges to~$0$ in $L^p(\R^3)$ for any $p>2$. By the Hardy-Littlewood-Sobolev inequality \cite{LL}, one has
\begin{equation}
D(\rho_{Q^\lambda},\rho_{Q^\lambda})\leq C\norm{\rho_{Q^\lambda}}_{L^{6/5}}^2
\end{equation}
which implies $D(\rho_{Q^\lambda},\rho_{Q^\lambda})\to0$ as $\lambda\to\ii$.
Eventually $\cE^0(Q^\lambda)\to \tilde E^0(q)$ and Proposition \ref{energy_free} is proved.\qed

\subsection{Density of finite rank operators in $\cK$}\label{sec_density}
This section is devoted to the generalization of results in \cite[Appendix B]{HLS3} concerning the density of finite rank operators, which will be useful for proving Theorem \ref{HVZ}.

\begin{lemma}\label{decomp_quelconque}
For any $Q\in\cK$ there exists an orthogonal projector $P$ such that $P-\PPi\in\cK$ and a trace-class operator $\delta\in\cQ$ such that $0\leq\delta\leq1$, $[P,\delta]=0$ and
$$Q=P-\PPi+\delta.$$ 
\end{lemma}
\begin{proof}
 This is an easy adaptation of \cite[Lemma 19]{HLS3}.
\end{proof}

We denote for simplicity $\gH_+=(1-\PPi) L^2(\R^3)$ and $\gH_-=\PPi L^2(\R^3)$.
\begin{proposition}\label{decomp_proj}
Let $P$ be an orthogonal projector on $L^2(\R^3)$ such that $Q=P-\PPi\in\cK$. Denote by $(f_1,...,f_N)\in (\gH_+\cap H^1(\R^3))^N$ an orthonormal basis of $E_{1}=\ker(P-\PPi-1)=\ker(\PPi)\cap\ker(1-P)$ and by $(g_1,...,g_M)\in (\gH_-\cap H^1(\R^3))^M$ an orthonormal basis of $E_{-1}=\ker(P-\PPi+1)=\ker(1-\PPi)\cap\ker(P)$. Then there exist an orthonormal basis $(v_i)_{i\geq1}\subset \gH_+\cap H^1(\R^3)$ of $(E_{1})^\perp$ in $\gH_+$, an orthonormal basis $(u_i)_{i\geq1}\subset \gH_-\cap H^1(\R^3)$ of $(E_{-1})^\perp$ in $\gH_-$, and a sequence $(\lambda_i)_{i\geq 1}\in \ell_2(\R^+)$ 
such that
\begin{equation}
P=\sum_{n=1}^N|f_n\rangle\langle f_n|+\sum_{i=1}^\ii\frac{|u_i+\lambda_iv_i\rangle\langle u_i+\lambda_iv_i|}{1+\lambda_i^2},
\label{reduc_P}
\end{equation} 
\begin{equation}
1-P=\sum_{m=1}^M|g_m\rangle\langle g_m|+\sum_{i=1}^\ii\frac{|v_i-\lambda_iu_i\rangle\langle v_i-\lambda_iu_i|}{1+\lambda_i^2}.
\label{reduc_PP}
\end{equation} 
Additionally
$\sum_{i\geq1}\lambda_i^2(\|\nabla u_i\|^2_{L^2}+\|\nabla v_i\|^2_{L^2})<\ii.$
\end{proposition}
\begin{proof}
 This is an obvious corollary of \cite[Theorem 7]{HLS3}.
\end{proof}

\begin{corollary}\label{finite_rank_dense}
Let $Q\in\cK$. Then there exists a sequence $\{Q_k\}_{k\geq1}$ of finite rank operators belonging to $\cK_{\rm r}$ such that $\norm{Q_k-Q}_{\cQ}\to0$ as $k\to\ii$ and for any $k\geq1$, $\tr_0(Q_k)=\tr_0(Q)$. 
\end{corollary}
\begin{proof}
Taking $\lambda_i=0$ for $i > k$ in the decomposition of Proposition \ref{decomp_proj}, one can approximate $P$ by another projector $P_k$ such that $P_k-\PPi\to P-\PPi$ in $\cQ$ as $k\to\ii$ and $P_k-\PPi$ is finite rank:
\begin{multline}
P_k-\PPi  =  \sum_{n=1}^N|f_n\rangle\langle f_n|-\sum_{m=1}^M|g_m\rangle\langle g_m|+\sum_{i=1}^k\frac{\lambda_i^2}{1+\lambda_i^2}\big(|v_i\rangle\langle v_i|-|u_i\rangle\langle u_i|\big)\\
+\sum_{i=1}^k\frac{\lambda_i}{1+\lambda_i^2}\big(|u_i\rangle\langle v_i|+|v_i\rangle\langle u_i|\big)\label{expand_P}.
\end{multline} 
It then suffices to approximate each function in \eqref{expand_P} by a smoother one, for instance by defining for $\epsilon\ll1$, $\tilde u_i:= \norm{R_\epsilon u_i}_{L^2}^{-1}R_\epsilon u_i$ 
and orthonormalizing these new functions, where $R_\epsilon$ was defined  previously in Equation \eqref{def_R_epsilon}.

Then for any $Q=P-\PPi+\delta$ of the form given by Lemma \ref{decomp_quelconque}, it remains to approximate $\delta$ by a finite rank operator $\delta_k$ such that $[P_k,\delta_k]=0$, which is done in the same way.
\end{proof}

\subsection{Proof of Theorem \ref{HVZ}: existence of minimizers under a charge constraint}\label{proof_HVZ}
The proof of Theorem \ref{HVZ} follows ideas of \cite{HLS3}. The proof that any minimizer solves \eqref{scf_defauts} is the same as before and will be omitted.

\subsubsection*{\bf Step 1: \textit{Large HVZ-type inequalities}}
Let us start by the following result, which indeed shows that \textbf{(b)}$\Rightarrow$\textbf{(c)}:
\begin{lemma}[Large HVZ-type inequalities] Let $Z \in \NN \setminus
  \left\{0\right\}$, $\nu\in\cC$ and assume that \textnormal{\textbf{(A1)}}
  holds. Then, for every $q,q'\in\R$, one has 
$$E^\nu(q)\leq E^\nu(q-q')+E^0(q').$$
If moreover there is a $q'\neq0$ such that $E^\nu(q)= E^\nu(q-q')+E^0(q')$, then there is a minimizing sequence of $E^\nu(q)$ which is not precompact.
\end{lemma}
\begin{proof}
Thanks to Corollary \ref{finite_rank_dense}, the proof is exactly the same as \cite[Prop. 6]{HLS3}.
\end{proof}

\subsubsection*{\bf Step 2: \textit{A necessary and sufficient condition for compactness}}
The following Proposition is the analogue of \cite[Lemma 8]{HLS3}:
\begin{proposition}[Conservation of charge implies
  compactness]\label{conservation} Let $Z\in \NN \setminus
  \left\{0\right\}$, $\nu\in\cC$, $q\in\R$ and assume that \textnormal{\textbf{(A1)}}
  holds. Assume that $(Q_n)_{n\geq1}$ is a minimizing sequence in $\cK_{\rm r}$
  for \eqref{eq:Emunu} such that $Q_n\wto Q\in\cK$ for the weak-$\ast$ topology of $\cQ$. Then $Q_n\to Q$ for the strong topology of $\cQ$ if and only if $\tr_0(Q)=q$.
\end{proposition}
\begin{proof}
Let $(Q_n)_{n\geq1}\subseteq\cK_{\rm r}$ be as stated and assume that $\tr_0(Q)=q$. We know from the proof of Theorem \ref{exists_min} that
\begin{equation}
\cE^\nu(Q)\leq \lim_{n\to\ii}\cE^\nu(Q_n)=E^\nu(q),
\label{ineg_wlsc}
\end{equation}
hence $Q\in\cK$ is a minimizer of $E^\nu(q)$. Therefore $Q$ satisfies the equation
$$Q=\chi_{(-\ii,{\epsilon_{F}})}(H_Q)-\gamma^0_{\rm per}+\delta$$
for some ${\epsilon_{F}}\in(\Sigma^+_Z,\Sigma^-_{Z+1})$ and where $\delta$ is a finite rank operator satisfying $0\leq\delta\leq1$ and ${\rm Ran}(\delta)\subseteq\ker(H_Q-{\epsilon_{F}})$. In particular $Q\in\cK_{\rm r}$ by Proposition \ref{regul_solution}. We now introduce
$$P:=\chi_{(-\ii,{\epsilon_{F}})}(H_Q),\quad P':=\chi_{({\epsilon_{F}},\ii)}(H_Q)\quad\text{and}\quad \pi:=\chi_{\{{\epsilon_{F}}\}}(H_Q).$$
Let us write
$$\cE^\nu(Q_n)=\cE^\nu(Q)+\tr_0(H_Q(Q_n-Q))+\frac12 D(\rho_{Q_n}-\rho_Q,\rho_{Q_n}-\rho_Q).$$
Now using \cite[Lemma 1]{HLS1} and the hypothesis $\tr_0(Q_n) =
\tr_0(Q)$, we obtain
\begin{eqnarray*}
\tr_0(H_Q(Q_n-Q)) & = & \tr_0((H_Q-\epsilon_F)(Q_n-Q)) =
\tr_P((H_Q-{\epsilon_{F}})(Q_n-Q)) \\ 
 & = & \tr(|H_Q-{\epsilon_{F}}|(P'(Q_n-Q)P'-P(Q_n-Q)P)),
\end{eqnarray*}
where we recall that by definition $\tr_P(A) = \tr(PAP+(1-P)A(1-P))$.
We have $-P-\delta \leq Q_n-Q\leq 1-P-\delta$ which yields
\begin{equation}
P(Q_n-Q)^2P+P'(Q_n-Q)^2P'\leq P'(Q_n-Q)P'-P(Q_n-Q)P
\label{estim_Q2_Q}
\end{equation}
and in particular $P'(Q_n-Q)P'\geq0$ and $P(Q_n-Q)P\leq0$, i.e.
$\tr_0(H_Q(Q_n-Q))\geq0.$
Since we know that $\lim_{n\to\ii}\cE^\nu(Q_n)=\cE^\nu(Q)$, we infer
\begin{equation}
 \lim_{n\to\ii}\tr(|H_Q-{\epsilon_{F}}|P'(Q_n-Q)P')=\lim_{n\to\ii}\tr(|H_Q-{\epsilon_{F}}|P(Q_n-Q)P)=0
\label{limit_diag_Q_n}
\end{equation}
and from \eqref{estim_Q2_Q}
$$\lim_{n\to\ii}\tr(|H_Q-{\epsilon_{F}}|P'(Q_n-Q)^2P')=\lim_{n\to\ii}\tr(|H_Q-{\epsilon_{F}}|P(Q_n-Q)^2P)=0.$$
On the one hand, it is easy to see that
$$P|H_Q-{\epsilon_{F}}|P\geq c P|H_Q-\kappa|P \quad \mbox{and} \quad 
P'|H_Q-{\epsilon_{F}}|P'\geq c P'|H_Q-\kappa|P'$$
for some small enough constant $c>0$ and some $\kappa\notin\sigma(H_Q)$ close enough to $\epsilon_F$. On the other hand, the weak
convergence of $(Q_n)$ and the fact that $\pi$ is a ``smooth'' finite rank
operator imply that
\begin{equation}
 \lim_{n\to\ii}\tr(|H_Q-\kappa|\pi(Q_n-Q)^2\pi)=\lim_{n\to\ii}\tr(|H_Q-\kappa|\pi(Q_n-Q)\pi)=0.
\label{limit_diag_Q_n_bis}
\end{equation}
It is then clear that this yields
\begin{equation}
\label{limit_Q_n_S2}
\lim_{n\to\ii}\tr(|H_Q-\kappa|(Q_n-Q)^2)=0. 
\end{equation}
As we have chosen $\kappa\notin\sigma(H_Q)$, we can mimic the proof of Lemma \ref{resolv_bd} and obtain that 
\begin{equation}
 c_1(1-\Delta)\leq|H_Q-\kappa|\leq c_2 (1-\Delta).
\label{estim_H_Q}
\end{equation}
Hence \eqref{limit_Q_n_S2} shows that $(1-\Delta)^{1/2}(Q_n- Q)\to0$ in $\gS_2$.

Writing now
\begin{multline}
(Q_n-Q)^{--} = P(Q_n-Q)P +(\gamma^0_{\rm per}-P)(Q_n-Q)\gamma^0_{\rm per}\\
-(P-\gamma^0_{\rm per})(Q_n-Q)(P-\gamma^0_{\rm per})+\gamma^0_{\rm per}(Q_n-Q)(\gamma^0_{\rm per}-P)
\label{decomp_--}
\end{multline}
and using \eqref{estim_H_Q}, \eqref{limit_diag_Q_n} and \eqref{limit_diag_Q_n_bis}, we easily see that $|H_Q-\kappa|^{1/2}(Q_n-Q)^{--}|H_Q-\kappa|^{1/2}\to0$  and $(1-\Delta)^{1/2}(Q_n-Q)^{--}(1-\Delta)^{1/2}\to0$ in $\gS_1$.
The proof is the same for $(Q_n-Q)^{++}$.
\end{proof}

\subsubsection*{\bf Step 3: \textit{Proof that} \textbf{(c)}$\Rightarrow$\textbf{(b)}} We argue by contradiction. Let $(Q_n)_{n\geq1}\subseteq\cK$ be a minimizing sequence for $E^\nu(q)$ which is not precompact for the topology of $\cQ$. By the density of $\cK_{\rm r}$ in $\cK$, we can further assume that each $Q_n\in\cK_{\rm r}$. The bound \eqref{coercive} on the energy tells us that $(Q_n)_{n\geq1}$ is bounded in $\cQ$. Then, up to extraction and by Proposition~\ref{conservation}, we can assume that $Q_n\wto Q\in\cK$ where $\tr_0(Q)\neq q$, and that $\rho_{Q_n}\wto\rho_Q$ weakly in $\cC$. We write $\tr_0(Q)=q-q'$ with $q'\neq0$. We now prove that
\begin{equation}
E^\nu(q)\geq E^\nu(q-q')+E^0(q')
\label{absurde_HVZ} 
\end{equation}
which will contradict \textbf{(c)}. To this end, we argue like in the
proof of \cite[Thm. 3]{HLS3}: consider a smooth  radial function $\chi$
with support in $B(0,1)$ such that $0\leq\chi\leq1$ and $\chi \equiv 1$
in $B(0,1/2)$; define $\chi_R(x):=\chi(x/R)$. Then let be $\eta_R:=\sqrt{1-\chi_R^2}$. Let us introduce the following localization operators
$$Y_R:=\gamma^0_{\rm per}\eta_R\gamma^0_{\rm per}+(\gamma^0_{\rm per})^\perp\eta_R(\gamma^0_{\rm per})^\perp,\qquad X_R=\sqrt{1-Y_R^2}.$$
\begin{lemma}\label{prop_XR_YR}
We have for all $3<p\leq\ii$,
\begin{equation}
\norm{[\eta_R,\gamma^0_{\rm per}]}_{\gS_p}=O(R^{-1+3/p}),\qquad \norm{Y_R-\eta_R}_{\gS_p}=O(R^{-1+3/p}).
\label{estim_commut_proj}
\end{equation}
Moreover $\norm{X_R^2-\chi_R^2}_{\gS_p}=O\left( R^{-1+3/p} \right)$.
\end{lemma}
\begin{proof}
By \eqref{Cauchy} and \eqref{identite_commutateurs}
$$[\gamma^0_{\rm per},\eta_R]= -\frac{1}{4 i \pi} \int_\curv (z-H^0_{\rm
  per})^{-1} \left[\Delta,\eta_R\right] (z-H^0_{\rm per})^{-1} \, dz$$
which yields
$\norm{[\eta_R,\gamma^0_{\rm per}]}_{\gS_p}\leq C\norm{(1-\Delta)^{-1}
  (\nabla\eta_R)}_{\gS_p}\leq C'\norm{\nabla\eta_R}_{L^p}$
by the Kato-Seiler-Simon inequality and following the proof of Lemma~\ref{lem:commut1}. Eventually we notice
$Y_R-\eta_R=\left(1-2\gamma^0_{\rm per}\right)[\gamma^0_{\rm per},\eta_R]$
and thus \eqref{estim_commut_proj} is proved since $\gamma^0_{\rm per}$ is bounded. The last inequality is a consequence of $\norm{Y_R}\leq1$, $\norm{\eta_R}\leq1$.
\end{proof}

\begin{lemma}\label{lemm_estim_commut_YR}
One has
\begin{equation}
\norm{\left[Y_R,|H^0_{\rm per}-\Sigma|^{1/2}\right]|H^0_{\rm per}-\Sigma|^{-1/2}}=O(R^{-1})
\label{estim_commut_YR}
\end{equation}
where we recall that $\Sigma=(\Sigma_Z^++\Sigma_{Z+1}^-)/2$ is the middle of the gap.
\end{lemma}
\begin{proof}
We use the well-known integral representation of the square root~\cite{Bhatia}
\begin{equation}
|H^0_{\rm per}-\Sigma|^{1/2}=\frac1\pi\int_0^\ii\frac{|H^0_{\rm per}-\Sigma|}{s+|H^0_{\rm per}-\Sigma|}\frac{ds}{\sqrt{s}}. 
\label{repr_square_root}
\end{equation}
Recall that $Y_R=\PPi\eta_R\PPi+(\PPi)^\perp\eta_R(\PPi)^\perp$. For
shortness, we only detail the estimation of the term involving $(\PPi)^\perp\eta_R(\PPi)^\perp$. Using that $|H^0_{\rm per}-\Sigma|$ commutes with $(\PPi)^\perp$ and that $(\PPi)^\perp$ is bounded, we see that it suffices to estimate
\begin{multline}
(\PPi)^\perp\int_0^\ii\left[\eta_R,\frac{|H^0_{\rm
 per}-\Sigma|}{s+|H^0_{\rm per}-\Sigma|}\right]|H^0_{\rm
 per}-\Sigma|^{-1/2}\frac{ds}{\sqrt{s}}(\PPi)^\perp \\ 
 = -\frac 1 2 \int_0^\ii\frac{(\PPi)^\perp}{s+|H^0_{\rm per}-\Sigma|}\left[\eta_R,-\Delta\right]|H^0_{\rm per}-\Sigma|^{-1/2}\frac{(\PPi)^\perp}{s+|H^0_{\rm per}-\Sigma|}\sqrt{s}ds.
\label{estim_formule_integrale}
\end{multline}
Then $\left[\eta_R,-\Delta\right]=(\Delta\eta_R)+2(\nabla\eta_R)\cdot\nabla$, hence
$\norm{\left[\eta_R,-\Delta\right]|H^0_{\rm per}-\Sigma|^{-1/2}}=O(R^{-1})$
where we have used that $\nabla|H^0_{\rm per}-\Sigma|^{-1/2}$ is a bounded operator by Lemma \ref{estim_H0per}. Then we use that $|H^0_{\rm per}-\Sigma|\geq \epsilon$ for some $\epsilon>0$ to estimate the right hand side of \eqref{estim_formule_integrale} in the operator norm by
$$\norm{\left[\eta_R,-\Delta\right]|H^0_{\rm per}-\Sigma|^{-1/2}}\times\int_0^\ii\frac{\sqrt{s}ds}{(\epsilon+s)^2}=O(R^{-1}).$$
\end{proof}

Notice now that $Y_RQ_nY_R\in\cK$ for all $R\geq1$ (the same is true for $X_RQ_nX_R$ but we shall actually not need it). To see this, notice for instance that
\begin{multline}
(1+|\nabla|) (Y_RQ_nY_R)=-(1+|\nabla|)|H^0_{\rm per}-\Sigma|^{-1/2}\bigg(\left[Y_R,|H^0_{\rm per}-\Sigma|^{1/2}\right]\times\\
\times|H^0_{\rm per}-\Sigma|^{-1/2} |H^0_{\rm per}-\Sigma|^{1/2}Q_nY_R-Y_R|H^0_{\rm per}-\Sigma|^{1/2}Q_nY_R\bigg)
\end{multline}
which belongs to $\gS_2$ since $|H^0_{\rm per}-\Sigma|^{1/2}Q_n\in\gS_2$ and $\left[Y_R,|H^0_{\rm per}-\Sigma|^{1/2}\right]|H^0_{\rm per}-\Sigma|^{-1/2}$ is bounded by Lemma \ref{lemm_estim_commut_YR}.
The proof that $(1+|\nabla|)(Y_RQ_nY_R)^{++}(1+|\nabla|)$ and $(1+|\nabla|)(Y_RQ_nY_R)^{--}(1+|\nabla|)$ are trace-class is similar. Eventually, the proof that $-\PPi\leq Y_RQY_R\leq 1-\PPi$ is easy, using the equivalent condition \eqref{equivalent_condition} and the fact that $\PPi$ commutes with $Y_R$.

We are now able to prove \eqref{absurde_HVZ} as announced. We write, following \cite{HLS2},
\begin{multline}
\cE^\nu(Q_n)  = \tr(X_R|H^0_{\rm per}-\Sigma|^{1/2}(Q_n^{++}-Q_n^{--})|H^0_{\rm per}-\Sigma|^{1/2}X_R)\\
+\tr(Y_R|H^0_{\rm per}-\Sigma|^{1/2}(Q_n^{++}-Q_n^{--})|H^0_{\rm per}-\Sigma|^{1/2}Y_R)+\Sigma\tr_0(X_RQ_nX_R)\\
+\Sigma\tr_0(Y_RQ_nY_R)+\frac12 D(\rho_{Q_n}-\nu,\rho_{Q_n}-\nu)-\frac12 D(\nu,\nu)
\label{decomp_energy_1}
\end{multline}
where we have used that $[\PPi,X_R]=[\PPi,Y_R]=0$ to infer
$\tr_0(Q_n)=\tr_0(X_RQ_nX_R)+\tr_0(Y_RQ_nY_R).$
Then, by Lemma \ref{lemm_estim_commut_YR} and using the fact that
$(Q_n)_{n \ge 1}$ is a bounded sequence in $\cQ$, we deduce that 
\begin{multline}
 \tr(Y_R|H^0_{\rm per}-\Sigma|^{1/2}(Q_n^{++}-Q_n^{--})|H^0_{\rm per}-\Sigma|^{1/2}Y_R)\\
\geq \tr(|H^0_{\rm per}-\Sigma|^{1/2}Y_R(Q_n^{++}-Q_n^{--})Y_R|H^0_{\rm per}-\Sigma|^{1/2})-C/R
\end{multline}
for some constant $C>0$. Arguing similarly for the other terms, we obtain
\begin{multline}
\cE^\nu(Q_n)  \geq\tilde\cE^0(Y_RQ_nY_R)+\tr(\chi_R|H^0_{\rm per}-\Sigma|^{1/2}(Q_n^{++}-Q_n^{--})|H^0_{\rm per}-\Sigma|^{1/2}\chi_R)\\
+\Sigma\tr(\chi_R(Q_n^{++}+Q_n^{--})\chi_R)
+\frac12 D(\rho_{Q_n}-\nu,\rho_{Q_n}-\nu)-\frac12
D(\nu,\nu)-C'/R
\label{decomp_energy_2}
\end{multline}
for some constant $C'$, where 
$$\tilde\cE^0(Q):=\tr(|H^0_{\rm per}-\Sigma|^{1/2}(Q^{++}-Q^{--})|H^0_{\rm per}-\Sigma|^{1/2})+\Sigma\tr_0(Q).$$
Recall (Proposition \ref{energy_free})
$$E^0(q)=\inf\{\tilde\cE^0(Q),\quad Q\in\cK,\quad \tr_0(Q)=q\}.$$
Thus, using
$$q=\tr_0(Q_n)=\tr_0(Y_RQ_nY_R)+\tr_0(X_RQ_nX_R),$$
and the fact that $q\mapsto E^0(q)$ is Lipschitz by Proposition \ref{energy_free}, \eqref{decomp_energy_2} yields
\begin{multline}
 \cE^\nu(Q_n)  \geq \tr(\chi_R|H^0_{\rm per}-\Sigma|^{1/2}(Q_n^{++}-Q_n^{--})|H^0_{\rm per}-\Sigma|^{1/2}\chi_R)\\
+\Sigma\tr(\chi_R(Q_n^{++}+Q_n^{--})\chi_R)+E^0\big(q-\tr(\chi_R(Q_n^{++}+Q_n^{--})\chi_R)\big)\\
+\frac12 D(\rho_{Q_n}-\nu,\rho_{Q_n}-\nu)-\frac12 D(\nu,\nu)-C'/R
\label{decomp_energy_3}
\end{multline}
Let us now pass to the limit $n\to\ii$. First we notice 
\begin{multline}
 \liminf_{n\to\ii}\tr(\chi_R|H^0_{\rm per}-\Sigma|^{1/2}(Q_n^{++}-Q_n^{--})|H^0_{\rm per}-\Sigma|^{1/2}\chi_R)\\
\geq \tr(\chi_R|H^0_{\rm per}-\Sigma|^{1/2}(Q^{++}-Q^{--})|H^0_{\rm per}-\Sigma|^{1/2}\chi_R),
\end{multline}
\begin{equation}
 \liminf_{n\to\ii} D(\rho_{Q_n}-\nu,\rho_{Q_n}-\nu)\geq D(\rho_{Q}-\nu,\rho_{Q}-\nu)
\end{equation}
by Fatou's Lemma and the weak convergence $\rho_{Q_n}\wto\rho_Q$ in $\cC$.
Then
$$\lim_{n\to\ii}\tr(\chi_RQ_n^{++}\chi_R)=\tr(\chi_RQ^{++}\chi_R),\  \lim_{n\to\ii}\tr(\chi_RQ_n^{--}\chi_R)=\tr(\chi_RQ^{--}\chi_R)$$
which is obtained by writing for instance
$$\tr(\chi_RQ_n^{++}\chi_R)=\tr(\chi_R(1+|\nabla|)^{-1}(1+|\nabla|)Q_n^{++}(1+|\nabla|)(1+|\nabla|)^{-1}\chi_R)$$
and using that $\chi_R(1+|\nabla|)^{-1}$ is compact (it belongs to $\gS_p$ for $p>3$ by the Kato-Seiler-Simon inequality) and that
$$(1+|\nabla|)Q_n^{++}(1+|\nabla|)\wto(1+|\nabla|)Q^{++}(1+|\nabla|)$$
for the weak-$\ast$ topology of $\gS_1$. Thus,
\begin{multline}
 E^\nu(q)=\liminf_{n\to\ii}\cE^\nu(Q_n)  \geq \tr(\chi_R|H^0_{\rm per}-\Sigma|^{1/2}(Q^{++}-Q^{--})|H^0_{\rm per}-\Sigma|^{1/2}\chi_R)\\
+\Sigma\tr(\chi_R(Q^{++}+Q^{--})\chi_R)+E^0\big(q-\tr(\chi_R(Q^{++}+Q^{--})\chi_R)\big)\\
+\frac12 D(\rho_{Q}-\nu,\rho_{Q}-\nu)-\frac12 D(\nu,\nu)-C'/R.
\label{decomp_energy_4}
\end{multline}
Passing now to the limit as $R\to\ii$, we obtain \eqref{absurde_HVZ}. This contradicts \eqref{HVZ} and shows that \textbf{(b)}$\Leftrightarrow$\textbf{(c)} in Theorem~\ref{HVZ}.

\subsubsection*{\bf Step 4: \textit{Characterization of the $q$'s such that {\rm \textbf{(c)}} holds}}
Because $q\mapsto E^\nu(q)$ is a convex function, it is classical that the set 
$I=\{q\in\R,\ \text{\textbf{(c)} holds}\}$
is a closed interval of $\R$, see e.g. \cite{rHF}. It is non empty
since it contains $\tr_0(\bar Q)$ for any minimizer $\bar Q$ of
$E^\nu_{\epsilon_{F}}$ obtained in Theorem \ref{exists_min},
for any ${\epsilon_{F}}$ in the gap $(\Sigma_Z^+,\Sigma_{Z+1}^-)$. Additionally,
$q\mapsto E^\nu(q)$ is linear on the connected components of $\R
\setminus I$ and $I$ is the largest interval on which this function is
strictly convex. 
Let us now state and prove the 
\begin{lemma}\label{2_min} Let $Z\in \NN \setminus \left\{0\right\}$,
  $\nu\in L^1(\R^3)\cap L^2(\R^3)$, and assume that \textnormal{\textbf{(A1)}} holds. Assume that $Q_1$
  and $Q_2$ are respectively two minimizers of $E^\nu(q_1)$ and
  $E^\nu(q_2)$ with $q_1\neq q_2$. Then $\rho_{Q_1}\neq\rho_{Q_2}$ and
  therefore $$E^\nu(tq_1+(1-t)q_2)\leq
  \cE^\nu(tQ_1+(1-t)Q_2)<tE^\nu(q_1)+(1-t)E^\nu(q_2).$$ 
\end{lemma}
\begin{proof}
Assume by contradiction that $\rho_{Q_1}=\rho_{Q_2}$. It is classical that the operators $Q_1$
and $Q_2$ satisfy the self-consistent equations
$$Q_1=\chi_{(-\ii,\epsilon_1)}(H_{Q_1})-\PPi+\delta_1,\qquad
Q_2=\chi_{(-\ii,\epsilon_2)}(H_{Q_2})-\PPi+\delta_2$$
where $0\leq\delta_k\leq 1$ and $\text{Ran}(\delta_k)\subseteq\ker(H_{Q_k}-\epsilon_k)$ for $k=1,2$. Necessarily
$\epsilon_1$ and $\epsilon_2$ are in $[\Sigma^+_Z,\Sigma^-_{Z+1}]$ otherwise
$Q_1$ and $Q_2$ would not be compact, which is not possible since every
operator of $\cK$ is compact. Since $H_{Q_1}=H_{Q_2}$ has only a point spectrum in the gap, we deduce that if $\epsilon_k\in(\Sigma^+_Z,\Sigma^-_{Z+1})$, then necessarily $\delta_k$ is finite rank. If $\epsilon_k\in\{\Sigma^+_Z,\Sigma^-_{Z+1}\}$, then it can easily be proved that at least $\delta_k\in\gS_1$.
Hence $Q_1$ and $Q_2$ differ by a trace-class operator: $Q_2=Q_1+\delta$,
$\tr|\delta|<\ii$. Now $0\neq
q_2-q_1=\tr(\delta)=\int\rho_\delta$ which contradicts our assumption
that $\rho_\delta=\rho_{Q_1} - \rho_{Q_2} = 0$. The rest follows from
the strict convexity of $\rho\mapsto D(\rho,\rho)$. 
\end{proof}

\begin{corollary}
There is no minimizer for $E^\nu(q)$ if $q\notin I$, the interval on which {\rm \textbf{(c)}} holds. Thus {\rm \textbf{(a)}} implies {\rm \textbf{(c)}}.
\end{corollary}
\begin{proof}
 Assume that there is a minimizer $Q_1$ for some $q_1\notin I$, for instance $q_1>\max I:=q_2$. Applying Lemma \ref{2_min} to $q_1$ and $q_2$ shows that $E^\nu(\cdot)$ cannot be linear on $[q_2,q_1]$ which contradicts the definition of $I$. 
\end{proof}

\subsection{Proof of Theorem \ref{thm:supercell_0}: thermodynamic limit of the supercell model for a perfect crystal}\label{sec:proof_TLSCP}
\subsubsection*{\bf Step 1} Let us first prove that $\dps \limsup_{L \to +\infty}
\frac{1}{L^3}I^0_{{\rm sc},L} \le I^0_{\rm per}$. 
Starting from the Bloch decomposition 
$$
\gamma^0_{\rm per} = \frac{1}{(2\pi)^3} \int_{\Gamma^\ast}
(\gamma^0_{\rm per})_\xi \, d\xi
$$
of $\gamma^0_{\rm per}$ it is possible to construct a convenient test function
$\widetilde{\gamma}_{{\rm sc},L}$ for (\ref{eq:supercell2}) as
follows: 
$$
\widetilde{\gamma}_{{\rm sc},L}(x,y)\,=\,\frac{1}{(2\pi)^3}
\sum_{\xi\,\in\,\frac{2\pi}{L}\ZZ^3 \cap \Gamma^\ast}
e^{i \xi x} \bigg( \int_{\xi+[-\frac{2\pi\eta}L,\frac{2\pi(1-\eta)}{L})^3}
e^{-i \xi' x} (\gamma^0_{\rm per})_{\xi'}(x,y) \; e^{i \xi' y} d\xi'
\bigg) \; e^{- i \xi y} ,
$$
with $\eta=0$ if $L$ is even and $\eta=1/2$ if $L$ is odd.
It is indeed easy to check that $\widetilde{\gamma}_{{\rm sc},L}$ is
in $\mathcal{ P}_{{\rm sc},L}$ and satisfies
$\rho_{\widetilde{\gamma}_{{\rm sc},L}} = \rho_{\gamma^0_{\rm per}}$. In
particular,
$$\int_{\Lambda_L}\rho_{\widetilde{\gamma}_{{\rm
      sc},L}}=\int_{\Lambda_L}\mu_{\rm per} = ZL^3,$$ 
and, since both $\rho_{\gamma^0_{\rm per}}$ and $\mu_{\rm per}$
      are $\Z^3$-periodic, 
$$
D_{G_L}\left(\rho_{\widetilde{\gamma}_{{\rm sc},L}}-\mu_{\rm
   per},  
\rho_{\widetilde{\gamma}_{{\rm sc},L}}-\mu_{\rm per}
    \right) 
 =  L^3 \, D_{G_1}\left(\rho_{\gamma^0_{\rm per}}-\mu_{\rm per},
\rho_{\gamma^0_{\rm per}}-\mu_{\rm per}  \right) .
$$
Besides,
\begin{multline*}
 \frac{1}{L^3} \, \tr_{L^2_{\rm per}(\Lambda_L)} \left( - \frac 1 2
  \Delta \widetilde \gamma_{{\rm sc},L} \right) 
 =  \frac 1 {(2\pi)^3} \int_{\Gamma^\ast} \tr_{L^2_{\xi}(\Gamma)} 
\left( - \frac 1 2 \Delta \left(\gamma^0_{\rm per}\right)_\xi \right)  \, d\xi  \\ 
  - \frac{1}{2(2\pi)^3}
  \sum_{\xi\,\in\,\frac{2\pi}{L}\ZZ^3 \cap \Gamma^\ast}
\int_{\xi+[-\frac{2\pi\eta}L,\frac{2\pi(1-\eta)}L)^3} |\xi-\xi'|^2  \tr_{L^2_{\xi'}(\Gamma)}
((\gamma^0_{\rm per})_{\xi'})  \, d\xi'\\
- \frac{i}{(2\pi)^3} \sum_{\xi\,\in\,\frac{2\pi}{L}\ZZ^3 \cap \Gamma^\ast}
\int_{\xi+[-\frac{2\pi\eta}L,\frac{2\pi(1-\eta)}L)^3} (\xi-\xi') \cdot
\tr_{L^2_{\xi'}(\Gamma)} (-i\nabla (\gamma^0_{\rm per})_{\xi'}) \, d\xi' .
\end{multline*}
It follows from the boundedness of $\int_{\Gamma^\ast}
\tr_{L^2_\xi(\Gamma)} ((1-\Delta) 
(\gamma^0_{\rm per})_\xi) \, d\xi$ and from the inequality $|-2i\nabla| \le
(1-\Delta)$ that the last two terms of the above expression go to zero,
hence that 
$$
\lim_{L \to +\infty} \frac{1}{L^3} \, \tr_{L^2_{\rm per}(\Lambda_L)}
\left( - \frac 1 2 
  \Delta \widetilde{\gamma}_{{\rm sc},L} \right) 
 = \frac 1 {(2\pi)^3} \int_{\Gamma^\ast} \tr_{L^2_\xi(\Gamma)} 
\left( - \frac 1 2 (\Delta \gamma^0_{\rm per})_\xi \right) \, d\xi. 
$$
Therefore
$
\lim_{L \to +\infty} L^{-3} \mathcal{ E}_{{\rm
    sc},L}^0(\widetilde{\gamma}_{{\rm sc},L}) = \mathcal{ E}^0_{\rm
  per}(\gamma^0_{\rm per}) = I^0_{\rm per}
$
and consequently
$$
\limsup_{L \to +\infty} \frac{1}{L^3} I^0_{{\rm sc},L} \le 
\lim_{L \to +\infty} \frac{1}{L^3} \mathcal{ E}_{{\rm
    sc},L}^0(\widetilde{\gamma}_{{\rm sc},L}) = I^0_{\rm per}.
$$

\subsubsection*{\bf Step 2} 
Let us now establish that $\dps \liminf_{L \to +\infty}
\frac{1}{L^3} I^0_{{\rm sc},L} \ge I^0_{\rm per}$.
First, the existence of a minimizer $\gamma_L$ for (\ref{eq:supercell2})
and the uniqueness of the corresponding density $\rho^0_{{\rm sc},L}$
follows from the same arguments as in the proof of
\cite[Thm 2.1]{CLL}. Note that, by symmetry, $\rho^0_{{\rm sc},L}$ is
$\Z^3$-periodic. We now define the operator $\gamma^0_{{\rm sc},L}$ as
\begin{equation*}
\dps
{\gamma}_{{\rm sc},L}^0 \,=\, \frac {1}{L^3} 
 \, \sum_{k \in \ZZ^3 \cap \Lambda_L} \tau_k^\ast \gamma_L
 \tau_k.
\end{equation*}
By simple periodicity arguments, it is clear that $\tau_k^\ast \gamma_L
      \tau_k$ is also a minimizer for~\eqref{eq:supercell2} for all
      $k\in\Z^3$. By convexity, so is ${\gamma}_{{\rm
      sc},L}^0$. Besides, ${\gamma}_{{\rm sc},L}^0$ commutes with
      the translations $\tau_k$ for all $k \in \Z^3$ so that
its kernel ${\gamma}_{{\rm sc},L}^0(x,y)$ satisfies 
$$\forall (x,y,z) \in \R^3 \times \R^3 \times \Z^3, \quad 
{\gamma}_{{\rm sc},L}^0(x+z,y+z)={\gamma}_{{\rm sc},L}^0(x,y).$$
The optimality conditions imply that $\gamma^0_{{\rm sc},L}$
can be expanded as follows
$$
\gamma^0_{{\rm sc},L}(x,y) = \frac{1}{L^3}
\sum_{\xi \in \frac{2\pi}L \ZZ^3 \cap
  \Gamma^\ast} \sum_{k \ge 1}  n_{k,\xi}^L e^{i\xi \cdot x} v_{k,\xi}^L(x) \, 
\overline{v_{k,\xi}^L(y)} e^{-i\xi \cdot y}
$$
where for any $\xi \in \frac{2\pi}L \ZZ^3 \cap
  \Gamma^\ast$, $(v_{k,\xi}^L)_{k \ge 1}$ is a Hilbert basis of $L^2_{\rm
    per}(\Gamma)$ consisting of eigenfunctions of the self-adjoint
  operator on $L^2_{\rm per}(\Gamma)$ defined by
$$
- \frac 1 2 \Delta - i \xi \cdot\nabla + (\rho^0_{{\rm sc},L}-\mu_{\rm
  per}) \star_{\Gamma} G_1 + \frac 1 2 |\xi|^2
$$
associated with eigenvalues $\lambda_{1}^L(\xi) \le \lambda_{2}^L(\xi) \le
\cdots$ 
The occupation numbers $n_{k,\xi}^L$ are in the range $[0,1]$ and such that
$$
\frac{1}{L^3} \sum_{\xi \in \frac{2\pi}L \ZZ^3 \cap
  \Gamma^\ast} \sum_{k \ge 1}  n_{k,\xi}^L = Z.
$$
Lastly, there exists a Fermi level $\epsilon_F^L \in \RR$ such that 
$n_{k,\xi}^L =1$ whenever $\lambda_{k}^L(\xi) < \epsilon_F^L$ and
$n_{k,\xi}^L =0$ whenever $\lambda_{k}^L(\xi) > \epsilon_F^L$.
One has
\begin{multline}
\frac{1}{L^3} I^0_{{\rm sc},L} = \frac{1}{L^3} \cE^0_{{\rm sc},L}
(\gamma_{{\rm sc},L}^0)  =  
\frac{1}{L^3} \sum_{\xi \in \frac{2\pi}L \ZZ^3 \cap
  \Gamma^\ast} \sum_{k \ge 1} \frac{n_{k,\xi}^L}2
\| (-i \nabla + \xi)v_{k,\xi}^L \|_{L^2_{\rm per}(\Gamma)}^2 \\ 
 + \frac{1}{2 \, L^3} D_{G_L}(\rho_{{\gamma}_{{\rm sc},L}^0}-\mu_{\rm
  per},\rho_{{\gamma}_{{\rm sc},L}^0}-\mu_{\rm per}).
\label{eq:estim_bi_1}
\end{multline}
We now introduce
$$
\widetilde \gamma_{{\rm sc},L}^0(x,y) =  \frac{1}{(2\pi)^3}
\int_{\Gamma^\ast}  \sum_{k \ge 1} n_{k,\beta_L(\xi)}^L e^{i\xi 
  \cdot x} v_{k,\beta_L(\xi)}^L(x) \, 
\overline{v_{k,\beta_L(\xi)}^L(y)} e^{-i\xi \cdot y} \, d\xi
$$
where $\beta_L(\xi)$ is the unique element of $\frac{2\pi}L \ZZ^3 \cap
  \Gamma^\ast$ such that $\xi \in \beta_L(\xi) +
  [-\frac{2\pi\eta}L,\frac{2\pi(1-\eta)}{L})^3$. 
 It is easy to check that $\tilde{\gamma}_{{\rm sc},L}^0\in\cP_{\rm
 per}^Z$. Thus $\tilde{\gamma}_{{\rm sc},L}^0\in\cP^Z_{\rm per}$ can be
 used as a test function for (\ref{def_I0per}). Therefore
\begin{equation} \label{eq:estim_bi_2}
\cE^0_{\rm per}(\tilde{\gamma}_{{\rm sc},L}^0) \ge I^0_{\rm per}.
\end{equation}
As $\rho_{\tilde{\gamma}_{{\rm sc},L}^0}=\rho_{{\gamma}_{{\rm
      sc},L}^0}$ is $\Z^3$-periodic, one has
\begin{equation} \label{eq:estim_bi_3}
D_{G_1}(\rho_{\tilde{\gamma}_{{\rm sc},L}^0}-\mu_{\rm
  per},\rho_{\tilde{\gamma}_{{\rm sc},L}^0}-\mu_{\rm per}) =
  \frac{1}{L^3}
D_{G_L}(\rho_{{\gamma}_{{\rm sc},L}^0}-\mu_{\rm
  per},\rho_{{\gamma}_{{\rm sc},L}^0}-\mu_{\rm per}).
\end{equation}
Besides,
\begin{multline} 
 \frac 1 {(2\pi)^3} \int_{\Gamma^\ast} \tr_{L^2_{\xi}(\Gamma)} 
\left( - \frac 1 2 \Delta \left(\gamma^0_{{\rm sc},L}\right)_\xi \right)
\, d\xi =   \frac{1}{(2\pi)^3}
\int_{\Gamma^\ast} \sum_{k \ge 1} \frac{n_{k,\beta_L(\xi)}^L}2
\| (-i \nabla + \xi)v_{k,\beta_L(\xi)}^L \|_{L^2_{\rm per}(\Gamma)}^2 
 \\
= \frac{1}{L^3} \sum_{\xi \in \frac{2\pi}L \ZZ^3 \cap
  \Gamma^\ast} \sum_{k \ge 1} \frac{n_{k,\xi}^L}2
\| (-i \nabla + \xi)v_{k,\xi}^L \|_{L^2_{\rm per}(\Gamma)}^2
+ R_L \label{eq:estim_bi_4}
\end{multline}
with
$$
R_L = \frac{1}{(2\pi)^3}
\int_{\Gamma^\ast} \sum_{k \ge 1} \frac{n_{k,\beta_L(\xi)}^L}2
\left(
\| (-i \nabla + \xi)v_{k,\beta_L(\xi)}^L \|_{L^2_{\rm per}(\Gamma)}^2
-  \| (-i \nabla + \beta_L(\xi))v_{k,\beta_L(\xi)}^L \|_{L^2_{\rm
    per}(\Gamma)}^2 \right).
$$
Putting \eqref{eq:estim_bi_1}--\eqref{eq:estim_bi_4} together, we end up
with
$\frac{1}{L^3} I^0_{{\rm sc},L} + R_L \ge I^0_{\rm per}.$
As
$$
|R_L| \le \left( \frac{6Z I^0_{{\rm sc},L}}{L^3} \right)^{1/2} \, 
  \frac{2\pi}L + \frac{3Z}2 \left(\frac{2\pi}L\right)^2,
$$
we finally obtain $\dps \liminf_{L \rightarrow + \infty}
\frac{1}{L^3} I^0_{{\rm sc},L} \ge I^0_{\rm per}$. 

\subsubsection*{\bf Step 3: \textit{Convergence of the density}}
A byproduct of Steps~1 and 2 is that $(\tilde{\gamma}^0_{{\rm sc},L})_{L \in
  \NN \setminus \left\{0\right\}}$ is a minimizing sequence for
$I^0_{\rm per}$. The convergence results on the density
$\rho_{{\gamma}^0_{{\rm sc},L}}=\rho_{\tilde{\gamma}^0_{{\rm sc},L}}$
immediately follow from the proof of \cite[Thm 2.1]{CLL}. 

\subsubsection*{\bf Step 4: \textit{Convergence of the mean-field Hamiltonian and its spectrum}}
One has
$H^0_{{\rm sc},L} - H^0_{\rm per} = \Phi_L$
where $\Phi_L$ solves the Poisson equation
$-\Delta \Phi_L = 4\pi (\rho_{\gamma^0_{{\rm sc},L}} - \rho_{\gamma^0_{\rm
    per}})$ on 
$\Gamma$ with periodic boundary conditions. As it follows from Step~3
that $(\rho_{\gamma^0_{{\rm sc},L}} - \rho_{\gamma^0_{\rm
    per}})$ converges to zero in $L^2_{\rm
  per}(\Gamma)$, we obtain that $\Phi_L$ converges to zero in $H^2_{\rm
  per}(\Gamma)$, hence in $L^\infty(\RR^3)$. Consequently,
\begin{equation}
 \norm{H^0_{{\rm sc},L} - H^0_{\rm per}} \leq \norm{\Phi_L}_{L^\ii}\to0
\label{limit_proof_Hamil}
\end{equation}
as $L\to\ii$.
This clearly implies, via the min-max principle, that 
$$
\sup_{n\geq1}\sup_{\xi\in\Gamma^*}\left| \lambda_{n}^L(\xi)-\lambda_{n}(\xi) \right| \le 
 \| \Phi_L\|_{L^\infty} \mathop{\longrightarrow}_{L \to \infty} 0
$$
where $(\lambda_{n}^L(\xi))_{n\geq1, \, \xi \in \Gamma^\ast}$ (resp. $(\lambda_{n}(\xi))_{n\geq1, \, \xi \in \Gamma^\ast}$) are the Bloch eigenvalues of $H^0_{{\rm sc},L}$ (resp. $H^0_{\rm per}$).

\subsubsection*{\bf Step 5: \textit{Uniqueness of $\gamma^0_{{\rm sc},L}$ for large
  values of $L$}}
In the remainder of the proof, we assume that \textnormal{\textbf{(A1)}} holds, i.e. that $H^0_{\rm per}$ has a gap.

The spectrum of $H^0_{{\rm sc},L}$ considered as an operator on
$L^2_{\rm per}(\Lambda_L)$ is given by 
$$
\sigma_{L^2_{\rm per}(\Lambda_L)}(H^0_{{\rm sc},L})=
\bigcup_{n \in \NN \setminus \left\{0\right\}} \bigcup_{\xi \in
  \frac{2\pi}L \Z^3\cap \Gamma^*} \lambda_{n}^L(\xi).
$$ 
It follows from Step~4 that there exists some $L_0 \in \NN \setminus
\left\{0\right\}$ such that for all $L \ge L_0$, the lowest $ZL^3$
eigenvalues of $H^0_{{\rm sc},L}$ (including multiplicities) are 
$$
\bigcup_{1 \le n \le Z} \bigcup_{\xi \in \frac{2\pi}L \Z^3\cap \Gamma^*} \lambda_{n}^L(\xi)
$$ 
and there is a gap between the $(ZL^3)$-th and the $(ZL^3+1)$-st
eigenvalues. As a consequence, $\gamma^0_{{\rm sc},L}$ is uniquely
defined: it is the spectral projector associated with the lowest $ZL^3$
eigenvalues of $H^0_{{\rm sc},L}$, considered as an operator on
$L^2_{\rm per}(\Lambda_L)$.

\subsubsection*{\bf Step~6} Let ${\epsilon_{F}} \in (\Sigma_Z^+,\Sigma_{Z+1}^-)$. For $L$ large
enough, 
$\gamma^0_{{\rm sc},L} = \chi_{(-\infty,{\epsilon_{F}}]}(H^0_{{\rm sc},L})
= \chi_{(-\infty,0]}(H^0_{{\rm sc},L}-{\epsilon_{F}})$
as operators acting on $L^2_{\rm per}(\Lambda_L)$. This means that $\gamma^0_{{\rm sc},L}$
satisfies the Euler-Lagrange equation associated
with $I^0_{{\rm sc},L,{\epsilon_{F}}}$, see \eqref{def_min_free_mu}. As the functional $\gamma \mapsto 
\mathcal{ E}^0_{{\rm sc},L}(\gamma) - {\epsilon_{F}} \tr_{L^2_{\rm per}(\Lambda_L)}(\gamma)$ is
convex on $\cP_{{\rm sc},L}$, $\gamma^0_{{\rm sc},L}$ is a minimizer of this functional. Its
uniqueness follows as usual from the uniqueness of the
minimizing density and from the fact that $0$ is not in 
the spectrum of $H^0_{{\rm sc},L}-{\epsilon_{F}}$.

\subsection{Proof of Theorem \ref{thermo_defect}: thermodynamic limit of
  the supercell model for a crystal with local defects}\label{sec:proof_TLSCD}

We follow the method of \cite{HLSo}. As in the previous section, we denote by $\gamma^0_{{\rm sc},L}$ the
minimizer of~\eqref{eq:supercell2}, which is unique for $L$ large enough
and is also the unique minimizer of~\eqref{def_min_free_mu}.
Let $\cK_L$ be the set of operators $Q_L$ on $L^2_{\rm
  per}(\Lambda_L)$ such that $\gamma^0_{{\rm sc},L} + Q_L \in \cP_{{\rm
    sc},L}$. In fact
\begin{multline*}
\cK_L  =  \big\{ Q_L \in \gS_1(L^2_{\rm per}(\Lambda_L)) \; | \; 
  Q_L^\ast=Q_L, \;
|\nabla| Q_L |\nabla| \in \gS_1(L^2_{\rm per}(\Lambda_L)), \;\\ 
\qquad \qquad  -\gamma^0_{{\rm sc},L} \leq Q_L \leq 1 - \gamma^0_{{\rm
    sc},L} \big\}.
\end{multline*}
We introduce $\cE^\nu_{{\rm sc},L,{\epsilon_{F}}}:=\cE^\nu_{{\rm sc},L}-{\epsilon_{F}}\tr_{L^2_{\rm per}(\Lambda_L)}$.
Let $Q_L \in \cK_L$, one has 
\begin{multline*}
\cE^\nu_{{\rm sc},L,{\epsilon_{F}}}(\gamma^0_{{\rm sc},L} + Q_L) - \cE^0_{{\rm
    sc},L,{\epsilon_{F}}}(\gamma^0_{{\rm sc},L}) = 
\tr_{L^2_{\rm per}(\Lambda_L)}(H^0_{{\rm sc},L} Q_L) -
    D_{G_L}(\rho_{Q_L},\nu_L)  \\ 
 + \frac 1 2
    D_{G_L}(\rho_{Q_L},\rho_{Q_L}) 
    - {\epsilon_{F}} \tr_{L^2_{\rm per}(\Lambda_L)}(Q_L)
 - D_{G_L}(\nu_L,\rho_{\gamma^0_{{\rm sc},L}}-\mu_{\rm per})
+ \frac 12 D_{G_L}(\nu_L,\nu_L).
\end{multline*}
Note that in the above expression, $H^0_{{\rm sc},L}$ is considered as
an operator on $L^2_{\rm per}(\Lambda_L)$. Using
Theorem~\ref{thm:supercell_0}, this equality can be rewritten, for $L$
large enough, as
\begin{multline}
\cE^\nu_{{\rm sc},L,\epsilon_F}(\gamma^0_{{\rm sc},L} + Q_L) - \cE^0_{{\rm
    sc},L,{\epsilon_{F}}}(\gamma^0_{{\rm sc},L})  = 
\tr_{L^2_{\rm per}(\Lambda_L)}(|H^0_{{\rm sc},L}-{\epsilon_{F}}|
    (Q_L^{++,L}-Q_L^{--,L}) )\\
 + \frac 1 2 D_{G_L}(\rho_{Q_L}-\nu_L,\rho_{Q_L}-\nu_L)  - D_{G_L}(\nu_L,\rho_{\gamma^0_{{\rm sc},L}}-\mu_{\rm per})
    \label{eq:LT_def_1} 
\end{multline}
where we have set
$$
Q_L^{++,L} = (1-\gamma^0_{{\rm sc},L}) Q_L (1-\gamma^0_{{\rm sc},L})
\quad \mbox{and} \quad
Q_L^{--,L} = \gamma^0_{{\rm sc},L} Q_L \gamma^0_{{\rm sc},L}.
$$
It follows from \eqref{eq:LT_def_1} that
\begin{multline*}
I^\nu_{{\rm sc},L,\epsilon_F} - I^0_{{\rm sc},L,{\epsilon_{F}}} =
\inf \left\{E_{{\rm sc},L}^\nu(Q_L) -  {\epsilon_{F}} \tr_{L^2_{\rm
      per}(\Lambda_L)}(Q_L), \; Q_L \in \cK_L \right\}  
 \\
 - D_{G_L}(\nu_L,\rho_{\gamma^0_{{\rm sc},L}}-\mu_{\rm per})
+ \frac 12 D_{G_L}(\nu_L,\nu_L).
\end{multline*}
where
\begin{multline*}
E_{{\rm sc},L}^\nu(Q_L) - {\epsilon_{F}} \tr_{L^2_{\rm per}(\Lambda_L)}(Q_L) :=- D_{G_L}(\rho_{Q_L},\nu_L)  + \frac 1 2 D_{G_L}(\rho_{Q_L},\rho_{Q_L})\\
+\tr_{L^2_{\rm per}(\Lambda_L)}\left(|H^0_{{\rm sc},L}-{\epsilon_{F}}|^{1/2}
    (Q_L^{++,L}-Q_L^{--,L}) |H^0_{{\rm sc},L}-{\epsilon_{F}}|^{1/2}\right).
\end{multline*}

\medskip

First, using $\nu$ being in $L^1(\R^3)\cap L^2(\R^3)$ and the convergence of $\Phi_L =
      (\rho_{\gamma^0_{{\rm 
      sc},L}}-\rho_{\gamma^0_{{\rm per}}}) \star_{\Gamma} G_1$ to zero
      in $L^\infty$ (see Step~4 of the proof of
      Theorem~\ref{thm:supercell_0} in Section~\ref{sec:proof_TLSCP}),
      we obtain
$$
- D_{G_L}(\nu_L,\rho_{\gamma^0_{{\rm sc},L}}-\mu_{\rm per})
+ \frac 12 D_{G_L}(\nu_L,\nu_L) \longrightarrow 
- \int_{\RR^3} \nu\big((\rho_{\gamma^0_{{\rm per}}}-\mu_{\rm per})
\star_\Gamma G_1\big)  + \frac 12 D(\nu,\nu).
$$
Our goal is to prove that 
\begin{equation}
\lim_{L \to \ii} E_{{\epsilon_{F}},L}^\nu
= E_{\epsilon_{F}}^\nu  
\label{limit_E_mu_L}
\end{equation}
where 
\begin{equation} \label{eq:def_EmuLnu}
E_{{\epsilon_{F}},L}^\nu = \inf \left\{E_{{\rm
      sc},L}^\nu(Q_L) -  {\epsilon_{F}} \tr_{L^2_{\rm 
      per}(\Lambda_L)}(Q_L), \; Q_L \in \cK_L \right\}  .
\end{equation}

\medskip

\subsubsection*{\bf Step 1: \textit{Preliminaries}} In the proof of \eqref{limit_E_mu_L}, we shall need several times to compare states living in $L^2_{\rm per}(\Lambda_L)$ with states living in $L^2(\R^3)$. To this end, we introduce the map
\begin{eqnarray*}
i_L:L^2(\R^3) & \to & L^2_{\rm per}(\Lambda_L)\\
\phi & \mapsto & \sum_{z\in\Z^3}(\1_{\Lambda_L}\phi)(\cdot-Lz).
\end{eqnarray*}
Notice that $(i_L)^*: L^2_{\rm per}(\Lambda_L)\to L^2(\R^3)$ is the operator which to any periodic function $\phi\in L^2_{\rm per}(\Lambda_L)$ associates the function $\1_{\Lambda_L}\phi\in L^2(\R^3)$. Remark that $i_L(i_L)^*=Id_{L^2_{\rm per}(\Lambda_L)}$ whereas $(i_L)^*i_L=\1_{\Lambda_L}$. Hence $i_L$ defines an isometry from $L^2(\Lambda_L)$ to $L^2_{\rm per}(\Lambda_L)$. The equality $(i_L)^*i_L\phi=\phi$ is only true when $\phi\in L^2(\R^3)$ has its support in $\Lambda_L$. When $\phi\in H^1(\R^3)$ satisfies $\mbox{Supp}(\phi)\subset\Lambda_L$, then one also has $\partial_{x_i}i_L(\phi)=i_L(\partial_{x_i}\phi)$. 

Notice in addition that if $A\in\gS_1(L^2_{\rm per}(\Lambda_L))$, then $(i_L)^*Ai_L\in\gS_1(L^2(\Lambda_L))\subseteq\gS_1(L^2(\R^3))$ and
$$\tr_{L^2_{\rm per}(\Lambda_L)}(A)=\tr_{L^2(\R^3)}\left((i_L)^*Ai_L\right).$$
Similarly if $A\in\gS_p(L^2_{\rm per}(\Lambda_L))$, 
\begin{equation}
 \norm{A}_{\gS_p(L^2_{\rm per}(\Lambda_L))}=\norm{(i_L)^*Ai_L}_{\gS_p(L^2(\R^3))}.
\label{egalite_box_space_S_p}
\end{equation}

Finally, we shall use that for any $\Z^3$-periodic bounded function $f$, $i_Lf=fi_L$, where we use the same notation $f$ to denote the multiplication operator by the function $f$ acting either on $L^2_{\rm per}(\Lambda_L)$ or on $L^2(\R^3)$. Similarly, the operator $H^0_{{\rm sc},L}=-\Delta/2+(\rho_{\gamma^0_{{\rm sc},L}}-\mu_{\rm per})\star_\Gamma G_1$ can be seen as acting on $L^2_{\rm per}(\Lambda_L)$ or on $L^2(\R^3)$ and we use the same notation in the two cases. Then we have for any $\phi\in\cC_0^\ii(\R^3)$ satisfying ${\rm Supp}(\phi)\subseteq\Lambda_L$
\begin{equation}
H^0_{{\rm sc},L}i_L\phi=i_LH^0_{{\rm sc},L}\phi. 
\label{commut_boite}
\end{equation}
Notice that one can also define $-i\nabla$ on $L^2(\R^3)$ or on $L^2_{\rm per}(\Lambda_L)$ and we shall adopt the same notation for these two operators.
We gather some useful limits in the following 
\begin{lemma} \label{lem:lim_gamma} Let be $\psi\in L^2(\R^3)$ and $\phi\in\cC_0^\ii(\R^3)$. Then we have as $L\to\ii$
\begin{enumerate}
\item $(i_L)^*\gamma^0_{{\rm sc},L}i_L\psi\rightarrow\gamma^0_{\rm per}\psi$ in $L^2(\R^3)$;
\item $(i_L)^*H^0_{{\rm sc},L}\gamma^0_{{\rm sc},L}i_L\phi\rightarrow H^0_{{\rm per}}\gamma^0_{\rm per}\phi$ in $L^2(\R^3)$;
\item $(i_L)^*\Delta\gamma^0_{{\rm sc},L}i_L\phi\rightarrow \Delta\gamma^0_{\rm per}\phi$ in $L^2(\R^3)$;
\item $(i_L)^*(1+|\nabla|)i_L\phi\rightarrow(1+|\nabla|)\phi$ in $L^2(\R^3)$;
\item $(i_L)^*|H^0_{{\rm sc},L}-{\epsilon_{F}}|^{1/2}i_L\phi\rightarrow |H^0_{{\rm per}}-{\epsilon_{F}}|^{1/2}\phi$ in $L^2(\R^3)$ for any fixed ${\epsilon_{F}}$ in the gap $(\Sigma_{Z}^+,\Sigma_{Z+1}^-)$.
\end{enumerate}
\end{lemma}

\begin{proof} 
The operator $(i_L)^*\gamma^0_{{\rm sc},L}i_L$ being uniformly bounded with respect to $L$, it suffices to prove the first assertion for a dense subset of $L^2(\R^3)$ like $\cC_0^\ii(\R^3)$. Hence we may assume that $\psi=\phi\in\cC_0^\ii(\R^3)$.

Let $K$ be a compact set in the resolvent set of $H^0_{\rm per}$. We are going to prove that
\begin{equation}
 \lim_{L\to\ii}(i_L)^*(z-H_{{\rm sc},L}^0)^{-1}i_L\phi \rightarrow_{L\to\ii}(z-H^0_{\rm per})^{-1}\phi
\label{limit_resolv_box_space}
\end{equation}
in $L^2(\R^3)$, uniformly for $z\in K$. To this end, we first notice that by Theorem \ref{thermo-lim}, $K$ is contained in the resolvent set of $H^0_{{\rm sc},L}$ for $L$ large enough and thus
$$\norm{(z-H_{{\rm sc},L}^0)^{-1}-(z-H_{{\rm per}}^0)^{-1}}_{\cB(L^2_{\rm per}(\Lambda_L))}\leq C(K)\norm{H_{{\rm per}}^0-H_{{\rm sc},L}^0}_{\cB(L^2_{\rm per}(\Lambda_L))}\to0$$
by \eqref{limit_proof_Hamil}. Hence it suffices to prove \eqref{limit_resolv_box_space} with $H_{{\rm sc},L}^0$ replaced by $H_{{\rm per}}^0$ (seen as an operator acting on $L^2_{\rm per}(\Lambda_L)$). Then we use its Bloch decomposition (detailed in Appendix, see \eqref{Bloch_dec_H_0}) and compute, assuming $L$ large enough for $\mbox{Supp}(\phi) \subset \Lambda_L$,
\begin{equation*}
 (i_L)^*(z-H_{{\rm per}}^0)^{-1}i_L\phi(x)=\!\!\!\!\!\sum_{k\in\frac{2\pi}L \Z^3\cap\Gamma^*}\sum_{n\geq1}\frac{L^{-3}}{z-\lambda_n(k)}\left(\int_{\R^3} \overline{e_n(k,\cdot)}\phi\right)\1_{\Lambda_L}(x)e_n(k,x),
\end{equation*}
\begin{equation*}
\norm{(i_L)^*(z-H_{{\rm per}}^0)^{-1}i_L\phi}_{L^2(\R^3)}^2 =\sum_{k\in\frac{2\pi}L\Z^3\cap\Gamma^*}\sum_{n\geq1}\frac{L^{-3}}{|z-\lambda_n(k)|^2}\left|\int_{\R^3} \overline{e_n(k,\cdot)}\phi\right|^2.
\end{equation*}
It is then easy to see that $(i_L)^*(z-H_{{\rm per}}^0)^{-1}i_L\phi\wto (z-H_{{\rm per}}^0)^{-1}\phi$ weakly in $L^2(\R^3)$ (one can take the scalar product against a function $\psi\in\cC_0^\ii(\R^3)$ to identify the limit) and that $\norm{(i_L)^*(z-H_{{\rm per}}^0)^{-1}i_L\phi}_{L^2(\R^3)}\to\norm{(z-H_{{\rm per}}^0)^{-1}\phi}_{L^2(\R^3)}$, yielding the strong convergence in $L^2(\R^3)$.

For the proof of $(1)$, it then suffices to choose a curve $\curv$ around the first $Z$ bands of $H^0_{\rm per}$ and use the above convergence of the resolvent in the Cauchy formula.  Assertion $(2)$ is an easy consequence of $(1)$ and \eqref{commut_boite}. Indeed, by Theorem \ref{thermo-lim}, we know that $\lim_{L\to\ii}\norm{H^0_{{\rm sc},L}-H^0_{\rm per}}_{\cB(L^2(\R^3))}=0$. Since $(i_L)^*\gamma^0_{{\rm sc},L}i_L$ is bounded, this implies that for $L$ large enough such that ${\rm Supp}(\phi)\subseteq\Lambda_L$
$$\norm{(i_L)^*\gamma^0_{{\rm sc},L}(H^0_{{\rm sc},L}i_L-i_LH^0_{\rm per})\phi}_{L^2(\R^3)}=\norm{(i_L)^*\gamma^0_{{\rm sc},L}i_L(H^0_{{\rm sc},L}-H^0_{\rm per})\phi}_{L^2(\R^3)}\to0$$
where we have used \eqref{commut_boite}. Then we notice that $H^0_{\rm per}\phi\in L^2(\R^3)$. Hence $(1)$ implies that $\lim_{L\to\ii}\norm{((i_L)^*\gamma^0_{{\rm sc},L}i_L-\gamma^0_{\rm per})H^0_{\rm per}\phi}_{L^2(\R^3)}=0$. The argument is exactly the same for the third assertion $(3)$.
Assertion $(5)$ can be proved in the same way, using \eqref{commut_boite} and the integral representation of the square root \eqref{repr_square_root}.

Finally, it remains to prove that $(4)$ is true, which is done by computing explicitly, for $L$ large enough such that ${\rm Supp}(\phi)\subseteq \Lambda_L$,
$$(i_L)^*(1+|\nabla|)i_L\phi=\sum_{k\in\frac{2\pi}L\Z^3}\frac{(2\pi)^{3/2}}{L^{3}}(1+|k|)\widehat{\phi}(k)e^{ik\cdot x}\1_{\Lambda_L}(x),$$
$$\norm{(i_L)^*(1+|\nabla|)i_L\phi}_{L^2(\R^3)}^2=\sum_{k\in\frac{2\pi}L\Z^3}\frac{(2\pi)^{3}}{L^{3}}|(1+|k|)\widehat{\phi}(k)|^2\to_{L\to\ii}\norm{(1+|\nabla|)\phi}_{L^2(\R^3)}^2.$$
The strong convergence is obtained as above.
\end{proof}

\begin{lemma}\label{lem:lim_op} Let $V\in\cC_0^\ii(\R^3)$. We have as $L\to\ii$
\begin{equation}
(i_L)^*(1-\Delta)^{-1}i_L(V)i_L\rightarrow(1-\Delta)^{-1} V ,
\label{limit_S_2_a}
\end{equation}
\begin{equation}
(i_L)^*(1+|\nabla|)^{-1}i_L(V)(1+|\nabla|)^{-1}i_L\rightarrow(1+|\nabla|)^{-1} V(1+|\nabla|)^{-1}
\label{limit_S_2_b}
\end{equation}
strongly in $\gS_2(L^2(\R^3))$.
\end{lemma}
\begin{proof}
For $L$ large enough, we have
\begin{align*}
\norm{(i_L)^*(1-\Delta)^{-1}i_L(V)i_L}_{\gS_2(L^2(\R^3))} &=  \norm{(1-\Delta)^{-1}i_L(V)}_{\gS_2(L^2_{\rm per}(\Lambda_L))}\\
 &\leq \frac{\norm{V}_{L^2(\R^3)}}{(2\pi)^{3/2}}\left(\sum_{k\in\frac{2\pi}L\Z^3}\frac{(2\pi/L)^{3}}{(1+|k|^2)^2}\right)^{1/2},
\end{align*}
which shows that $(i_L)^*(1-\Delta)^{-1}i_L(V)i_L$ is bounded in $\gS_2(L^2(\R^3))$ since
\begin{equation}
\lim_{L\to\ii}\sum_{k\in\frac{2\pi}L\Z^3}\frac{(2\pi/L)^{3}}{(1+|k|^2)^2}=\int_{\R^3}|g(p)|^2dp,\qquad g(p)=(1+|p|^2)^{-1}.
\label{limit_Riemann}
\end{equation}
Arguing as in the proof of the fourth assertion of Lemma \ref{lem:lim_gamma}, we can prove that \eqref{limit_S_2_a} holds in the strong sense, hence the convergence holds weakly in $\gS_2(L^2(\R^3))$ towards $(1-\Delta)^{-1}V$. Now
$$\lim_{L\to\ii}\norm{(i_L)^*(1-\Delta)^{-1}i_L(V)i_L}_{\gS_2(L^2(\R^3))}=\frac{\norm{V}_{L^2(\R^3)}\norm{g}_{L^2(\R^3)}}{(2\pi)^{3/2}}=\norm{(1-\Delta)^{-1}V}_{\gS_2(L^2(\R^3))}$$
and the limit holds strongly in $\gS_2(L^2(\R^3))$.

The argument is the same for \eqref{limit_S_2_b}, noticing that
\begin{align*}
 &\norm{(i_L)^*(1+|\nabla|)^{-1}i_L(V)(1+|\nabla|)^{-1}i_L}_{\gS_2(L^2(\R^3))}^2\\
 &\qquad \qquad \qquad = \tr_{L^2_{\rm per}(\Lambda_L)}\left((1+|\nabla|)^{-2}i_L(V)(1+|\nabla|)^{-2}i_L(V)\right)\\
 &\qquad \qquad \qquad = (2\pi)^{-3/2}\iint_{(\Lambda_L)^2}|h_L(x-y)|^2V(x)V(y)dx\,dy\\
 &\qquad \qquad \qquad \rightarrow_{L\to\ii}\norm{(1+|\nabla|)^{-1}V(1+|\nabla|)^{-1}}_{\gS_2(L^2(\R^3))}^2
\end{align*}
where we have used that
$$h_L(x):=\sum_{k\in\frac{2\pi}L\Z^3}\frac{(2\pi)^{3/2}}{L^3(1+|k|)^2}e^{ik\cdot x}$$
converges to the Fourier inverse $\cF^{-1}(h)$ of $h(p)=(1+|p|)^{-2}$, strongly in $L^2_{\rm loc}(\R^3)$.
\end{proof}

\subsubsection*{\bf Step 2: \textit{Upper bound}} We prove here that $\limsup_{L \to \ii} E_{{\epsilon_{F}},L}^\nu \le  E_{\epsilon_{F}}^\nu$.
Let $\epsilon > 0$.  Using Lemma~\ref{decomp_quelconque}, Proposition~\ref{decomp_proj},
  Corollary~\ref{finite_rank_dense}, and the notation therein, one can
  find a finite rank 
operator $Q \in \cK_{\rm r}$ such that  
\begin{equation} \label{eq:diff_epsilon}
E^\nu_{\epsilon_{F}} \leq \cE^\nu(Q)-{\epsilon_{F}} \tr_0(Q) \leq E^\nu_{\epsilon_{F}} +\epsilon,
\end{equation}
of the form
\begin{multline}
Q  =  \sum_{m=-M}^{-1}|v_m\rangle\langle v_m|-\sum_{n=-N}^{-1}|u_n\rangle\langle u_n|+\sum_{i=0}^k\frac{\lambda_i^2}{1+\lambda_i^2}\big(|v_i\rangle\langle v_i|-|u_i\rangle\langle u_i|\big)\\
+\sum_{i=0}^k\frac{\lambda_i}{1+\lambda_i^2}\big(|u_i\rangle\langle
v_i|+|v_i\rangle\langle u_i|\big) + \delta' \quad \mbox{with} \quad
\delta' = \sum_{j=1}^J n_j |w_j\rangle \langle
w_j| \label{expand_P_2}.
\end{multline} 
Let $0 < \eta <\!\!< 1$. It is possible to choose a family of orthonormal
functions $u_n^\eta$, $v_m^\eta$, $w_j^\eta$ in
$C^\infty_0(\R^3)$ such that  
\begin{equation} \label{eq:eta}
\|  u_n^\eta -u_n\|_{H^2} \le \eta, \quad 
\|  v_m^\eta -v_m\|_{H^2} \le \eta, \quad 
\|  w_j^\eta -w_j\|_{H^2} \le \eta.
\end{equation}
for all $n=-N...k$, $m=-M...k$ and $j=1...J$.
Let us define the Gram matrices
$$(S_-^\eta)_{i,j}:=\pscal{\gamma^0_{\rm per}u_i^\eta,u_j^\eta},\qquad (S_+^\eta)_{i,j}:=\pscal{(1-\gamma^0_{\rm per})v_i^\eta,v_j^\eta}$$
which, by \eqref{eq:eta} satisfy $S_+^\eta=Id_{M+k+1}+o(1)_{\eta\to0}$ and $S_-^\eta=Id_{N+k+1}+o(1)_{\eta\to0}$. We also introduce the orthogonal projector $\Pi^\eta$ on ${\rm Span}\{\gamma^0_{\rm per}u^\eta_n,(1-\gamma^0_{\rm per})v^\eta_m\}$ and define 
$(S_{w}^\eta)_{i,j}:=\pscal{(1-\Pi^\eta)w_i^\eta,w_j^\eta}$. Clearly $S_{w}^\eta=Id_{J}+o(1)_{\eta\to0}$.

Now we introduce a new orthonormal system in $L^2_{\rm per}(\Lambda_L)$
$$u_{i,L}^{\eta}:=\sum_{n=-N}^k(S^{-1/2}_{-,L})_{i,n}{\gamma^0_{{\rm sc},L}} i_Lu_{n}^\eta,\quad v_{i,L}^{\eta}:=\sum_{m=-M}^k(S^{-1/2}_{+,L})_{i,m}(1-\gamma^0_{{\rm sc},L}) i_Lv_{m}^\eta,$$
$$(S_{-,L})_{i,j}=\pscal{{\gamma^0_{{\rm sc},L}} i_Lu_{i}^\eta,i_Lu_{j}^\eta}_{L^2_{\rm per}(\Lambda_L)},\  (S_{+,L})_{i,j}=\pscal{(1-\gamma^0_{{\rm sc},L}) i_Lv_{i}^\eta,i_Lv_{j}^\eta}_{L^2_{\rm per}(\Lambda_L)}.$$
Notice that by the first assertion of Lemma \ref{lem:lim_gamma}, $\lim_{L\to\ii}S_{\pm,L}=S^\eta_\pm$. Finally, we introduce the projector $\Pi_L$ on ${\rm Span}(u_{n,L}^{\eta},v_{m,L}^{\eta})$
and define
$$ w_{j,L}^{\eta}:=\sum_{\ell=1}^J(S^{-1/2}_{w,L})_{j,\ell}(1-\Pi_L) i_Lw_{\ell}^\eta,\quad  (S_{w,L})_{i,j}=\pscal{(1-\Pi_L) i_Lw_{i}^\eta,i_Lw_{j}^\eta}.$$

We now define a state in $\cK_L$ by 
\begin{multline*}
Q_L^{\eta}  =  \sum_{m=-M}^{-1}|v_{m,L}^{\eta}\rangle\langle  v_{m,L}^{\eta}|-\sum_{n=-N}^{-1}| u_{n,L}^{\eta}\rangle\langle u_{n,L}^{\eta}|+\sum_{i=0}^k\frac{\lambda_i^2}{1+\lambda_i^2}\big(|v_{i,L}^{\eta}\rangle\langle v_{i,L}^{\eta}|-|u_{i,L}^{\eta}\rangle\langle u_{i,L}^{\eta}|\big)\\
 +\sum_{i=0}^k\frac{\lambda_i}{1+\lambda_i^2}\big(|
u_{i,L}^{\eta}\rangle\langle 
v_{i,L}^{\eta}|+|v_{i,L}^{\eta}\rangle\langle 
u_{i,L}^{\eta}|\big) + \sum_{j=1}^J n_j |w_{j,L}^{\eta}\rangle \langle w_{j,L}^{\eta}| .
\end{multline*} 
By Lemma \ref{lem:lim_gamma}, we have 
\begin{equation}
(i_L)^*u_{n,L}^{\eta}\to_{L\to\ii}\tilde u_n^\eta,\ (i_L)^*v_{m,L}^{\eta}\to_{L\to\ii}\tilde v_m^\eta\  \text{and}\ (i_L)^*w_{j,L}^{\eta}\to_{L\to\ii}\tilde w_j^\eta 
\label{limit_box_space}
\end{equation}
in $L^2(\R^3)\cap L^\ii(\R^3)$, where the limits are defined by
$$\tilde u^{\eta}_i:=\sum_{n=-N}^k(S_{-}^\eta)^{-1/2}_{i,n}{\gamma^0_{{\rm per}}} u_{n}^\eta,\quad \tilde v_{i}^{\eta}:=\sum_{m=-M}^k(S_{+}^\eta)^{-1/2}_{i,m,}(1-\gamma^0_{{\rm per}}) v_{m}^\eta,$$
$$ \tilde w_{j}^{\eta}:=\sum_{\ell=1}^J(S_{w}^\eta)^{-1/2}_{j,\ell}(1-\Pi^\eta) w_{\ell}^\eta.$$
By Lemma \ref{lem:lim_gamma}, we know that for any fixed $\phi,\psi\in\cC_0^\ii(\R^3)$,
$$\lim_{L\to\ii}\pscal{i_L^*(H^0_{{\rm sc},L}-{\epsilon_{F}})\gamma^0_{{\rm sc},L}i_L\phi,(i_L)^*\gamma^0_{{\rm sc},L}i_L\psi}=\pscal{(H^0_{{\rm per}}-{\epsilon_{F}})\gamma^0_{{\rm per}}\phi,\gamma^0_{{\rm per}}\psi}.$$
Hence, inserting the definition of $Q_L^\eta$ in the kinetic energy and using the convergence of the Gram matrices, we obtain
$$\lim_{L\to\ii}\tr_{L^2_{\rm per}(\Lambda_L)}\left((H^0_{{\rm sc},L}-{\epsilon_{F}})Q_L^\eta \right)=\tr_{L^2(\R^3)}\left((H^0_{{\rm per}}-{\epsilon_{F}})\tilde Q^\eta \right)$$
where $\tilde Q^\eta$ is defined similarly as $Q^\eta_L$ but with the functions $(\tilde u_n^\eta,\tilde v_m^\eta,\tilde w_j^\eta)$ instead of $(u_{n,L}^\eta,v_{m,L}^\eta,w_{j,L}^\eta)$.

Let us now prove that
$$\lim_{L\to\ii}D_{G_L}(\rho_{Q_L^\eta},\rho_{Q_L^\eta})=D(\rho_{\tilde Q^\eta},\rho_{\tilde Q^\eta}).$$
The convergence \eqref{limit_box_space} implies that $\1_{\Lambda_L}\rho_{Q_L^\eta}$ converges to $\rho_{\tilde Q^\eta}$ in particular in $L^1(\R^3)\cap L^2(\R^3)$. Notice the definition of $D_{G_L}(\cdot,\cdot)$ implies that 
\begin{equation}
\forall \rho\in L^1_{\rm per}(\Lambda_L)\cap L^2_{\rm per}(\Lambda_L),\qquad D_{G_L}(\rho,\rho)\leq C\left(\norm{\rho}_{L^1_{\rm per}(\Lambda_L)}^2+\norm{\rho}_{L^2_{\rm per}(\Lambda_L)}^2\right) 
\label{estim_D_L12}
\end{equation}
for a constant $C$ independent of $L$. Let us now write $\rho_{Q^\eta_L}=\rho_{1,L}+\rho_{2,L}$ where $\rho_{1,L}$ is the periodic function which equals $\1_{B(0,L/4)}\rho_{Q^\eta_L}$ on $\Lambda_L$. The convergence of $\1_{\Lambda_L}\rho_{Q_L^\eta}$ towards $\rho_{\tilde Q^\eta}$ in $L^1(\R^3)\cap L^2(\R^3)$ and \eqref{estim_D_L12} give that 
$$\lim_{L\to\ii}\norm{\rho_{2,L}}_{L^1_{\rm per}(\Lambda_L)}=\lim_{L\to\ii}\norm{\rho_{2,L}}_{L^2_{\rm per}(\Lambda_L)}=\lim_{L\to\ii}D_{G_L}(\rho_{2,L},\rho_{2,L})=0.$$
Hence, it remains to show that 
\begin{equation}
\lim_{L\to\ii}D_{G_L}(\rho_{1,L},\rho_{1,L})=D(\rho_{\tilde Q^\eta},\rho_{\tilde Q^\eta}). 
\label{limit_D_rho}
\end{equation}
To this end we use the estimate \cite{LS2}
$$\sup_{x\in\Lambda_L}\left|G_L(x)-\frac{1}{|x|}\right|=O(L^{-1}),$$
to obtain
$$D_{G_L}(\rho_{1,L},\rho_{1,L})=\iint_{(\Lambda_L)^2}G_L(x-y)\rho_{1,L}(x)\rho_{1,L}(y)dx\,dy=D(\1_{\Lambda_L}\rho_{1,L},\1_{\Lambda_L}\rho_{1,L})+O(L^{-1})$$
where we have used that $\norm{\rho_{1,L}}_{L^1_{\rm per}(\Lambda_L)}$ is uniformly bounded and that $x-y\in\Lambda_L$ for any $x,y\in B(0,L/4)$, the support of $\rho_{1,L}$.
The convergence of $\1_{\Lambda_L}\rho_{1,L}$ towards $\rho_{\tilde Q^\eta}$ in $L^1(\R^3)\cap L^2(\R^3)$ then proves \eqref{limit_D_rho}. Using the same argument for the term $D_{G_L}(\rho_{Q^\eta_L},\nu_L)$ we obtain
$$\lim_{L\to\ii}\left(-D_{G_L}(\rho_{Q_L^\eta},\nu_L)+\frac12 D_{G_L}\left(\rho_{Q_L^\eta},\rho_{Q_L^\eta}\right)\right)=-D(\rho_{\tilde Q^\eta},\nu)+\frac12 D\left(\rho_{\tilde Q^\eta},\rho_{\tilde Q^\eta}\right).$$
Finally 
\begin{equation} \label{eq:CV_energy_eta}
\lim_{L \to \ii} 
E_{{\rm sc},L}^\nu(Q^\eta_L) - {\epsilon_{F}} \tr_{L^2_{\rm
    per}(\Lambda_L)}(Q^\eta_L)
 = \cE^\nu(\tilde Q^\eta)-{\epsilon_{F}}\tr(\tilde Q^\eta) .
\end{equation}
Passing to the limit as $\eta\to0$ using \eqref{eq:eta} and the convergence of the Gram matrices $S_\pm^\eta$ and $S_w^\eta$, we eventually obtain
$$\limsup_{L\to\ii}E_{{\epsilon_{F}},L}^\nu \leq \cE^\nu(Q)-{\epsilon_{F}}\tr(Q)\leq E^\nu_{\epsilon_{F}}+\epsilon.$$

\subsubsection*{\bf Step 3. \textit{Lower bound}} We end the proof by showing that $\liminf_{L \to \ii} E_{{\epsilon_{F}},L}^\nu \ge
  E_{\epsilon_{F}}^\nu$.
As $\Lambda_L$ is bounded for any fixed $L$, the existence of a minimizer $Q_L$ of
$$
\inf \left\{E_{{\rm sc},L}^\nu(Q_L) -  {\epsilon_{F}} \tr_{L^2_{\rm
      per}(\Lambda_L)}(Q_L), \; Q_L \in \cK_L \right\}  
$$
is straightforward. In addition, the spectrum of $H^0_{{\rm
    sc},L}$, considered as an operator on $L^2_{{\rm per}}(\Lambda_L)$,
being purely discrete and bounded below, $Q_L$ is finite rank.

Using \eqref{eq:gap_unif} and reasoning as in the proof of Lemma~\ref{estim_H0per} 
(see Section~\ref{proof_estim_H0per}), we prove that there exists a
constant $c>0$ (independent of $L$) such that 
$|H^0_{{\rm sc},L}-{\epsilon_{F}}| \ge c (1-\Delta)$
on $L^2_{\rm per}(\Lambda_L)$, for $L$ large enough. The following
uniform bounds follow from Step~2:
\begin{align}
 & &\tr_{L^2_{\rm per}(\Lambda_L)}(|H^0_{{\rm
    sc},L}-{\epsilon_{F}}|^{1/2}(Q_L^{++,L}-Q_L^{--,L})|H^0_{{\rm
    sc},L}-{\epsilon_{F}}|^{1/2}) \le C, \label{eq:bound_01} \\
 & & \tr_{L^2_{\rm
    per}(\Lambda_L)}((1+|\nabla|)(Q_L^{++,L}-Q_L^{--,L})(1+|\nabla|))
    \le C,  \label{eq:bound_02} \\
 & & \tr_{L^2_{\rm
    per}(\Lambda_L)}((1+|\nabla|)Q_L^2(1+|\nabla|)) \le C,
    \label{eq:bound_03} \\
 & & D_{G_L}(\rho_{Q_L}-\nu_L,\rho_{Q_L}-\nu_L) \le C, \label{eq:bound_04}
\end{align}
with $C$ independent of $L$. 

Consider now the sequence of operators $\tilde Q_L:=(i_L)^*Q_Li_L$ acting on $L^2(\R^3)$. It is bounded in $\gS_2(L^2(\R^3))$ by \eqref{eq:bound_03} and since
$\tr_{L^2(\R^3)}(\tilde Q_L^2)=\tr_{L^2_{\rm per}(\Lambda_L)}(Q_L^2)$ by \eqref{egalite_box_space_S_p}.
Hence $\tilde Q_L$ weakly converges, up to
extraction, to some $Q \in \gS_2(L^2(\R^3))$. Similarly, the 
Hilbert-Schmidt operator $R_L:=(i_L)^*Q_L (1+|\nabla|)i_L$ weakly converges 
up to extraction to some $R$ in $\gS_2(L^2(\R^3))$. Let $\phi$ and $\psi$ be in
$C^\infty_0(\R^3)$ and assume that $\mbox{Supp}(\phi) \cup\mbox{Supp}(\psi) \subset \Lambda_L$. Then 
\begin{align*}
\pscal{(i_L)^*Q_L (1+|\nabla|)i_L\phi,\psi}_{L^2(\R^3)} & = \pscal{\tilde Q_L (i_L)^*(1+|\nabla|)i_L\phi,\psi}_{L^2(\R^3)}\\
 & \to_{L\to\ii} \pscal{Q (1+|\nabla|)\phi,\psi}_{L^2(\R^3)}, 
\end{align*}
where we have used that $\tilde Q_L\wto Q$ weakly in $\gS_2$ and that $(i_L)^*(1+|\nabla|)i_L\phi\to(1+|\nabla|)\phi$ strongly in $L^2(\R^3)$ by the third assertion of Lemma \ref{lem:lim_gamma}. Hence $Q(1+|\nabla|)=R \in \gS_2(L^2(\RR^3))$.

Similarly, define the operator $S_L:=(i_L)^*Q_L^{--,L}i_L$ which is nonpositive and yields a bounded sequence in $\gS_1(L^2(\R^3))$ by \eqref{eq:bound_02}. Up to extraction, we may assume that $(S_L)$ converges for the
weak-$\ast$ topology to some $S \in \gS_1(L^2(\R^3))$. To identify the
limit $S$, we compute as above for $\phi,\psi\in L^2(\R^3)$,
\begin{eqnarray*}
\pscal{S_L\phi,\psi}_{L^2(\R^3)} & = & \pscal{(i_L)^*\gamma^0_{{\rm sc},L}Q_L\gamma^0_{{\rm sc},L}i_L\phi,\psi}_{L^2(\R^3)}\\
 & = & \pscal{\tilde Q_L(i_L)^*\gamma^0_{{\rm sc},L}i_L\phi,(i_L)^*\gamma^0_{{\rm sc},L}i_L\psi}_{L^2(\R^3)}. 
\end{eqnarray*}
Using now the first assertion of Lemma \ref{lem:lim_gamma} we obtain
$\lim_{L\to\ii}\pscal{S_L\phi,\psi}_{L^2(\R^3)}=\pscal{Q^{--}\phi,\psi}_{L^2(\R^3)}.$
Hence $Q^{--}=S\in\gS_1$. The same arguments allow to conclude that in fact, $Q \in \cK$. 

Now, let $T_L:=(i_L)^*|H^0_{{\rm sc},L}-{\epsilon_{F}}|^{1/2}Q_L^{--,L}|H^0_{{\rm sc},L}-{\epsilon_{F}}|^{1/2}i_L$ which also defines a bounded sequence in $\gS_1(L^2(\R^3))$. Up to extraction, we may assume that $T_L\wto T$ for the weak-$\ast$ topology of $\gS_1$. Arguing as above and using Lemma \ref{lem:lim_gamma}, we deduce that $T=|H^0_{{\rm per}}-{\epsilon_{F}}|^{1/2}Q^{--}|H^0_{{\rm per}}-{\epsilon_{F}}|^{1/2}$. Now, Fatou's Lemma yields
\begin{multline*}
\liminf_{L\to\ii}\tr_{L^2_{\rm per}(\Lambda_L)}\left(|H^0_{{\rm sc},L}-{\epsilon_{F}}|^{1/2}(-Q_L^{--,L})|H^0_{{\rm sc},L}-{\epsilon_{F}}|^{1/2} \right)\\
 = \liminf_{L\to\ii}\tr_{L^2(\R^3)}(-T_L)\geq \tr_{L^2(\R^3)}\left(|H^0_{{\rm per}}-{\epsilon_{F}}|^{1/2}(-Q^{--})|H^0_{{\rm per}}-{\epsilon_{F}}|^{1/2}\right)
\end{multline*}
This proves that
$$\liminf_{L\to\ii}\tr_{L^2_{\rm per}(\Lambda_L)}\left((H^0_{{\rm sc},L}-{\epsilon_{F}})Q_L\right)\geq \tr_0(H^0_{\rm per}Q)-{\epsilon_{F}}\tr_0(Q).$$

We now study the term involving the density $\rho_{Q_L}$. First, following the proof of Proposition \ref{cont_rho} and using the bounds \eqref{eq:bound_01}--\eqref{eq:bound_03}, we can prove that there exists a constant $C$ such that for all $L$ large enough
$\norm{\rho_{Q_L}}_{L^2_{\rm per}(\Lambda_L)}\leq C.$
Hence, up to extraction, we have $\1_{\Lambda_L}\rho_{Q_L}\wto \rho$ weakly in $L^2(\R^3)$ for some function $\rho\in L^2(\R^3)$. We now introduce an auxiliary function $\rho_L\in L^2(\R^3)$ defined in Fourier space as follows:
$$\widehat{\rho_L}:=\sum_{k\in\frac{2\pi}L\Z^3\setminus\{0\}}\frac{c_{k,L}(\rho_{Q_L})}{|B_k|^{1/2}}\1_{B_k}+\frac{c_{0,L}(\rho_{Q_L})}{|B_0|^{1/2}}\1_{B_0}$$
where for any $k\in(2\pi/L)\Z^3\setminus\{0\}$, $B_k:=B\left(k+\frac{k}{10L|k|},\frac{1}{10L}\right)$ which is chosen to ensure that $1/|k'|\leq1/|k|$ for any $k'\in B_k$, and $B_0:=B\left(0,\frac{1}{10L}\right)$.

Notice that $\rho_L$ is bounded in $L^2(\R^3)$ as we have by definition
$$\int_{\R^3}\rho_L^2=\int_{\R^3}|\widehat{\rho_L}|^2=\sum_{k\in\frac{2\pi}L\Z^3}|c_{k,L}(\rho_{Q_L})|^2=\int_{\Lambda_L}\rho_{Q_L}^2.$$
On the other hand (up to extraction) $\rho_L\wto \rho$ weakly in $L^2(\R^3)$, the same weak limit as $\1_{\Lambda_L}\rho_{Q_L}$. This is easily seen by considering a scalar product against a fixed function $\phi\in\cC_0^\ii(\R^3)$.
Now by the choice of the balls $B_k$, we also have for $L\gg1$
$$D(\rho_L,\rho_L)=4\pi\int\frac{|\widehat{\rho_L}(k')|^2}{|k'|^2}dk'\leq D_{G_L}(\rho_{Q_L},\rho_{Q_L})\leq C.$$
Hence, up to extraction we may assume that $\rho_L\wto \rho$ weakly in $\cC$. Using the regularity of $\widehat{\nu}$, we also deduce that 
$$\liminf_{L\to\ii}\left(-D_{G_L}(\rho_{Q_L},\nu_L)+\frac12 D_{G_L}(\rho_{Q_L},\rho_{Q_L})\right)\geq -D(\rho,\nu)+\frac12 D(\rho,\rho).$$
What remains to be proved is that $\rho=\rho_Q$ where $Q$ is the weak limit of $(i_L)^*Q_Li_L$ obtained above. This will clearly show
$$\liminf_{L\to\ii}E_{{\epsilon_{F}},L}^\nu\geq \cE^\nu(Q)-\epsilon_{F}\tr_0(Q)\geq E^\nu_{\epsilon_{F}}$$
and end the proof of Theorem \ref{thermo_defect}. We identify the limit of $\1_{\Lambda_L}\rho_{Q_L}$ using its weak convergence to $\rho$ in $L^2(\R^3)$.

We start with $\rho_{Q^{++,L}_L}$ and write, fixing some $V\in\cC_0^\ii(\R^3)$ and assuming $L$ large enough for ${\rm Supp}(V)\subset\Lambda_L$,
\begin{equation*}
\int_{\Lambda_L}\rho_{Q^{++,L}_L}V  = \tr_{L^2_{\rm per}(\Lambda_L)}(Q^{++,L}_Li_L(V))= \tr_{L^2(\R^3)}(A_LB_L)\quad \text{with}
\end{equation*}
$$A_L:=(i_L)^*(1+|\nabla|)Q^{++,L}_L(1+|\nabla|)i_L,\quad B_L:=(i_L)^*(1+|\nabla|)^{-1}i_L(V)(1+|\nabla|)^{-1}i_L.$$
The sequence $(A_L)$ is bounded in $\gS_1(L^2(\R^3))$, hence in $\gS_2(L^2(\R^3))$, by \eqref{eq:bound_02} and converges (up to extraction) towards $(1+|\nabla|)Q^{++}(1+|\nabla|)$ weakly in $\gS_2(L^2(\R^3))$ (we proceed as above to identify the weak limit using the fourth assertion of Lemma~\ref{lem:lim_gamma}). By Lemma~\ref{lem:lim_op}, $B_L$ converges  towards $(1+|\nabla|)^{-1}V(1+|\nabla|)^{-1}$ strongly in $\gS_2(L^2(\R^3))$.  We thus obtain
$$\lim_{L\to\ii}\int_{\Lambda_L}\rho_{Q^{++,L}_L}V =\tr_{L^2(\R^3)}(Q^{++}V)=\int_{\R^3}\rho_{Q^{++}}V.$$
Likewise, it can be proved that the weak limit of $\rho_{Q^{--,L}_L}$ is $\rho_{Q^{--}}$.

Let us now treat $\rho_{Q^{+-,L}_L}$ (the other case $\rho_{Q^{-+,L}_L}$ being similar). Following the proof of Proposition \ref{cont_rho}, we write
\begin{align*}
&\int_{\Lambda_L}\rho_{Q^{+-,L}_L}V = \tr_{L^2_{\rm per}(\Lambda_L)}\left(Q^{+-,L}_L[\gamma^0_{{\rm sc},L},i_L(V)]\right) \\
& \quad=  -\frac{1}{4i\pi}\int_{\curv}dz\tr_{L^2_{\rm per}(\Lambda_L)}\left(Q^{+-,L}_L(z-H^0_{{\rm sc},L})^{-1}(\Delta i_L(V)- i_L(V)\Delta)(z-H^0_{{\rm sc},L})^{-1}\right).
\end{align*}
We only detail the argument to pass to the limit in
\begin{align*}
 &\tr_{L^2_{\rm per}(\Lambda_L)}\left(Q^{+-,L}_L(z-H^0_{{\rm sc},L})^{-1} i_L(V)\Delta(z-H^0_{{\rm sc},L})^{-1}\right)\label{form_trace_density}\\
 & \qquad \qquad= \tr_{L^2_{\rm per}(\Lambda_L)}\left(\Delta(z-H^0_{{\rm sc},L})^{-1}Q^{+-,L}_L(z-H^0_{{\rm sc},L})^{-1}(1-\Delta) (1-\Delta)^{-1}i_L(V)\right)\\
 & \qquad\qquad = \tr_{L^2(\R^3)}\left(C_L(i_L)^*(1-\Delta)^{-1}i_L(V)i_L\right)
\end{align*}
with $C_L:=(i_L)^*\Delta(z-H^0_{{\rm sc},L})^{-1}Q^{+-,L}_L(z-H^0_{{\rm sc},L})^{-1}(1-\Delta) i_L$.
One has, up to extraction, $C_L\wto \Delta(z-H^0_{{\rm per}})^{-1}Q^{+-}(z-H^0_{{\rm per}})^{-1}(1-\Delta)$ weakly in $\gS_2(L^2(\R^3))$. To see this, one first remarks that $C_L$ is bounded in $\gS_2(L^2(\R^3))$ and then identifies the weak limit by passing to the limit in $\pscal{C_L\phi,\psi}$ for some fixed $\phi,\psi\in\cC_0^\ii(\R^3)$, using the uniform convergence of the resolvent for $z\in\curv$, as shown in the proof of Lemma \ref{lem:lim_gamma}. Then by Lemma \ref{lem:lim_op} we know that $(i_L)^*(1-\Delta)^{-1}i_L(V)i_L$ converges towards $(1-\Delta)^{-1}V$ strongly in $\gS_2(L^2(\Lambda_L))$, hence we can pass to the limit in the above expression, uniformly in $z\in\curv$. We conclude that
$$\lim_{L\to\ii}\int_{\R^3}\1_{\Lambda_L}\rho_{Q_L}V=\int_{\R^3}\rho_QV$$
for any $V\in\cC_0^\ii(\R^3)$, thus $\rho=\rho_Q$.\qed

\appendix
\section{Proof of Theorem~\ref{thm_per}}\label{proof_thm_periodic}
Our proof uses classical ideas for Hartree-Fock theories. See \cite[Section 4]{LSY} for a very similar setting.
Let us consider a minimizer $\gamma^0_{\rm per}$ of $I^0_{\rm
  per}$ (it is known to exist by \cite[Thm 2.1]{CLL}). First we note
that the periodic potential $V_{\rm per}:=(\rho_{\gamma^0_{\rm per}}-\mu_{\rm
  per}) \star_\Gamma G_1$ is in $L^2_{\rm loc}(\R^3)$. Thus $V_{\rm per}$ 
defines a $\Delta$-bounded operator on $L^2(\R^3)$ with relative bound
  zero (see \cite[Thm XIII.96]{RS}) and therefore $H^0_{\rm
  per}=-\Delta/2+V_{\rm per}$ is self-adjoint on
  $\mathcal{D}(-\Delta)=H^2(\R^3)$ with form domain
  $H^1(\R^3)$. Besides, the spectrum of $H^0_{\rm per}$ is purely
  absolutely continuous, composed of bands as stated in \cite[Thm 1-2]{Thomas} and \cite[Thm XIII.100]{RS}. The Bloch eigenvalues $\lambda_k(\xi)$, $k\geq1$, $\xi\in\Gamma^*$ are known to be real analytic in each fixed direction and cannot be constant with respect to the variable $\xi$. Hence the function
$$C:\kappa\mapsto \sum_{k\geq1} \left|\{\xi\in\Gamma^*\ |\ \lambda_k(\xi)\leq\kappa\}\right|$$
is continuous and nondecreasing on $\R$. The operator $H^0_{\rm per}$ being bounded from below, we have $C\equiv0$ on $(-\ii,\inf\lambda_k(\Gamma^*))$ and it is known \cite[Lemma A-2]{Thomas} that $\lim_{\kappa\to\ii}C(\kappa)=\ii$. We can thus choose a Fermi level ${\epsilon_{F}}$ such that
\begin{equation}
 Z=C({\epsilon_{F}})=\sum_{k\geq1} \left|\{\xi\in\Gamma^*\ |\ \lambda_k(\xi)\leq{\epsilon_{F}}\}\right|.
\label{def_Z}
\end{equation}

Considering a variation $(1-t)\gamma^0_{\rm per}+t\gamma$ for any $\gamma\in\cP^Z_{\rm per}$ and $t\in[0,1]$, we deduce that $\gamma^0_{\rm per}$ minimizes the following linear functional
$$\gamma\in\cP^Z_{\rm per}\mapsto \frac{1}{(2\pi)^3}\int_{\Gamma^*} \tr_{L^2_\xi(\Gamma)}\left((H^0_{\rm per})_\xi\gamma_\xi\right)d\xi,$$
where $H^0_{\rm per}$ is the mean-field operator defined in \eqref{def_H0_per}. 
We subtract the chemical potential ${\epsilon_{F}}$ defined above and introduce the functional 
$$\gamma\in\cP_{\rm per}\mapsto F(\gamma):=\frac{1}{(2\pi)^3}\int_{\Gamma^*} \tr_{L^2_\xi(\Gamma)}\left((H^0_{\rm per}-{\epsilon_{F}})_\xi\gamma_\xi\right)d\xi.$$
Notice that since $\frac{1}{(2\pi)^3}\int_{\Gamma^*} \tr_{L^2_\xi(\Gamma)}\left(\gamma_\xi\right)d\xi=Z$ for any $\gamma\in\cP_{\rm per}^Z$, then $\gamma^0_{\rm per}$ also minimizes $F$ on $\cP_{\rm per}^Z$.

For any $\xi\in\Gamma^*$, we can find orthonormal functions $e_k(\xi,\cdot)\in L^2_\xi(\Gamma)$ such that
\begin{equation}
 (H^0_{\rm per})_\xi=\sum_{k\geq1}\lambda_k(\xi)|e_k(\xi,\cdot)\rangle\langle e_k(\xi,\cdot)|,
\label{Bloch_dec_H_0}
\end{equation}
each function $(\xi,x)\mapsto e_k(\xi,x)$ being measurable on $\Gamma^* \times\Gamma$. Let us now define $\gamma^0\in \cP_{\rm per}$ by
$$(\gamma^0)_\xi(x,y)=\sum_{k\geq1}\delta_k(\xi)e_k(\xi,x) \overline{e_k(\xi,y)},\qquad 
\delta_k(\xi)=\left\{\begin{array}{ll}
1 & \text{if } \lambda_k(\xi)\leq{\epsilon_{F}}\\
0 & \text{if } \lambda_k(\xi)>{\epsilon_{F}}.
\end{array}\right.$$
Saying differently
$\gamma^0=\chi_{(-\ii,{\epsilon_{F}}]}(H^0_{\rm per}).$ Notice ${\epsilon_{F}}$ was chosen to ensure $\gamma^0\in\cP^Z_{\rm per}$.

We now prove that $\gamma^0$ is the unique minimizer of the function $F$ defined above, on the set $\cP_{\rm per}$ without a charge constraint. Since $\gamma^0\in\cP_{\rm per}^Z$, this will prove that $\gamma^0_{\rm per}=\gamma^0$ and that $\gamma^0_{\rm per}$ is the unique minimizer of $F$ on $\cP_{\rm per}$. We write
\begin{eqnarray*}
F(\gamma)-F(\gamma^0) &=& (2\pi)^{-3}\int_{\Gamma^\ast}  \tr_{L^2_\xi(\Gamma)}\left((H^0_{\rm per}-{\epsilon_{F}})_\xi(\gamma-\gamma^0)_\xi\right)d\xi\\
 & =  & \sum_{k\geq1}(2\pi)^{-3}\int_{\Gamma^\ast}  (\lambda_k(\xi)-{\epsilon_{F}})\big(\pscal{\gamma_\xi e_k(\xi),e_k(\xi)}_\xi-\delta_k(\xi)\big)d\xi,
\end{eqnarray*}
where $\pscal{\cdot,\cdot}_\xi$ is the usual inner product of
$L^2_\xi(\Gamma)$. 
Since $0\leq\gamma\leq1$ in $L^2(\R^3)$, we have that $0\leq
\gamma_\xi\leq 1$ on $L^2_\xi(\Gamma)$ and thus $\pscal{\gamma_\xi
  e_k(\xi,\cdot),e_k(\xi,\cdot)}\in[0,1]$, for almost every
$\xi\in\Gamma^*$. Hence, using the definition of $\delta_k(\xi)$,
\begin{equation*}
F(\gamma)-F(\gamma^0) = \sum_{k\geq1}(2\pi)^{-3}\int_{\Gamma^\ast}   |\lambda_k(\xi)-{\epsilon_{F}}|\times\big|\pscal{\gamma_\xi e_k(\xi,\cdot),e_k(\xi,\cdot)}-\delta_k(\xi)\big|d\xi\geq0.
\end{equation*}
This shows that $\gamma^0$ minimizes $F$ on $\cP^Z_{\rm per}$. If now $F(\gamma)=F(\gamma^0)$, then necessarily $\pscal{\gamma_\xi e_k(\xi,\cdot),e_k(\xi),\cdot}=\delta_k(\xi)$ for almost every $\xi\in\Gamma^*$ and any $k\geq1$, the set $\{\xi\in\Gamma^*\ |\ \exists k,\ \lambda_k(\xi)={\epsilon_{F}}\}$
having a Lebesgue measure equal to zero by \cite[Lemma 2]{Thomas}. Using now that the operators $\gamma_\xi$ and $(1-\gamma)_\xi$ are nonnegative, we infer that $\gamma_\xi e_k(\xi)=\delta_k(\xi)e_k(\xi)$ for all $k\geq1$ and almost all $\xi\in\Gamma^*$. Hence $\gamma=\gamma^0$ and $\gamma^0$ is the unique minimizer of $F$. In particular $\gamma^0_{\rm per}=\gamma^0$, i.e. $\gamma^0_{\rm per}$ solves the self-consistent equation \eqref{SCF_per}.

Consider now another minimizer $\gamma$ of the energy $\cE_{\rm per}^0$ on $\cP^Z_{\rm per}$, we recall that $\rho_\gamma=\rho_{\gamma^0_{\rm per}}$ as was shown in \cite{CLL}. Hence the operators $H^0_{\rm per}$ and $\gamma^0$ defined above do not depend on the chosen minimizer. The above argument applied to $\gamma$ shows that $\gamma=\gamma^0=\gamma^0_{\rm per}$, i.e. $\gamma^0_{\rm per}$ is unique.\qed


\subsubsection*{Acknowledgment} We are thankful to \'Eric Séré for useful advice on the model, to Claude Le Bris and Isabelle Catto for helpful comments on the manuscript. We all acknowledge support from the INRIA project \textit{MICMAC} and from the ACI program \textit{SIMUMOL} of the French Ministry of Research. M.L. acknowledges support from the ANR project ``ACCQUAREL''. A.D. has been supported by a grant from Région Ile-de-France.

\end{document}